\documentclass[12pt, a4paper]{article}
\usepackage{amssymb, amsmath, latexsym, epsfig, a4wide, url}

\numberwithin{equation}{section}

\newcommand{\Ker}{\mbox{Ker~}}
\newcommand{\Vhat}{\hat{V}}
\newcommand{\gLtilde}{\tilde{\Lambda}}
\newcommand{\Atilde}{\tilde{A}}

\newcommand{\Etilde}{\tilde{E}}
\newcommand{\Ltilde}{\tilde{L}}
\newcommand{\Ttilde}{\tilde{T}}
\newcommand{\Qtilde}{\tilde{Q}}

\newcommand{\xhat}{\hat{x}}
\newcommand{\cObar}{\overline{\mathcal O}}
\newcommand{\cQbar}{\overline{\mathcal Q}}

\newcommand{\ga}{\alpha}
\newcommand{\gb}{\beta}
\newcommand{\gam}{\gamma}
\newcommand{\gd}{\delta}
\newcommand{\eps}{\epsilon}

\newcommand{\gt}{\theta}

\newcommand{\gk}{\kappa}
\newcommand{\gl}{\lambda}

\newcommand{\go}{\omega}

\newcommand{\Gam}{\Gamma}
\newcommand{\gD}{\Delta}

\newcommand{\gL}{\Lambda}
\newcommand{\gS}{\Sigma}
\newcommand{\gO}{\Omega}

\newcommand{\Abar}{\overline{A}}
\newcommand{\Bbar}{\overline{B}}

\newcommand{\Dbar}{\overline{D}}

\newcommand{\Gbar}{\overline{G}}

\newcommand{\Jbar}{\overline{J}}

\newcommand{\Lbar}{\overline{L}}

\newcommand{\Qbar}{\overline{Q}}

\newcommand{\Tbar}{\overline{T}}

\newcommand{\Xbar}{\overline{X}}
\newcommand{\Ybar}{\overline{Y}}

\newcommand{\abar}{\overline{a}}
\newcommand{\bbar}{\overline{b}}
\newcommand{\cbar}{\overline{c}}

\newcommand{\ibar}{\overline{\imath}}
\newcommand{\jbar}{\overline{\jmath}}
\newcommand{\kbar}{\overline{k}}
\newcommand{\lbar}{\overline{l}}
\newcommand{\mbar}{\overline{m}}
\newcommand{\nbar}{\overline{n}}

\newcommand{\tbar}{\overline{t}}

\newcommand{\vbar}{\overline{v}}

\newcommand{\zbar}{\overline{z}}

\newcommand{\gtbar}{\overline{\theta}}

\newcommand{\mubar}{\overline{\mu}}

\newcommand{\phibar}{\overline{\phi}}

\newcommand{\psibar}{\overline{\psi}}

\newcommand{\Phibar}{\overline{\Phi}}

\newcommand{\gObar}{\overline{\Omega}}

\newcommand{\delbar}{\overline{\partial}}

\newcommand{\cA}{\mathcal A}

\newcommand{\cH}{\mathcal H}

\newcommand{\cM}{\mathcal M}

\newcommand{\cO}{\mathcal O}

\newcommand{\cQ}{\mathcal Q}

\newcommand{\cS}{\mathcal S}

\newcommand{\bC}{\mathbb C}

\newcommand{\bR}{\mathbb R}

\newcommand{\bZ}{\mathbb Z}

\renewcommand{\Im}{\mbox{Im}}

\renewcommand{\Re}{\mbox{Re~}}

\newcommand{\Tr}{\mbox{Tr}}

\newcommand{\be}{\begin{equation}}
\newcommand{\bea}{\begin{eqnarray}}

\renewcommand{\d}{\partial}
\newcommand{\ee}{\end{equation}}
\newcommand{\eea}{\end{eqnarray}}
\newcommand{\half}{\frac{1}{2}}

\newcommand{\ret}{\nonumber \\}
\newcommand{\sk}{\vspace{1 em} \noindent}

\title{A mini-course on topological strings}
\author{Marcel Vonk}
\date{\today}

\begin{document}
 \begin{titlepage}
  \begin{flushright}
   UUITP-06/05\\
   hep-th/0504147
  \end{flushright}
  \vspace{1cm}
  \begin{center}
   {\huge \bf A mini-course on topological strings}
  \end{center}
  \vspace{1cm}
  \begin{center}
   {\large Marcel Vonk} \\[3mm]
   Department of Theoretical Physics \\
   Uppsala University \\
   Box 803 \\
   SE-751 08 Uppsala \\
   Sweden \\[3mm]
   {\tt marcel.vonk@teorfys.uu.se}
  \end{center}
  \vspace{1cm}
  \begin{center}
   {\large \bf Abstract}
  \end{center}
  \noindent
  These are the lecture notes for a short course in topological string
  theory that I gave at Uppsala University in the fall of 2004. The notes are
  aimed at PhD students who have studied quantum field theory and general
  relativity, and who have some general knowledge of ordinary string theory. The
  main purpose of the course is to cover the basics: after a review of the
  necessary mathematical tools, a thorough discussion of the construction of the
  $A$- and $B$-model topological strings from twisted $N=(2,2)$ supersymmetric
  field theories is given. The notes end with a brief discussion on some
  selected applications.
  \vfill
  \begin{flushleft}
   April 2005
  \end{flushleft}
 \end{titlepage}
 
\tableofcontents

 \section{Introduction}
  \subsection{Motivation}
   When asked about the use of topological field theory and topological string
   theory, different string theorists may give very different answers. Some will
   point at the many interesting mathematical results that these theories have
   led to. Others will tell you that topological string theory is a very useful
   toy model to understand more about properties of ordinary string theory in a
   simplified setting. And finally, there will be those pointing at the many and
   often unexpected ways in which topological string theory can be applied to
   derive results in ordinary string theories and other related fields.

   \sk
   Which opinion one prefers is of course a matter of taste, but it is a fact
   that all three of these reasons for studying topological strings have been a
   motivation for people to learn more about the subject in the past
   fifteen years or so. This has of course also resulted in a
   lot of research, and as an inevitable consequence the subject has become
   quite large. For those that have been following the field for fifteen years,
   this is not so much of a problem, but new graduate students and other
   interested physicists and mathematicians can easily lose track in the
   vast amount of literature.

   \sk
   When a subject reaches this stage in its development, automatically the first
   reviews and textbooks start to appear. This has also happened for topological
   field theory and topological string theory; below, I will list a number of good
   references. So what does the current text have to add to this? Hopefully: the
   basics. Many of the texts to be mentioned below are excellent introductions
   for the more advanced string theorists, but topological string theory could
   (and should) be accessible to a much larger audience, including beginning
   graduate students who may not even have seen that much of ordinary string
   theory yet.
  
   \sk
   The aim of this mini-course is to allow those people to discover the subject
   in roughly the same way as it was discovered historically in the first few
   years of its development. The results obtained in this period are often the
   starting point of more advanced texts, so hopefully this review will
   give the reader a solid basis for studying the many later developments in
   the field.
  
  \subsection{Background assumed}
   Of course, no introduction can be completely self-contained, so let me say a
   few words about the background knowledge the reader is assumed to have. These
   notes are written with a mathematically interested physicist reader in mind,
   which probably means that the physically interested mathematician reader
   will find too much dwelling upon mathematical minutiae and too many vague
   statements about physical constructions for his liking. On the other hand,
   even though a lot of the mathematical background is reviewed in some detail
   in the next section, a reader having no familiarity whatsoever with these
   subjects will have a hard time keeping track of the story. In particular,
   some background in differential geometry and topology can be quite useful. A
   reader knowing at least two or three out of four of the concepts
   ``differential form'', ``cohomology class'', ``vector bundle'' and
   ``connection'' is probably at exactly the right level for this course.
   Readers knowing less will simply have to put in some more work; readers
   knowing much more might be looking at the wrong text, but are of course still
   invited to read along. (And for those in this category: comments are very
   welcome!) I also assume some background in linear algebra; in particular, the
   reader should be familiar with the concept of the dual of a vector space.
   Finally, I assume the reader knows what a projective space (such as $\bC
   P^1$) is. If necessary, explanations of such concepts can be found in the
   references that I mention in the section ``Literature'' below.
  
   \sk
   As a physics background, this text requires at least some familiarity with
   quantum field theory, both in the operator and in the path integral formalism
   -- and some intuition about how the two are related. Not many advanced
   technical results will be used; the reader should just have a feeling for
   what a quantum field theory {\em is}. Furthermore, one needs
   some familiarity with fermion fields and the rules of integration and
   differentiation of anti-commuting (Grassmann) variables. Some knowledge of
   supersymmetry is also highly recommended. From general relativity, the
   reader should know the concepts of a metric and the Riemann and Ricci tensors.
   Apart from the last section, the text does not really require any knowledge
   about string theory, though of course some familiarity with this subject
   will make the purpose of it all a lot more transparent. In particular, in
   chapter \ref{sec:topologicalstrings}, it is very useful to know some
   conformal field theory, and to be familiar with how scattering diagrams in
   string theory at arbitrary loop order should (in principle) be calculated.
   
   \sk
   The last section of these notes, section \ref{sec:applications}, deals with
   the applications of topological string theory. Of course, if one does not
   know the subjects that the theory is applied to, one will not be very
   impressed by its applications. Therefore, to appreciate this section, the
   reader needs a bit more knowledge of the current state of affairs in string
   theory. This section falls outside the main aim of the lectures outlined
   above; it is intended as an a posteriori rationale for it all, and as an
   appetizer for the reader to start reading the more advanced introductions and
   research papers on the subject.

  \subsection{Overview}
   In section \ref{sec:mathematics}, I start by reviewing the necessary
   mathematical background. After a short introduction about topology, serving
   as a motivation for the whole subject, I discuss the important technical
   concepts that will be needed throughout the text: homology and cohomology,
   vector bundles and connections. This section is only intended to cover the
   basics, so the more mathematically inclined reader may safely skip it. More
   advanced mathematical concepts, in particular the ones related to Calabi-Yau
   manifolds and their moduli spaces, will be discussed later in the text.

   \sk
   Section \ref{sec:topfieldth} gives a general introduction to topological
   field theories -- starting from a general ``physicist's definition'' and the
   example of Chern-Simons theory, and then defining the special class of
   two-dimensional cohomological field theories that we will be studying for the
   rest of the course.
   
   \sk
   The aim is then to construct actual examples of these cohomological field
   theories, but for this some mathematical background about Calabi-Yau
   threefolds and their moduli spaces is needed, which is introduced in section
   \ref{sec:calabiyau}. Section \ref{sec:twisting} then discusses twisted
   $N=(2,2)$ supersymmetric two-dimensional field theories and the two ways to
   ``twist'' these theories, leading to two very important classes of
   cohomological field theories.
   
   \sk
   In section \ref{sec:topologicalstrings}, the theories obtained in the
   previous section are coupled to gravity -- a procedure which has a scary ring
   to it, but which in the topological case turns out to be quite an easy step
   -- in order to construct topological string theories. The operators of these
   theories are discussed, and the important fact that their results are nearly
   holomorphic in their coupling constants (the so-called holomorphic anomaly
   equation) is explained.
   
   \sk
   After having worked his way through this part of the text, the reader has a
   knowledge which roughly corresponds to the state of the art in 1993. Ten
   years in string theory is a long time, so this can hardly be called an
   up-to-date knowledge, but it should certainly serve as a solid background.
   However, at least a bird's-eye perspective on what followed should be given,
   and this is the aim of section \ref{sec:applications}. Several
   applications are mentioned here, but almost unavoidably, even more are not
   mentioned at all. Still, the choice was simple: I have just mentioned those
   applications of which I had some understanding,
   either because I have worked on them or because I have tried to and
   failed. With my apologies, both to the reader and to all of those physicists
   whose interesting work is not mentioned, I strongly encourage the reader not
   to take this section too seriously, but to see it as an invitation to start
   reading some of the texts I will mention below.
   
   \sk
   Of course, even apart from the applications, I had to omit many subjects in
   keeping the size of these notes within reasonable limits. In my opinion, the
   four most notable parts of the basic theory that would have deserved a more
   thorough treatment are Landau-Ginzburg models, open topological strings,
   topological gravity and the calculation of numerical topological invariants
   of Calabi-Yau spaces. The reader is strongly encouraged to study the
   literature mentioned below to learn more about these intriguing subjects.
   
  \subsection{Literature}
   When mentioning important topological field theory and string theory results
   in these lectures, I will try to include a reference to their original
   derivation. In the current section I will only mention some general texts on
   the subject which can be useful for further or parallel study.
   
   \sk
   Physicist readers wishing to brush up their mathematical background knowledge
   or look up certain mathematical results I mention can have a look in one of
   the many ``topology/geometry for physicists'' books. One example which
   contains all the background needed for this course is the book by M.~Nakahara
   \cite{Nakahara:1990th}. 
   
   \sk
   Unfortunately, there is no such thing as a crash
   course in string theory, but the necessary background can of course be found
   in the classic two-volume monographs by M.~Green, J.~Schwarz and E.~Witten
   \cite{Green:1987sp} and by J.~Polchinski \cite{Polchinski:1998rq}. A good
   review about $N=(2,2)$ supersymmetry in two dimensions can be found in
   chapter 12 of the book by K.~Hori et.\ al \cite{Hori:2003ms}. A very good
   introduction to conformal field theory is given in the lecture notes
   \cite{Schellekens:1996tg} by B.~Schellekens. A more detailed and
   extensive treatment of this subject can also be found in the famous book
   \cite{DiFrancesco:1997nk} by P.~Di Francesco, P.~Mathieu and
   D.~S\'en\'echal.
   
   \sk 
   My story will roughly follow the development of the basics of topological
   string theory as they were laid out in a series of beautiful papers by
   E.~Witten\footnote{Actually, a complete list of these papers would be at least
   twice as long; one can get a good impression by searching on ``Witten'' and
   ``topological'' on SPIRES. The papers mentioned above more or less cover the
   material contained in this course. As a second remark, I only mention a
   single author for pedagogical reasons; the reader should not get the false
   impression that there were no contributions by others in the early days! See
   the book by K.~Hori et al.\ \cite{Hori:2003ms} for for a more historical
   overview of the literature.} \cite{Witten:1988xi} - \cite{Witten:1992fb}, and
   more or less completed in a  seminal paper by M.~Bershadsky, S.~Cecotti,
   H.~Ooguri and C.~Vafa \cite{Bershadsky:1993cx}. The papers by Witten are very
   readable, and I recommend all serious students to read them to find many of
   the details that I had to leave out here. Of the ones mentioned,
   \cite{Witten:1990bs} and \cite{Witten:1991zz} give a nice general overview.
   The paper by Bershadsky et al., though more technical, is also a
   classic, and many of the open questions in it still serve as a motivation for
   research in topological string theory. Together, these papers should serve as
   a more thorough substitute for the material covered below.

   \sk
   Many reviews about topological field theories exist; the ones I used in
   preparing this course are by S.~Cordes, G.~Moore and S.~Ramgoolam 
   \cite{Cordes:1994fc}, R.~Dijkgraaf \cite{Dijkgraaf:1991qh} and J.~Labastida
   and C.~Lozano \cite{Labastida:1997pb}. Of course, reviews about topological
   {\em string} theory 
   usually also contain a discussion of topological field theory. The first one
   of these, as far as I am aware, and still a very useful reference, is the
   review by R.~Dijkgraaf, E.~Verlinde and H.~Verlinde \cite{Dijkgraaf:1990qw},
   written in 1990. From the same period, there are also the papers
   by E.~Witten mentioned above. The reader interested in more recent
   applications of topological string theories may consult the two recent
   reviews by M.~Mari\~no \cite{Marino:2004eq} and by A.~Neitzke and C.~Vafa
   \cite{Neitzke:2004ni}. The first is a very readable review about the
   relations between topological string theories and matrix models which I will
   briefly touch upon in section \ref{ssec:matrixmodels}; the second gives a
   good overview of the developments in the field after 1993, and may serve as a
   good starting point for the reader who has finished reading my notes and is
   hungry for more. Finally, a book about topological strings and several of
   its applications by M.~Mari\~no \cite{Marino:2005bk} will appear in the 
   course of this year.
   
   \sk
   For those who still want to dig deeper, there is the 900-page
   book \cite{Hori:2003ms} by K.~Hori, S.~Katz, A.~Klemm, R.~Pandharipande,
   R.~Thomas, C.~Vafa, R.~Vakil and E.~Zaslow, which discusses topological
   string theory from the point of view of mirror symmetry. This book is a must
   for anyone who wants to have a complete knowledge of the field. In these
   notes, I have tried to limit the number of references by only including
   those that I thought would be useful for the reader in understanding the
   main story line. A much more complete overview of the literature can also be
   found in the book by Hori et al.
   
  \subsection{Acknowledgements}
   These notes are based on a series of five two-hour blackboard lectures (with
   a break, don't worry) for the PhD students at Uppsala University, and on a
   condensed three times forty-five minute powerpoint version\footnote{See
   \url{http://www.teorfys.uu.se/people/marcel} for the powerpoint slides.}
   during the 19th Nordic Network Meeting in Uppsala. I want to thank both
   audiences for many useful comments and questions, and the organizers of the
   Nordic Network Meeting for the opportunity to speak at this conference.
   Furthermore, Niklas Johansson proof-read the entire manuscript, and Arthur
   Greenspoon edited the final version of these notes. I am very grateful to
   both of them for many useful comments.

 \section{Mathematical background}
  \label{sec:mathematics}
  In this section, I review some mathematical concepts which will be
  important in the rest of these notes. The material will be very basic; the
  more advanced mathematical topics will be explained in the later sections
  when they are needed for the first time. Readers with a background in
  differential geometry and topology may therefore safely skip to section
  \ref{sec:topfieldth} and return to this section for reference and notation if
  necessary.

  \subsection{Topological spaces}
   The notion of ``continuity'' is a very important concept, both in physics
   and mathematics. Let us recall when a function $f: \bR \to \bR$ is
   continuous. Intuitively speaking, $f$ is continuous near a point $x$ in its
   domain if its value does not jump there. That is, if we just take $\gd x$
   to be small enough, the two function values $f(x)$ and $f(x + \gd x)$ should
   approach each other arbitrarily closely. In more rigorous terms, this leads
   to the following definition:

   \begin{quote}
    {\em A function $f: \bR \to \bR$ is continuous at $x \in \bR$ if for all
    $\eps > 0$, there exists a $\gd > 0$ such that for all $y \in \bR$ with
    $|y-x|<\gd$, we have that $|f(y) - f(x)|< \eps$. The whole function is
    called continuous if it is continuous at every point $x$.}
   \end{quote}

   \noindent
   For the reader who has never before seen this definition in its full glory,
   it might take some time to absorb it. However, it then becomes quite clear
   that this is nothing but a very precise way of formulating the above
   intuitive idea. Still, there is something strange about this definition.
   Note that it depends crucially on the notion of {\em distance}: both in
   $|y-x|$ and in $|f(y)-f(x)|$, the distance between two points in $\bR$
   appears. In fact, the whole reason for using $|y-x|$ in the definition,
   instead of the perhaps more intuitive $x + \gd x$, is that in
   this form the definition can easily be generalized to maps between spaces
   where addition and subtraction are not defined -- simply by replacing
   $|y-x|$ and $|f(y) - f(x)|$ by some other notions of distance on the
   domain and image of the map $f$.

\begin{figure}[ht]
    \begin{center}
     \includegraphics[height=5cm]{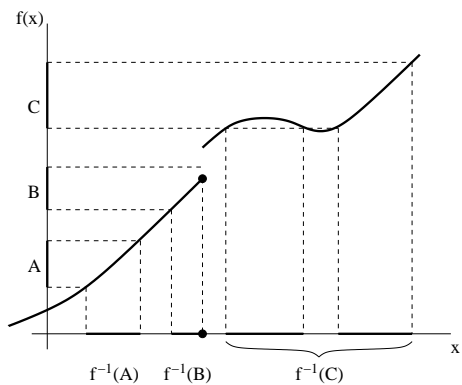}
    \end{center}
    \caption{A discontinuous function, with three open intervals $A, B$, and $C$
    and their inverse images. Note that the inverse image $f^{-1}(B)$ is not an
    open interval: the end point marked by the dot is included in the interval.}
    \label{fig:function}
   \end{figure}

   \sk
   But do we really need a notion of distance to understand continuity? Our
   intuition seems to indicate otherwise: a continuous map is a map where one
   does not have to ``tear apart'' the domain to obtain the image, and this
   seems to be a notion which is not intimately related to the notion of a
   distance. Can we cook up another definition of continuity where the notion
   of distance is not used? We can! Recall that an {\em open} subset of $\bR$
   is a collection of intervals $(a,b)$, with $a<b$, where the end points $a$
   and $b$ are not included, and that a {\em closed} subset consists of intervals
   $[a,b]$, with $a \leq b$, where the end points {\em are}
   included\footnote{The mathematically inclined reader may note that these
   definitions are not quite correct if we consider collections of infinitely
   many of those intervals, but for our purposes it will be enough to only
   consider the finite case.}. Of course, not every subset of $\bR$ is of one
   of those types; consider for example the ``half-open'' interval $[a,b)$. Now
   consider the function $f: \bR \to \bR$ which is drawn in figure
   \ref{fig:function}.

   \sk Note that the two open sets $A$ and $C$ in the image of $f$ also have
   open inverse images $f^{-1}(A)$ and $f^{-1}(C)$. On the other hand, as a
   consequence of the discontinuity of $f$, the open set $B$ has an inverse
   image $f^{-1}(B)$ which is {\em not} an open set! With a little thought, one
   can convince oneself that this is in fact a general property, and thus we
   arrive at the following definition of continuity:

   \begin{quote}
    {\em A function $f: \bR \to \bR$ is continuous if all open sets $S \subset
    \bR$ have an inverse image $f^{-1}(S)$ which is also an open set.}
   \end{quote}

   \noindent It is a nice exercise to try and prove that the above two
   definitions are indeed equivalent. That is, assuming that a certain function
   $f$ satisfies the first definition, prove that it satisfies the second
   definition, and conversely.

   \sk The moral of the story above is that it is possible to construct a
   mathematical theory of continuous mappings {\em without} the need for a
   notion of distance on the spaces involved. This mathematical theory is what
   is called {\em topology}, and it forms the underpinning of everything we
   will have to say in these notes. In topology, the world is made of a very
   flexible kind of rubber: what matters is not the exact shape or size of
   objects, but only whether they can continuously be deformed into one another
   or not. Topologically, there is no difference between a garden hose and the
   ring around your finger, or between a football and a closed suitcase. The
   first two objects are topologically different from the last two, though.

   \sk Mathematicians like to have their definitions wide and applicable. This
   especially holds for the definition of what is called a ``topological
   space''. As we saw above, all that we need for a definition of continuity is
   the notion of a set and its open subsets. Therefore, the mathematical
   definition of a topological space is that it can be {\em any} set $X$
   equipped with a suitable collection of subsets which by definition are called
   its ``open sets''. Here, ``suitable'' means that the collection of open
   subsets has to satisfy certain axioms, such as the rule that the union of any
   two open subsets is again in the collection of open subsets, etc.

   \sk This definition is much too broad for our purposes, and anyway, we
   will not be taking a mathematically rigorous, axiomatic approach towards
   topological string theory. In these notes, the topological spaces that we
   will encounter will always be real or complex manifolds or orbifolds -- that
   is, spaces which locally look like $\bR^n$, $\bR^n/\Gam$, $\bC^m$ or
   $\bC^m/\Gam$, with $\Gam$ some finite group such as $\bZ_k$. For such spaces,
   the everyday intuition of continuity of maps will suffice.

  \subsection{Manifolds}
\begin{figure}[ht]
    \begin{center}
     \includegraphics[height=5cm]{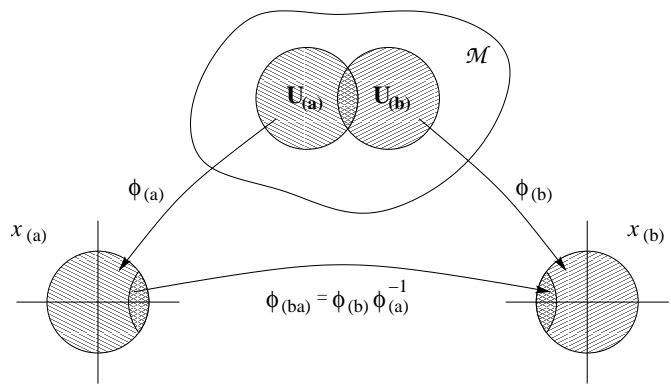}
    \end{center}
    \caption{The manifold $\cM$ is covered by patches $U_{(a)}$, parameterized
    by coordinates $x_{(a)} \in \bR^n$. To go from one coordinate to another,
    one uses the transition functions $\phi_{(ba)}$.}
    \label{fig:manifold}
   \end{figure}

   It might be useful to very briefly remind the reader of the definition of a
   manifold. An $n$-dimensional real manifold is a topological space which
   locally looks like $\bR^n$. That is, one can cover the space with open
   subsets $U_{(a)}$ called (local) charts, or in the physics literature
   ``patches'', and map these patches to open subsets of $\bR^n$ by
   continuous and invertible maps $\phi_{(a)}: U_{(a)} \to \bR^n$.
   Prescribing such a map can be viewed as defining coordinates on the patch
   $U_{(a)}$. We will also denote these coordinates by $x^i_{(a)}$, with $1 \leq
   i \leq n$. On overlaps $U_{(a)} \cap U_{(b)}$, one can go from coordinates
   $x_{(a)}$ to $x_{(b)}$ by
   \be
    x_{(b)} = \phi_{(b)} \phi^{-1}_{(a)} (x_{(a)}).
   \ee
   This is illustrated in figure \ref{fig:manifold}. We denote the ``transition
   functions'' $\phi_{(b)} \phi^{-1}_{(a)}$ by $\phi_{(ba)}$. Clearly,
   $\phi_{(ab)} = \phi_{(ba)}^{-1}$.

   \sk Conversely, one can construct a manifold by giving a parametrization
   $x_{(a)}$ for a set of local charts $U_{(a)}$, and giving the transition
   functions $\phi_{(ba)}$ between the charts. However, these transition
   functions cannot be arbitrary functions. The reason is that in regions where
   three charts overlap, say $U_{(a)}, U_{(b)}$ and $U_{(c)}$, one should be
   able to go from coordinates $x_{(a)}$ to coordinates $x_{(b)}$, then from
   $x_{(b)}$ to $x_{(c)}$, and finally from $x_{(c)}$ to $x_{(a)}$ again. Since
   in each step we are simply switching to different coordinates for the same
   point $P$ on the manifold, in the end we should be back at the same
   coordinates as the ones we started with. That is, one should have
   \be
    \phi_{(ac)} \phi_{(cb)} \phi_{(ba)} = 1
    \label{eq:cocycle}
   \ee
   on triple overlaps $U_{(a)} \cap U_{(b)} \cap U_{(c)}$. This condition on the
   transition functions is called the {\em cocycle condition}, and it can be
   shown that this is the only condition one needs to put on the transition
   functions to obtain a well-defined manifold. (In particular, there are no
   extra conditions on overlaps of four or more patches, since all of these
   follow from (\ref{eq:cocycle}).)

   \sk Of course, the above construction only guarantees that the transition
   functions we find are continuous and invertible. In practice, we would like
   to be able to differentiate functions on our manifold as often as we like,
   and hence one should require the transition functions to be $C^{\infty}$
   (that is, infinitely many times differentiable) as well. The manifolds
   that allow such transition functions are called $C^{\infty}$-manifolds; we
   will always assume this property when we speak of a manifold.

   \sk All of this is straightforwardly generalized to complex manifolds: here,
   the coordinate functions $\phi_{(a)}$ should map $U_{(a)}$ to $\bC^m$, and
   the transition functions must be holomorphic -- that is, on an overlap
   $U_{(a)} \cap U_{(b)}$ the coordinates $z_{(b)}$ should be analytic
   functions of $z_{(a)}$, and hence be independent of $\zbar_{(a)}$. This
   condition ensures that holomorphic objects -- that is, globally defined
   objects such as functions which only depend on $z$ but not on $\zbar$ -- can
   be well-defined over the whole manifold.

  \subsection{Topological invariants}
   The main mathematical question that topology addresses is: when are two
   spaces topologically equivalent? That is: given a topological space $A$ and
   a topological space $B$, is there a continuous and invertible map $f: A \to
   B$? If such a map exists, it is called a {\em homeomorphism}, and the two
   spaces $A$ and $B$ are called {\em homeomorphic}. In the case that these maps
   are differentiable, as we will usually assume, we speak of {\em
   diffeomorphisms}.

\begin{figure}[ht]
    \begin{center}
     \includegraphics[height=5cm]{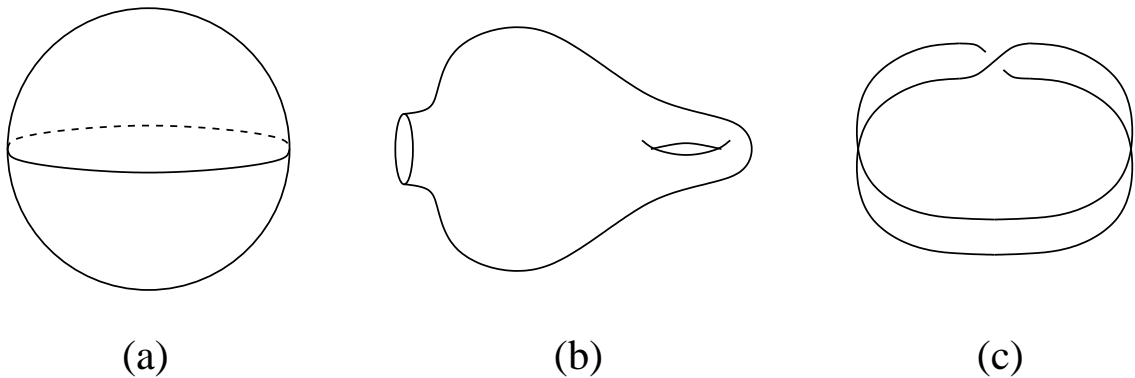}
    \end{center}
    \caption{(a) The sphere is an orientable manifold without handles or
    boundaries. (b) An orientable manifold with one boundary and one handle. (c)
    The M\"obius strip: an unorientable manifold with one boundary and no
    handles.}
    \label{fig:handles}
   \end{figure}

   \sk In general, it is incredibly hard to decide whether or not two spaces are
   homeomorphic. This statement may come as a surprise, but anyone who has ever
   tried to put up the lights on a Christmas tree knows that topological issues
   can be highly nontrivial. Restricting to the topology of $d$-dimensional,
   compact (that is, closed and bounded) and connected manifolds, the only cases
   in which we have a complete understanding of topology are $d=0,1$ and $2$.
   The only compact and connected $0$-dimensional manifold is a point. A
   $1$-dimensional compact and connected manifold can either be a line element
   or a circle, and it is intuitively clear (and can quite easily be proven)
   that these two spaces are topologically different. In two dimensions, there
   is already an infinite number of different topologies: a two-dimensional,
   compact and connected surface can have an arbitrary number of handles and
   boundaries, and can either be orientable or non-orientable (see figure
   \ref{fig:handles}). Again, it is intuitively quite clear that two surfaces
   are not homeomorphic if they differ in one of these respects.  On the other
   hand, it can be proven that any two surfaces for which these data are the
   same\footnote{To be precise, instead of the orientability of the manifold,
   one has to include the number of ``crosscaps'' in the data. A crosscap is
   inserted in a manifold by cutting out a small circular piece and identifying
   opposite points on this circle, making the manifold unorientable. It can be
   shown that inserting three crosscaps is equivalent to inserting a handle and
   a single crosscap, so for a unique description one should only consider
   surfaces with 0, 1 or 2 crosscaps.} can be continuously mapped to one
   another, and hence this gives a complete classification of the possible
   topologies of such surfaces.

   \sk A quantity such as the number of boundaries of a surface is called a
   {\em topological invariant}. A topological invariant is a number, or more
   generally any type of structure, which one can associate to a topological
   space, and which does not change under continuous mappings. Topological
   invariants can be used to distinguish between topological spaces: if two
   surfaces have a different number of boundaries, they can certainly not be
   topologically equivalent. On the other hand, the knowledge of a topological
   invariant is in general not enough to decide whether two spaces are
   homeomorphic: a torus and a sphere have the same number of boundaries
   (zero), but are clearly not homeomorphic. Only when one has some {\em
   complete set} of topological invariants, such as the number of handles,
   boundaries and crosscaps (see footnote) in the two-dimensional case, is it
   possible to determine whether or not two topological spaces are homeomorphic.

   \sk In more than two dimensions, many topological invariants are known, but
   for no dimension larger than two has a complete set of topological invariants
   been found. In fact, in four or more dimensions\footnote{In three
   dimensions, it is generally believed that a finite number of countable
   invariants would suffice for compact manifolds, but as far as I am aware this
   is not rigorously proven, and in particular there is at present no generally
   accepted construction of a complete set. A very interesting and intimately
   related problem is the famous Poincar\'e conjecture, stating that if a
   three-dimensional manifold has a certain set of topological invariants called
   its ``homotopy groups'' equal 
   to those of the three-sphere $S^3$, it is actually homeomorphic to the
   three-sphere.} such a set would consist of an uncountably infinite number of
   invariants! (I assume here that such an invariant is itself in a countable
   set, such as the set of all integers.) A general classification of topologies
   is therefore very hard to obtain, but even without such a general
   classification, each new invariant that can be constructed gives us a lot of
   interesting new information. For this reason, the construction of topological
   invariants is one of the most important issues in topology. The nature of
   these topological invariants can be quite diverse: in the case of two
   dimensions, we saw two topological invariants that were positive numbers (the
   numbers of handles and boundaries), and one that was an element of the set
   $\{0,1,2\}$: the number of necessary crosscaps. But topological invariants
   can also be polynomials, groups, or elements of a given group, for example.

   \sk As we will see in these notes, topological field and string theories can
   be used to construct certain topological invariants. Conversely,
   topologically invariant structures play an important role in the
   construction of topological field and string theories. The main examples of
   this last statement are homology and cohomology: two topological invariants
   which assign a group to a certain manifold. In the next section, we will
   briefly review these structures.

  \subsection{Homology and cohomology}
   \subsubsection{Differential forms}
    Let us begin by recalling the notion of a $p$-form. A $p$-form is the
    mathematical equivalent of an antisymmetric tensor field living on a
    manifold $\cM$. Consider a patch of $\cM$ parameterized by $n$ real
    coordinates $x^i$. A $p$-form can now be written as\footnote{To make the
    formulas more readable and to focus on the structure, we will ignore factors
    of $p!$ in this section. Also, here and everywhere else in these notes we
    use the summation convention, where we sum over repeated indices.}
    \be
     B(x) \equiv B_{i_1 \ldots i_p}(x) \, dx^{i_1} \wedge dx^{i_2} \wedge \cdots
     \wedge dx^{i_p}.
    \ee
    Here, the $dx^i$ are so-called {\em cotangent vectors} at the point $x$.
    (For the definition of the $\wedge$-product, see below.) Formally, a
    cotangent vector is a linear map from the tangent vector space at $x$ to the
    real numbers. The cotangent vector $dx^i$ is the specific linear map that
    maps the unit tangent vector in the $x^i$-direction to 1, and all tangent
    vectors in the other coordinate directions to 0. From this, we see that a
    general linear map from the tangent vector space at $x$ to $\bR$ can be written as
    \be
     A(x) = A_i(x) \, dx^i.
    \ee
    In other words, the space of all one-forms is precisely the space of all
    cotangent vector fields! Note that at every point $x$, the possible values of
    a one-form make up an $n$-dimensional vector space with coordinates
    $A_i(x)$. As we will see in the next section, when we glue together all of
    these vector spaces for different $x$, we get what is called a vector bundle
    -- in this particular case the ``cotangent bundle''.

    \sk The $dx^i$-notation of course suggests a relation to integrals. This can
    be made precise as follows. Suppose we have a one-dimensional closed curve
    $\gam$ inside $\cM$. Let us show that a one-form $A(x)$ on $\cM$ can 
    then naturally be integrated along $\gam$. Parameterize $\gam$ by
    a parameter $t$, so that its coordinates are given by $x^i(t)$. At time $t$,
    the velocity $dx(t)/dt$ is a tangent vector to $\cM$ at $x(t)$. One
    can insert this tangent vector into the linear map $A(x)$ to get a real
    number. By definition, inserting the vector $dx(t)/dt$ into the linear map $dx^i$ simply gives
    the component $dx^i(t)/dt$. Doing this for every $t$, we can then integrate over $t$:
    \be
     \int \left( A_i(x(t)) \frac{dx^i}{dt} \right) dt
     \label{eq:1dint}
    \ee
    It is clear that this expression is independent of the parametrization in
    terms of $t$. Moreover, from the way that tangent vectors transform, one can
    deduce how the linear maps $dx^i$ should transform, and from this how the
    coefficients $A_i(x)$ should transform. Doing this, one sees that the above
    expression is also invariant under changes of coordinates on $\cM$.
    Therefore, a one-form can unambiguously be integrated over a
    curve in $\cM$. We write such an integral as
    \be
     \int_\gam A_i(x) \, dx^i
     \label{eq:1dintshorter}
    \ee
    or even shorter, as
    \be
     \int_\gam A
    \ee
    Of course, when $\cM$ is itself one-dimensional, (\ref{eq:1dint}) gives
    precisely the ordinary integration of a function $A(x)$ over $x$, so the
    notation (\ref{eq:1dintshorter}) is indeed natural.

    \sk Similarly, one would like to define a two-form as something which can
    naturally be integrated over a two-dimensional surface in $\cM$. At a
    specific point $x$, the tangent plane to such a surface is spanned
    by a pair of tangent vectors, $(v^1, v^2)$. So to generalize the
    construction of a one-form, we should give a bilinear map from such a pair
    to $\bR$. The most general form of such a map is
    \be
     B_{ij}(x) \, dx^i \otimes dx^j,
     \label{eq:almosttwoform}
    \ee
    where the tensor product of two cotangent vectors acts on a pair of vectors
    as follows:
    \be
     dx^i \otimes dx^j \, (v^1, v^2) = dx^i(v^1) \, dx^j(v^2). \ee On the right hand
    side of this equation, one multiplies two ordinary numbers obtained by
    letting the linear map $dx^i$ act on $v^1$, and $dx^j$ on $v^2$.

    \sk The bilinear map (\ref{eq:almosttwoform}) is slightly too general to give
    a good integration procedure, though. The reason is that we would like the
    integral to change sign if we change the orientation of integration, just
    like in the one-dimensional case. In two dimensions, changing the
    orientation simply means exchanging $v^1$ and $v^2$, so we want our bilinear
    map to be antisymmetric under this exchange. This is achieved by defining a
    two-form to be
    \bea
     B & = & B_{ij}(x) \, \left( dx^i \otimes dx^j - dx^j \otimes dx^i \right)
     \ret
     & \equiv & B_{ij}(x) \, dx^i \wedge dx^j
    \eea
    We now see why a two-form corresponds to an {\em antisymmetric} tensor
    field: the symmetric part of $B_{ij}$ would give a vanishing contribution to
    $B$. Now, parameterizing a surface $\gS$ in $\cM$ with two coordinates $t_1$
    and $t_2$, and reasoning exactly like we did in the case of a one-form, one
    can show that the integration of a two-form over such a surface is indeed
    well-defined, and independent of the parametrization of both $\gS$ and
    $\cM$.

    \sk For $p$-forms of higher degree $p$, the construction goes in exactly the
    same way, where now the $\wedge$-product is a totally antisymmetric tensor
    product of $p$ cotangent vectors. In fact, one can use the wedge product to
    multiply an arbitrary $p$-form with an arbitrary $q$-form in the same way:
    \be
     B^{(1)} \wedge B^{(2)} = B^{(1)}_{[i_1 \ldots i_p} \, B^{(2)}_{i_{p+1}
     \ldots i_{p+q}]} \, dx^{i_1} \wedge \cdots \wedge dx^{i_{p+q}},
    \ee
    where the square brackets denote total antisymmetrization in the indices.
    (Of course, we do not strictly need this antisymmetrization in the formula,
    since the wedge product of the $dx^i$ is already antisymmetric.) Note that
    because of the antisymmetry, the maximal degree that a nonzero form can have
    is the dimension $n$ of $\cM$.

   \subsubsection{De Rham cohomology}
   There is a natural notion of taking derivatives of $p$-forms. Since
   taking an $x^i$-derivative of an antisymmetric tensor adds an extra lower
   index, it is natural to construct a derivative that maps $p$-forms to
   $(p+1)$-forms. This derivative $d$ is called the {\em exterior derivative},
   and it is defined as:
   \be
    d B \equiv \frac{\d B_{i_1 \ldots i_p}}{\d x^j} \, dx^j \wedge dx^{i_1} \wedge
    \cdots \wedge dx^{i_p}.
   \ee
   Because of the antisymmetry properties of the wedge product, we have that
   \be
    d^2 = 0.
   \ee
   This simple formula leads to the important notion of {\em cohomology}. Let
   us try to solve the equation $dB = 0$ for a $p$-form $B$. A trivial solution
   is $B=0$. From the above formula, we can actually easily find a much larger
   class of trivial solutions: $B = dA$ for a $(p-1)$-form $A$. More generally,
   if $B$ is any solution to $dB=0$, then so is $B + dA$. We want to consider
   these two solutions as equivalent:
   \be
    B \sim B + B' \qquad \mbox{if}  \qquad B' \in \Im~ d,
   \ee
   where $\Im~d$ is the image of $d$, that is, the collection of all $p$-forms
   of the form $dA$. (To be precise, the image of $d$ of course contains
   $q$-forms for any $0 < q \leq n$, so we should restrict this image to the
   $p$-forms for the $p$ we are interested in.) The set of all $p$-forms which
   satisfy $dB=0$ is called the 
   kernel of $d$, denoted $\Ker d$, so we are interested in $\Ker d$ up to the
   equivalence classes defined by adding elements of $\Im~d$. (Again, strictly
   speaking, $\Ker d$ consists of $q$-forms for several values of $q$, so we
   should restrict it to the $p$-forms for our particular choice of $p$.) This
   set of equivalence classes is called $H^p(\cM)$, the $p$-th cohomology group
   of $\cM$:
   \be
    H^p(\cM) = \frac{\Ker d}{\Im~d}.
    \label{eq:cohomology}
   \ee
   Clearly, $\Ker d$ is a group under addition: if two forms $B^{(1)}$ and
   $B^{(2)}$ satisfy $dB^{(1)}=dB^{(2)}=0$, then so does $B^{(1)}+B^{(2)}$.
   Moreover, if we change $B^{(i)}$ by adding some $dA^{(i)}$, the result of
   the addition will still be in the same cohomology class, since it differs
   from $B^{(1)}+B^{(2)}$ by $d(A^{(1)}+A^{(2)})$. Therefore, we can view this
   addition really as an addition of cohomology classes: $H^p(\cM)$ is itself
   an additive group. Also note that if $B^{(3)}$ and $B^{(4)}$ are in the same
   cohomology class (that is, their difference is of the form $dA^{(3)}$), then
   so are $c B^{(3)}$ and $c B^{(4)}$ for any constant factor $c$. In other
   words, we can multiply a cohomology class by a constant to obtain another
   cohomology class: cohomology classes actually form a {\em vector space}.

   \sk Now, we come to the reason why cohomology groups are so interesting for
   us. One important mathematical result is that for compact manifolds, the
   vector spaces $H^p(\cM)$ are in fact finite-dimensional. Their dimensions
   are called the {\em Betti numbers} $b^p$. Since there are no $p$-forms for
   $p>n$, we immediately see that $H^p(\cM) = 0$ and hence $b^p=0$ for $p>n$,
   so this gives us a set of $n+1$ numbers which are possibly nonzero. Just
   like for integrals of $p$-forms, one can show that the exterior derivative
   of a $p$-form is independent of the coordinates we use, and hence of the
   notion of a distance on the manifold. As a result, the whole construction of
   the cohomology groups does not depend on such a notion: the cohomology
   groups are topological invariants! Since any two vector spaces of the same
   dimension are isomorphic, the only nontrivial information contained in
   $H^p(\cM)$ is really its dimension $b^p$, so one can just as well consider
   the set of Betti numbers $b^p$ to be a set of $n+1$ topological invariants.

   \sk Note that so far, the whole construction only depended on the following
   properties:
   \begin{itemize}
    \item
     $d^2 = 0$
    \item
     $d$ acts linearly on a {\em graded} vector space. That is, it maps objects
     of a certain degree $p$ to objects of degree $p+1$.
    \item
     The objects on which $d$ acts, and the way in which it acts, are
     independent of the choice of coordinates on $\cM$. (Of course, the
     construction should also be independent of any other structure on $\cM$,
     and in particular it should not involve a choice of metric.)
   \end{itemize}
   Therefore, for any other operator with these properties, we can create a
   similar structure and derive topological invariants from it. Many
   constructions of this type are known in mathematics and in physics, and they
   all go by the name of cohomology. The particular type of cohomology we
   introduced above is called ``de Rham cohomology''. Of course, the
   requirement that $d$ maps objects of degree $p$ to objects of degree $p+1$
   can just as well replaced by the requirement that it maps them to objects of
   degree $p-1$.

   \sk In general cohomology theories, objects of the form $dA$ are called
   ``exact'', and objects for which $dB = 0$ are called ``closed''. We will use
   this terminology throughout these notes.

   \subsubsection{Homology}
   \label{sec:homology}
\begin{figure}[ht]
    \begin{center}
     \includegraphics[height=5cm]{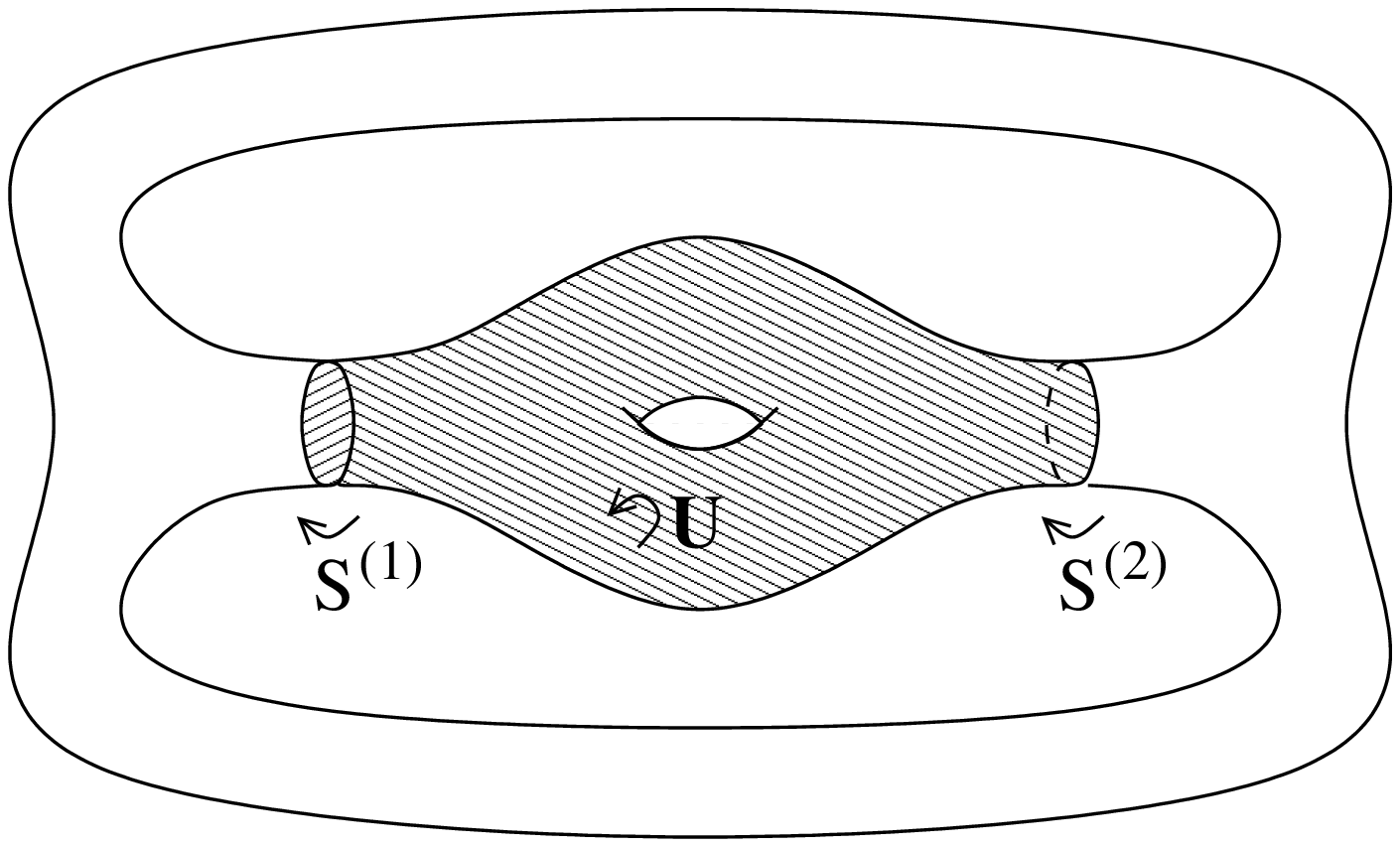}
    \end{center}
    \caption{The 1-dimensional submanifolds $S^{(1)}$ and $S^{(2)}$ represent
    the same homology class, since their difference is the boundary of $U$.}
    \label{fig:homology}
   \end{figure}

   Another operator of the kind introduced above is the boundary operator
   $\gd$, which maps compact submanifolds of $\cM$ to their
   boundary\footnote{Again, there is a more precise mathematical way of dealing
   with this boundary operator, called ``singular homology'', but for our
   purposes the intuitive picture of a boundary of a compact submanifold
   suffices.}. Here, $\gd S = 0$ means that a submanifold $S$ of $\cM$ has no
   boundary, and $S = \gd U$ means that $S$ is itself the boundary of some
   submanifold $U$. It is intuitively clear, and not very hard to prove, that
   $\gd^2=0$: the boundary of a compact submanifold does not have a boundary
   itself. That the objects on 
   which $\gd$ acts are independent of its coordinates is also clear. So is the
   grading of the objects: the degree $p$ is simply the dimension of the
   submanifold $S$. (Note that here we have an example of an operator that maps
   objects of degree $p$ to objects of degree $p-1$ instead of $p+1$.) What is less
   clear is that the collection of submanifolds actually forms a vector space,
   but one can of course always {\em define} this vector space to consist of
   formal linear combinations of submanifolds, and this is precisely how one
   proceeds. The $p$-dimensional elements of this vector space are called
   ``$p$-chains''. One should think of $-S$ as $S$ with its orientation
   reversed, and of the sum of two disjoint sets, $S^{(1)} + S^{(2)}$, as their
   union. The equivalence classes constructed from $\gd$ are called {\em
   homology} classes.
   For example, in figure \ref{fig:homology}, $S^{(1)}$ and $S^{(2)}$ both
   satisfy $\gd S = 0$, so they are elements of $\Ker \gd$. Moreover, it is
   clear that neither of them separately can be viewed as the boundary of
   another submanifold, so they are not in the trivial homology class $\Im~\gd$.
   However, the boundary of $U$ is $S^{(1)} - S^{(2)}$. (The minus sign in front
   of $S^{(2)}$ is a result of the fact that $S^{(2)}$ itself actually has the
   wrong orientation to be considered a boundary of $U$.) This can be written as
   \be
    S^{(1)} - S^{(2)} = \gd U,
   \ee
   or equivalently
   \be
    S^{(1)} = S^{(2)} + \gd U,
   \ee
   showing that $S^{(1)}$ and $S^{(2)}$ are in the same homology
   class\footnote{I chose this example because it is a counterexample to the
   common misconception that $S^{(1)}$ and $S^{(2)}$ are homologous if and only
   if they can be continuously deformed into each other.}.

   \sk The cohomology groups for the $\gd$-operator are called {\em homology
   groups}, and denoted by $H_p(\cM)$, with a lower index. (Of course
   historically, as can be seen from the terminology, homology came first and
   cohomology was related to it in the way we will discuss below. However,
   since the cohomology groups have a more natural additive structure, it is
   the name ``cohomology'' which is actually used for generalizations.) The
   $p$-chains $S$ that satisfy $\gd S = 0$ are called {\em $p$-cycles}. Again,
   the $H_p(\cM)$ only exist for $0 \leq p \leq n$.

   \sk There is an interesting relation between cohomology and homology groups.
   Note that we can construct a bilinear map from $H^p(\cM) \times H_p(\cM) \to
   \bR$ by
   \be
    \left( [B], [S] \right) \mapsto \int_S B,
    \label{eq:deRham}
   \ee
   where $[B]$ denotes the cohomology class of a $p$-form $B$, and $[\gS]$ the
   homology class of a $p$-cycle $\gS$. Using Stokes' theorem, it is easily seen
   that the result does not depend on the representatives for either $B$ or
   $S$:
   \bea
    \int_{S + \gd U} B + d A & = & \int_{S} B + \int_{S} d A + \int_{\gd U}
    B + d A \ret
    & = & \int_{S} B + \int_{\gd S} A + \int_{U} d(B + d A) \ret
    & = & \int_{S} B,
   \eea
   where we used that by the definition of (co)homology classes, $\gd S=0$ and
   $dB = 0$. As a result, the above map is indeed well-defined on homology and
   cohomology classes. A very important theorem by de Rham says that this map is
   nondegenerate. This means that if we take some $[B]$ and we know the result
   of (\ref{eq:deRham}) for all $[S]$, this uniquely determines $[B]$, and
   similarly if we start by picking an $[S]$. This in particular means that the
   vector space $H^p(\cM)$ is the dual vector space of $H_p(\cM)$, and hence
   that the Betti number $b^p$ is also the dimension of $H_p(\cM)$.

   \subsubsection{The Hodge star and harmonic forms}
   Another important operator on $p$-forms is the Hodge star operator. It is
   defined as
   \be
    *B = {\eps^{i^1 \ldots i^p}}_{j_1 \ldots j_{n-p}} \, B_{i^1 \ldots i^p} \,
    dx^{j_1} \wedge \cdots \wedge dx^{j_{n-p}}.
   \ee
   Note that the definition of this operator {\em does} use a metric on $\cM$:
   it is used to lower indices of the $\eps$-tensor. It is therefore important
   to realize that constructions involving this operator do not automatically
   lead to topologically invariant results -- though as we will see, in certain
   cases they do.

   \sk Since the Hodge star maps $p$-forms to $(n-p)$-forms, the wedge product
   of a $p$-form $B$ with the Hodge dual of another $p$-form $C$ can naturally
   be integrated over the whole manifold $\cM$. In fact, it can be shown that
   \be
    (B, C) \equiv \int_\cM B \wedge *C
   \ee
   defines a nondegenerate and (on Euclidean manifolds) positive definite inner
   product on the space $\gO^p(\cM)$ of $p$-forms for some fixed $p$. This can
   also be viewed as an inner product on the space $\gO^*(\cM)$ of all forms if one
   defines the inner product of a $p$-form and a $q$-form for $p \neq q$ to
   vanish -- that is, one takes the $\gO^p(\cM)$ for different 
   $p$ to be orthogonal to each other. We can then also define the adjoint
   operator of the exterior derivative with respect to this inner product. By
   definition, the adjoint operator $d^*$ is the operator which satisfies
   \be
    (B, d C) = (d^*B, C).
   \ee
   That this uniquely defines the way in which the $d^*$-operator acts can be
   seen as follows. If we fix $B$ and require the above relation for all $C$,
   this relation tells us what $(d^* B, C)$ is for every $C$. Since the inner
   product is nondegenerate, this uniquely fixes $d^*B$ itself. By comparing
   the degrees in the above formulas, one sees that if $B$ is a $p$-form, $d^*
   B$ will be a $(p-1)$-form, so $d^*$ acts in the ``opposite direction''
   compared to $d$.

   \sk It turns out that there is also a more direct expression for the adjoint
   operator acting on $p$-forms: it can be written as
   \be
    d^* =(-1)^{pn+p+1} *d*,
   \ee
   Note that, also from this expression, it is clear that $d^*$ lowers the degree
   of a form by 1.

   \sk Now, let us make the simple observation that for a fixed $B$,
   \bea
    0 & = & (B, d d C) \ret
      & = & (d^* d^* B, C)
   \eea
   for any $C$. Using once again the fact that the inner product is
   nondegenerate, this means that $(d^*)^2 B = 0$. Since $B$ was arbitrary,
   this means that
   \be
    (d^*)^2 = 0.
   \ee
   Does this mean that using the $d^*$-cohomology, we can define another
   set of topological invariants? At first sight, this is not clear, since we
   did not define $d^*$ in a metric-independent way. We will now argue that in
   fact the $d^*$-cohomology {\em is} a topological invariant, but that it
   gives exactly the same information as the $d$-cohomology.

   \sk Note that the operators $d d^*$ and $d^* d$ map $p$-forms to $p$-forms.
   One can show by writing out the definitions, including the omitted factors of
   $p!$, that
   \be
    d d^* + d^* d = \gD,
   \ee
   where $\gD$ is the Laplacian acting on the components $B_{i_1 \ldots
   i_p}$. We can now look for solutions to the equations $dB = d^* B = 0$. An
   important theorem by Hodge states that each $d$-cohomology class contains
   exactly one such form $B_h$, which by the previous remark is a harmonic
   form: $\gD B_h = 0$. In other words, the $d$-cohomology classes are in
   one-to-one correspondence to the harmonic forms on $\cM$. However, the Hodge
   theorem is symmetric, in the sense that it also predicts a unique harmonic
   form in each $d^*$-cohomology class. Therefore, both $H^p_d(\cM)$ and
   $H^p_{d^*}(\cM)$ can be identified with the vector space of harmonic
   $p$-forms, and hence we find the promised result that the two are equivalent.

   \subsubsection{Dolbeault cohomology}
   So far, we have been discussing {\em real} manifolds. However, there is
   an interesting generalization of all of this to {\em complex} manifolds,
   which is called ``Dolbeault cohomology''. On complex $m$-dimensional
   manifolds, we have local coordinates $z^i$ and $\zbar^i$. One can now study
   $(p,q)$-forms, which are forms containing $p$ factors of $dz^i$ and $q$
   factors of $d\zbar^j$:
   \be
    B = B_{i_1 \ldots i_p, j_1 \ldots j_q}(z,\zbar) \, dz^{i_1} \wedge \cdots
    \wedge dz^{i_p} \wedge d \zbar^{j_1} \wedge \cdots \wedge d \zbar^{j_q}.
   \ee
   Moreover, one can introduce {\em two} exterior derivative operators $\d$ and
   $\delbar$, where $\d$ is defined by
   \be
    \d B \equiv \frac{\d B_{i_1 \ldots i_p, j_1 \ldots j_q}}{\d z^k} \, dz^k
    \wedge dz^{i_1} \wedge \cdots \wedge dz^{i_p} \wedge d \zbar^{j_1} \wedge
    \cdots \wedge d \zbar^{j_q},
   \ee
   and $\delbar$ is defined similarly by differentiating with respect to
   $\zbar^k$ and adding a factor of $d\zbar^k$. Again, both of these operators
   square to zero. We can now construct two cohomologies -- one for each of
   these operators -- but as we will see, in the cases that we are interested
   in, the information contained in them is the same. Conventionally, one uses
   the cohomology defined by the $\delbar$-operator.

   \sk For complex manifolds, Hodge's theorem also holds: each cohomology class
   $H^{p,q}(\cM)$ contains a unique harmonic form. Here, a harmonic form $B_h$
   is a form for which the ``Laplacian''
   \be
    \gD_{\delbar} = \delbar \delbar^* + \delbar^* \delbar
   \ee
   has a zero eigenvalue: $\gD_{\delbar} B_h = 0$. In general, this operator does
   not equal the ordinary Laplacian, but one can prove that in the case where
   $\cM$ is a K\"ahler manifold\footnote{K\"ahler manifolds will be defined in
   chapter \ref{sec:calabiyau}. They are manifolds with a particular, symmetric type of
   metric. The manifolds we will study as target spaces for the topological
   string will always be K\"ahler manifolds, for reasons that will become clear
   as we go along.},
   \be
    \gD = 2 \gD_{\delbar} = 2 \gD_\d.
   \ee
   In other words, on a K\"ahler manifold the notion of a harmonic form is the
   same, independently of which exterior derivative one uses. As a first
   consequence, we find that the vector spaces $H^{p,q}_\d(\cM)$ and
   $H^{p,q}_{\delbar}(\cM)$ both equal the vector space of harmonic $(p,q)$-forms,
   so the two cohomologies are indeed equal. Moreover, every $(p,q)$-form is of
   course a $(p+q)$-form in the de Rham cohomology, and by the above result we
   see that a {\em harmonic} $(p,q)$-form can also be viewed as a de Rham
   harmonic $(p+q)$-form. Conversely, any de Rham $p$-form can be written as a
   sum of Dolbeault forms:
   \be
    B_p = B_{p,0} + B_{p-1, 1} + \ldots + B_{0,p}.
    \label{eq:deRhamDolbeault}
   \ee
   Acting on this with the Laplacian, one sees that for a harmonic $p$-form,
   \bea
    0 = \gD B_p = \gD_{\delbar} B_p = \gD_{\delbar} B_{p,0} + \gD_{\delbar} B_{p-1,
    1} + \ldots + \gD_{\delbar} B_{0,p}.
   \eea
   Since $\gD_{\delbar}$ does not change the degree of a form, $\gD_{\delbar}
   B_{p_1,p_2}$ is also a $(p_1,p_2)$-form. Therefore, the right hand side can
   only vanish if each term vanishes separately, so all the terms on the right
   hand side of (\ref{eq:deRhamDolbeault}) must be harmonic forms. Summarizing,
   we have shown that the vector space of harmonic de Rham $p$-forms is a
   direct sum of the vector spaces of harmonic Dolbeault $(p_1, p_2)$-forms
   with $p_1 + p_2 = p$. Since the harmonic forms represent the cohomology
   classes in a one-to-one way, we find the important result that for K\"ahler
   manifolds,
   \be
    H^p(\cM) = H^{p,0}(\cM) \oplus H^{p-1,1}(\cM) \oplus \cdots \oplus
    H^{0,p}(\cM).
   \ee
   That is, the Dolbeault cohomology can be viewed as a refinement of the de
   Rham cohomology. In particular, we have
   \be
    b^p = h^{p,0} + h^{p-1,1} + \ldots + h^{0,p},
   \ee
   where $h^{p,q} = \dim H^{p,q}(\cM)$ are the so-called {\em Hodge numbers} of
   $\cM$.
   
   \subsubsection{Relations between Hodge numbers and Poincar\'e duality}
    \label{sec:relationshodgenumbers}
    The Hodge numbers of a K\"ahler manifold give us several topological
    invariants, but not all of them are independent. In particular, the
    following two relations hold:
    \bea
     h^{p,q} & = & h^{q,p} \ret
     h^{p,q} & = & h^{m-p, m-q}.
     \label{eq:poincareduality}
    \eea
    The first relation immediately follows if we realize that $B \mapsto \Bbar$
    maps $\d$-harmonic $(p,q)$-forms to $\delbar$-harmonic $(q,p)$-forms, and
    hence can be viewed as an invertible map between the two respective
    cohomologies. As we have seen, the $\d$-cohomology and the
    $\delbar$-cohomology coincide on a K\"ahler manifold, so the first of the
    above two equations follows.
    
    \sk The second relation can be proven using the map
    \be
     (A, B) \mapsto \int_\cM A \wedge B
    \ee
    from $H^{p,q} \times H^{m-p, m-q}$ to $\bC$. It can be shown that this map
    is nondegenerate, and hence that $H^{p,q}$ and $H^{m-p,m-q}$ can be viewed
    as dual vector spaces. (Compare the discussion at the end of section
    \ref{sec:homology}.) In particular, it follows that these vector spaces have
    the same dimension, which is the statement in the second line of
    (\ref{eq:poincareduality}).
    
    \sk Note that the last argument of course also holds for de Rham cohomology,
    in which case we find the relation $b^p = b^{n-p}$ between the Betti
    numbers. We also know that $H^{n-p}(\cM)$ is dual to
    $H_{n-p}(\cM)$, so combining these statements we find an identification
    between the vector spaces $H^p(\cM)$ and $H_{n-p} (\cM)$. This
    identification between $p$-form cohomology classes and $(n-p)$-cycle
    homology classes is known as {\em Poincar\'e duality}. Intuitively, it can
    be thought of as follows. Take a certain $(n-p)$-cycle $\gS$ representing a
    homology class in $H_{n-p}$. One can now try to define a ``delta function''
    $\gd(\gS)$ which is localized on this cycle. Locally, $\gS$ can be
    parameterized by setting $p$ coordinates equal to zero, so $\gd(\gS)$ is a
    ``$p$-dimensional delta function'' -- that is, it is an object which is
    naturally integrated over $p$-dimensional submanifolds: a $p$-form. This
    intuition can be made precise, and one can indeed view the cohomology class of the
    resulting ``delta-function'' $p$-form as the Poincar\'e dual to $\gS$.

    \sk Going back to the relations (\ref{eq:poincareduality}), we see that the
    Hodge numbers of a K\"ahler manifold can be nicely written in a so-called
    {\em Hodge-diamond}:
    \be
     \begin{array}{ccccccc}
      &&& h^{0,0} \\
      && h^{1,0} && h^{0,1} \\
      &\mbox{\reflectbox{$\ddots$}} &&&& \ddots \\
      h^{m,0} &&& \cdots &&& h^{0,m} \\
      &\ddots &&&& \mbox{\reflectbox{$\ddots$}} \\
      && h^{m,m-1} && h^{m-1,m} \\
      &&& h^{m,m}      
     \end{array}
    \ee
    The integers in this diamond are symmetric under the reflection in its
    horizontal and vertical axes.

  \subsection{Vector bundles}
   \label{sec:vectorbundles}
   \subsubsection{Definition}
    Vector bundles are the mathematical objects that describe vector fields in
    physics\footnote{Here, we mean ``vector'' in the mathematical sense, so such
    a field can for example also be a scalar field, a tensor field, a spinor
    field, or a matrix-valued field.}. A vector bundle $E$ over a manifold
    $\cM$, denoted $E \to \cM$, is a manifold which can be projected onto $\cM$:
    \be
     \pi: E \to \cM
    \ee
    in such a way that the {\em fiber} $\pi^{-1}(x)$ at every point $x$ of
    $\cM$ has the structure of a (fixed) vector space $W$. Locally -- that is,
    on small enough patches $U_{(a)}$ for $\cM$ -- the vector bundle must look
    like $U_{(a)} \times W$, but this is not necessarily globally the case: the
    vector bundle does not need to be topologically equivalent to $\cM \times
    W$. For example, a cylinder can be viewed as a vector bundle over the
    circle $S^1$, where the fiber $W$ is simply a line $\bR$. This is a trivial
    vector bundle: the cylinder can be written as $S^1 \times \bR$. An example
    of a nontrivial vector bundle is the ``infinite M\"obius strip'', which can
    be obtained by taking the trivial vector bundle over an interval, e.\ g.\
    $[0,1] \times \bR$, and gluing $(0,y)$ to $(1, -y)$. This is again a vector
    bundle over the circle $S^1$, where this circle can be viewed as the set of
    points $(x,0)$, with $(0,0)$ and $(1,0)$ identified. The projection $\pi$
    maps the fiber, consisting of the points $(x_0, y)$ at fixed $x_0$, to the
    point $(x_0, 0)$ on the circle.

\begin{figure}[ht]
     \begin{center}
      \includegraphics[height=5cm]{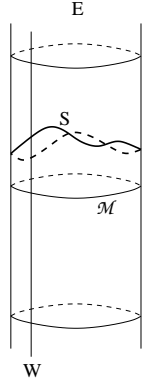}
     \end{center}
     \caption{A simple example of a vector bundle $E$: the cylinder. The circle
     $\cM = S^1$ is the base space (the other two circles are only drawn to
     visualize the geometry), and the fibers $W$ are lines. $S$ is a section of
     this vector bundle.}
     \label{fig:vectorbundle}
    \end{figure}

    \sk A {\em section} of a vector bundle is what a physicist would call a
    ``field configuration'': it is a map from $\cM$ to $E$ such that every point
    of $\cM$ is mapped to a point in the fiber $W$ living at that point. See
    figure \ref{fig:vectorbundle} for a graphical representation of these ideas.

    \sk Just as for manifolds, the easiest way to construct a vector bundle is
    to do it patch by patch. That is, one considers a set of charts $\{ U_{(a)}
    \}$ covering $\cM$, and constructs the so-called {\em local trivializations}
    \be
     U_{(a)} \times W.
    \ee
    Now, one glues together the $U_{(a)}$ using the transition functions
    $\phi_{(ba)}$ of the manifold, but of course one also needs to glue the
    vectors $V_{(a)}(x_{(a)})$ in the fibers $W$ on the patch $U_{(a)}$ to the
    vectors $V_{(b)}(x_{(b)})$ of the fibers $W$ on the patch $U_{(b)}$. This is
    done by choosing an extra set of transition functions $\gL_{(ba)}$:
    \be
     V_{(b)}(x_{(b)}) = \gL_{(ba)}(x_{(a)}) \, V_{(a)} (x_{(a)}).
    \ee
    Of course, on triple overlaps, these transition functions also have to
    satisfy the cocycle condition:
    \be
     \gL_{(ac)} \gL_{(cb)} \gL_{(ba)} = 1.
    \ee
    Again, this single condition on the transition functions turns out to be
    enough to make sure that one constructs a well-defined vector bundle.

    \sk If $W$ is an $k$-dimensional vector space, the transition functions
    $\gL_{(ba)}(x)$ are elements of $GL(k, \bR)$. However, one can of course
    study classes of vector bundles where the transition functions take values
    in a subgroup $G \subset GL(k, \bR)$. For example, if $W$ is equipped with
    an inner product (that is, if one has a natural notion of the length of a
    vector) then it is natural to consider only length-preserving transition
    functions, so one would take $G = O(k)$. In such a case, $G$ is called the
    {\em structure group} of the manifold.

   \subsubsection{Examples}
    \label{sec:vbexamples}
    In this section, we want to mention several examples of vector bundles which
    will be important to us in what follows.

    \sk {\em The principal bundle}

    \sk Our first example is in fact not a vector bundle, since its fiber is not
    a vector space, but a group. (A topological space of this more general kind
    is called a {\em fiber bundle}.).

    \sk Suppose we have a vector bundle $E$ over a manifold $\cM$
    with patches $U_{(a)}$ and fiber transition functions $\gL_{(ba)}$. Since
    $\gL_{(ba)}(x)$ is an element of the structure group $G$, we can construct a
    new fiber bundle with the group $G$ as a fiber, by taking local
    trivializations $U_{(a)} \times G$, and gluing them together with the same
    transition functions $\gL_{(ba)}$:
    \be
     g_{(b)}(x_{(b)}) = \gL_{(ba)}(x_{(a)}) \, g_{(a)} (x_{(a)}).
     \label{eq:glueprincipal}
    \ee
    The fiber bundle constructed in this way is called the {\em principal
    bundle associated to $E$}, and we will denote it by $\Etilde$. Note that if
    we know how the structure group acts on the fiber $W$, we can also go in
    the opposite direction and reconstruct $E$ from $\Etilde$.
    More generally, by choosing a different representation of $G$ -- that is, a
    different vector space $W'$ on which $G$ naturally acts -- we can
    construct another vector bundle $E'$ from $\Etilde$ by taking the action of
    the new transition functions $\gL'_{(ba)}$ to be the action of $\gL_{(ba)}$
    in the representation $W'$. This is the reason for the name ``principal
    bundle'': from it, we can construct a vector bundle for each representation
    of $G$. We will see examples of this below when we discuss tangent bundles
    and spin bundles.

    \sk {\em The tangent bundle}

    \sk The tangent bundle to a manifold is the vector bundle whose fiber at a
    point $x$ is the tangent space at $x$. Suppose we have two overlapping
    patches, $U_{(a)}$ and $U_{(b)}$. The coordinates are related by a
    transition function:
    \be
     x^i_{(b)} = \phi^i_{(ba)}(x_{(a)}).
    \ee
    Using standard mathematical terminology, let us denote the unit tangent
    vector at $x$ in the $x^i_{(a)}$-direction by $\d / \d x^i_{(a)}$. A
    general tangent vector on the patch $U_{(a)}$ can now be expressed as
    \be
     v = v^i_{(a)}(x) \, \frac{\d}{\d x^i_{(a)}}.
    \ee
    The reason for this notation is that one can identify the tangent vector $v$
    with a directional derivative in the direction $v$, or from a more physical
    point of view, that tangent vectors transform in the same way as directional
    derivatives. However, from our point of view, the notation will be purely
    symbolic: we will not think of the $\d / \d x^i_{(a)}$ as actually acting as
    derivatives on anything.

    \sk The transformation rule for the components of a vector under a change of
     coordinates from $x_{(a)}$ to $x_{(b)}$ is well known:
    \be
     v^i_{(b)}(x_{(b)}) = \frac{\d \phi^i_{(ba)}}{\d x^j_{(a)}} \,
     v^j_{(a)}(x_{(a)}).
    \ee
    That is, the transition functions of the fibers are given by
    \be
     \left( \gL_{(ba)} \right)^i_{~j} = \frac{\d \phi^i_{(ba)}}{\d x^j_{(a)}}
     = \frac{\d x^i_{(b)}}{\d x^j_{(a)}}.
     \label{eq:transfntangent}
    \ee
    At every point $x$, this defines an element $\gL_{(ba)}$ of $GL(n, \bR)$.
    Moreover, if we require the coordinate transition functions $\phi_{(ba)}$ to
    be length-preserving, $\gL_{(ba)}$ is actually an element of $O(n)$,
    and if our base manifold $\cM$ is orientable and we require that
    $\phi_{(ba)}$ preserves orientation, $\gL_{(ba)}$ is an element of $SO(n)$.
    We will always be considering this last situation in what follows.

    \sk As we explained before, we can construct the principal bundle $\Etilde$
    associated to the tangent bundle by replacing the fibers $\bR^n$ by the
    structure group $G = SO(n)$, and gluing the fibers with the same
    transition functions (\ref{eq:transfntangent}). This bundle is also called
    the {\em frame bundle}. Now, let us assume we have a different
    representation $W'$ of $SO(n)$. For example, instead of vectors $v^i$, the
    elements of $W'$ may be tensors with two indices, $T^{ij}$, so $W'$ has
    dimension $n^2$. From the way that tensors transform under coordinate
    transformations, we see that the new transition functions of the bundle $E'$
    will be
    \be
     \left( \gL'_{(ba)} \right)^{ij}_{~~kl} = \left( \gL_{(ba)} \right)^{i}_{~k}
     \left( \gL_{(ba)} \right)^{j}_{~l}.
    \ee
    Similarly, from the frame bundle $\Etilde$ we can define vector bundles for
    tensor fields with an arbitrary number of upper and lower indices. In
    particular, since $p$-forms have $p$ lower indices and are antisymmetric in
    these indices, we can view a $p$-form as a section of a vector bundle with
    $\binom{n}{p}$-dimensional fibers.

    \sk Since an $m$-dimensional complex manifold can be viewed as a
    $2m$-dimensional real manifold, it has a $2m$-dimensional tangent bundle.
    The most natural way to write the elements of this tangent bundle in terms
    of coordinates is of course
    \be
     v = v^i_{(a)}(z, \zbar) \, \frac{\d}{\d z^i_{(a)}} + v^{\ibar}_{(a)}(z,
     \zbar) \, \frac{\d}{\d \zbar^i_{(a)}}.
     \label{eq:holtangent}
    \ee
    Here, $v^i_{(a)}$ and $v^{\ibar}_{(a)}$ are complex conjugates. However,
    since we are working with complex coordinates, it is now natural to also
    study the {\em complexified} tangent bundle, where we allow the coordinates
    $v^i_{(a)}$ and $v^{\ibar}_{(a)}$ to be arbitrary, unrelated complex
    numbers. Note that this construction effectively doubles the dimension: a
    complex manifold with $m$ complex dimensions has a complexified tangent
    bundle with $2m$ complex-dimensional fibers. Also, note that in this
    construction we have not changed the transition functions: these still take  values
    in\footnote{Or rather in its subgroup $U(m)$, as follows from the
    requirement that the transition functions $\phi_{(ba)}$ for the manifold are
    holomorphic.} $SO(2m, \bR)$.
    The only difference is that the elements of $SO(2m, \bR)$ now act on {\em
    complex} 2$m$-dimensional vectors.

    \sk A particular phenomenon appears in the case of a complex manifold of
    complex dimension 1. The structure group is then $SO(2) = U(1)$. The
    irreducible representations of $U(1)$ are one-dimensional, and therefore the
    vector representation in this case is not irreducible. This can easily be seen
    by writing an arbitrary vector as in (\ref{eq:holtangent}), where the index
    $i$ now only takes a single value: clearly, the first term in
    (\ref{eq:holtangent}) has charge 1 under $U(1)$, and the second term has
    charge $-1$. (Of course, this statement is true in any dimension, but it is
    only in two dimensions that the $U(1)$ is the full structure group of the
    tangent bundle.) Since the split-up into holomorphic and anti-holomorphic
    coordinates is globally defined, the whole two-dimensional complexified
    tangent bundle can be split up into the sum of two one-dimensional
    subbundles with complex conjugate transition functions. We denote these two
    bundles by $T^{+}\cM$ and $T^-\cM$.

    \sk {\em Spin bundles}

    \sk Since we now know how to construct vector bundles corresponding to
    arbitrary representations of $SO(n)$, a natural question is: what about
    spinor representations? Strictly speaking of course, spinors do not
    transform in a representation of $SO(n)$, but in a representation of
    $Spin(n)$, which is a double cover of the group $SO(n)$. The best known
    example of this is the case of $Spin(3) = SU(2)$ which is a double cover of
    $SO(3)$. Therefore, we would like to construct a principal bundle with
    fibers $Spin(3)$ from the tangent bundle. However, the transition functions
    $\gL_{(ba)}(x)$ are elements of $SO(n)$, and not of $Spin(n)$. More
    precisely, there are {\em two} elements in $Spin(n)$ corresponding to each
    single element in $SO(n)$, and these two elements differ by a sign.
    Therefore, to ``lift'' the transition functions $\gL_{(ba)}$ of the tangent
    (or frame) bundle $T\cM$ to transition functions $\gLtilde_{(ba)}$ for a
    spin bundle\footnote{We will be somewhat sloppy, and denote both the
    principal $Spin(n)$-bundle and the associated vector bundle constructed from
    the spin 1/2 representation by the term ``spin bundle''. As we have seen,
    both of these bundles contain the same information.}, we need to make a
    choice of sign for each pair $(ba)$. This leads to a large arbitrariness in
    the construction, but of course many of these arbitrary choices may lead to
    topologically equivalent bundles. What is more important, however, is that
    some choices of signs are simply not allowed. The reason is that to result in a
    well-defined vector bundle, the transition functions $\gLtilde_{(ba)}$ have to
    satisfy the cocycle condition:
    \be
     \gLtilde_{(ac)} \gLtilde_{(cb)} \gLtilde_{(ba)} = 1.
     \label{eq:cocyclespin}
    \ee
    By choosing arbitrary signs, we might get $-1$ on the left hand side of
    this equation for some triple overlaps of patches, so this gives a certain
    number of conditions on the signs we choose. It may even happen that no
    choice of signs is consistent with the requirement (\ref{eq:cocyclespin}),
    so it will be impossible to define a spin bundle! An example of a
    manifold where this happens is the two-dimensional complex projective space
    $\bC P^2$. The
    manifolds on which it is possible to define a spin bundle are called ``spin
    manifolds''. In general, such a manifold can allow several inequivalent
    choices of signs for $\gLtilde_{(ba)}$, and these different choices are
    called {\em spin structures}. As an example, on a compact, orientable
    Riemann surface, for each homology one-cycle we have a free choice of sign,
    and so there are $2^{b^1}$ spin structures, and hence the same number of
    different spin bundles over the manifold.

    \sk It is possible to construct topological invariants on a manifold $\cM$
    which determine if a spin structure exists (the so-called Stiefel-Whitney
    classes), but we will not go into this. From now on, we will simply assume
    that a spin structure exists on our manifolds. In particular, as we
    mentioned, for two-dimensional orientable manifolds this is no restriction
    at all.

    \sk Another important property of the two-dimensional case is that, since
    the tangent bundle splits into two bundles $T^+ \cM$ and $T^- \cM$, one can
    also split the spin bundle (for a given choice of spin structure) into two
    bundles with complex conjugate transition functions. These bundles are
    denoted $S^+(\cM)$ and $S^-(\cM)$. As it is usually assumed that a spin
    structure on $\cM$ is chosen once and for all, the dependence on a choice
    of spin structure is not reflected in this notation.

    \sk {\em Lie algebra valued one-forms}

    \sk Finally, we want to construct a bundle which will be very important
    when we study connections in the next section. It is a bundle of
    Lie algebra-valued\footnote{The mathematical terminology ``X-valued objects
    of type Y'' means that the coefficients of the objects of type Y are not
    ordinary numbers, but are objects of type X. So in this particular case, we
    have one-forms $A_i dx^i$ where the $A_i$ will be elements of a
    Lie algebra.} one-forms. Recall that an element of the Lie algebra
    $\mathfrak g$ of a
    group $G$ is a tangent vector to the group near the identity element. That is,
    if we have a curve $g(t)$ inside the group manifold such that $g(0) = e$,
    where $e$ is the identity element of the group, then the ``velocity'' at
    $t=0$,
    \be
     \left. \frac{d g}{d t} \right|_{t=0},
    \ee
    is an element of the Lie algebra. More generally, at arbitrary ``time'' $t$ one
    can construct a Lie algebra element by ``translating back to the origin'':
    both
    \be
     g^{-1} \frac{d g}{d t} \qquad \mbox{and} \qquad \frac{d g}{d t} g^{-1}
    \ee
    are elements of the Lie algebra for any time $t$. Note that there are two
    inequivalent ways of translating back to the origin; it turns out that, for
    our purposes, the second one is the right one. For notational reasons, we
    ``partially integrate'' this expression by subtracting $0 = d(g g^{-1})/dt$,
    and write it as
    \be
     - g \frac{d g^{-1}}{d t}
     \label{eq:liealg}.
    \ee
    We could of course have removed the minus sign by a different choice of
    convention, but this will also turn out to be notationally convenient in
    what follows.

    \sk Now assume we have a vector bundle $E$ with structure group $G$ and
    principal bundle $\Etilde$. Consider a section $g(x)$ of the principal
    bundle\footnote{One can quite easily show that it is only possible to define such a
    section globally if $E$ is a trivial bundle, i.\ e.\ a direct product $\cM
    \times G$. However, for our purpose of finding a natural transition
    function, it is enough to define the above section on two neighboring
    patches.} -- that is, a group-valued ``function'' on the base manifold $\cM$.
    On a certain patch $U_{(a)}$, we can now construct a Lie algebra
    valued one-form by considering
    \be
     - g_{(a)} \, d g^{-1}_{(a)} = - g_{(a)} \frac{d g^{-1}_{(a)}}{d x^i} \, dx^i.
    \ee
    This is a one-form on $\cM$, and by comparing with (\ref{eq:liealg}), it is
    clear that its coefficients are elements of the Lie algebra $\mathfrak g$. Of
    course, there are many Lie algebra valued one-forms that cannot be
    constructed in this way, but the reason we constructed these particular
    ones is that we can read off from them what the transition functions for
    objects of this type are. To see this, suppose we now have a
    neighboring patch $U_{(b)}$. We know from (\ref{eq:glueprincipal}) how
    $g_{(b)}$ is constructed out of $g_{(a)}$. Inserting this in the above
    expression, we find that
    \bea
     - g_{(b)} \, d g^{-1}_{(b)} & = & - \gL_{(ba)} g_{(a)} \, d \left( g^{-1}_{(a)}
     \gL^{-1}_{(ba)} \right) \ret
     & = & - \gL_{(ba)} g_{(a)} \, d g^{-1}_{(a)} \gL^{-1}_{(ba)} - \gL_{(ba)} d
     \gL^{-1}_{(ba)}.
    \eea
    That is, the particular Lie algebra valued one-forms we have constructed
    naturally live in a fiber bundle with transition functions
    \be
     A_{(b)} = \gL_{(ba)} A_{(a)} \gL^{-1}_{(ba)} - \gL_{(ba)} d
     \gL^{-1}_{(ba)}.
     \label{eq:connectiontransform}
    \ee
    Note that strictly speaking, this is not a vector bundle: even though
    $A(x)$ can be viewed as an element of a vector space, the transition
    functions are not linear in $A(x)$. Also note that many other bundles of
    Lie algebra valued $p$-forms can be constructed. In particular, one might
    take transition functions of the form
    \be
     B_{(b)} = \gL_{(ba)} B_{(a)} \gL^{-1}_{(ba)}
     \label{eq:liealgpform}
    \ee
    to construct a bundle of $p$-forms for arbitrary $p$. Since these
    transition functions are linear in $B$, contrary to the example above, this
    bundle is a vector bundle.

   \subsubsection{Connections}
    An important extra structure one can put on a vector bundle is that of a
    {\em connection}. A connection tells us how to parallel transport a vector
    at a point $x \in \cM$ to a vector at a point $x' \in \cM$ along a curve
    $\gam$. Note that, a priori, there is no well-defined way to compare the
    fibers at two different points of $\cM$, so one indeed needs some extra
    structure to make this parallel transport well-defined.

    \sk Of course, to parallel transport vectors along curves, it is enough if
    we can define parallel transport under an infinitesimal displacement: given
    a vector $V$ at $x$, we would like to define its parallel transported
    version $\Vhat$ after an infinitesimal displacement by $\eps v$, where $v$
    is a tangent vector to $\cM$. By the definition of the cotangent vector
    $dx^i$, the new coordinates will then be $\xhat^i = x^i + \eps dx^i(v)$. We could of
    course write $dx^i(v)$ as $v^i$, but it will be useful to keep the one-form
    notation explicit. The infinitesimal change in the vector will be linear in
    the displacement and in $V$, so it will be of the form
    \be
     \Vhat^I (\xhat) = V^I(x) + \eps A^I_{Ji}(x) V^J(x) dx^i(v).
     \label{eq:paralleltrans}
    \ee
    If we take $A^I_{Ji}(x) dx^i$ to be a one-form, this expression is independent
    of our choice of coordinates for $\cM$. However, we would like it to be independent of
    the choice of coordinates on the fiber $W$ as well. To achieve this, let us
    act on the above equation with a position-dependent group element
    $g^I_J(x)$. To linear order in $\eps$, the left hand side then becomes
    \bea
     \mbox{LHS} & \to &g^I_J(\xhat) \, \Vhat^J (\xhat) \ret
     & = & g^I_J(x) \Vhat^J (\xhat) + \eps \frac{dg^I_J}{dx^i}(x) \,
     dx^i(v) \Vhat^J (\xhat) \ret
     & = & g^I_J(x) \Vhat^J (\xhat) + \eps \, dg^I_J(v) \, V^J (x)
    \eea
    where in the last line we inserted (\ref{eq:paralleltrans}) and the
    definition of $dg$. Acting on the right hand side of
    (\ref{eq:paralleltrans}) with $g^I_J(x)$ gives
    \bea
     \mbox{RHS} & \to & g^I_J(x) V^J(x) + \eps \Atilde^I_{Ji}(x) g^J_{~K}(x)
     V^K(x)\, dx^i(v),
    \eea
    where we denoted the transformed version of $A$ by $\Atilde$. Equating the
    last two results, and multiplying by $g^{-1}(x)$, we find that
    \be
     \Vhat(\xhat) = V(x) + \eps \left( g^{-1} \Atilde(v) \, g - g^{-1} d g(v)
     \right) V(x),
    \ee
    where we left out the $I$-indices and wrote $\Atilde$ as a one-form to keep the formula readable.
    When comparing this to (\ref{eq:paralleltrans}), we see that $A$ should
    transform as
    \be
     \Atilde = g A g^{-1} - g d g^{-1}.
     \label{eq:gaugefieldtransform}
    \ee
    Note that this is precisely the transformation
    (\ref{eq:connectiontransform})! That is, for the parallel transport to be
    independent of the coordinates on $W$, we need $A$ to be a section of the
    bundle of Lie algebra valued one-forms defined by that equation\footnote{The
    reader might object that (\ref{eq:connectiontransform}) is a transformation
    between patches, whereas (\ref{eq:gaugefieldtransform}) is a transformation
    on a single patch. However, the latter can of course be viewed as a special
    case of the former, where $U_{(a)} = U_{(b)}$ and we use the same
    $x$-coordinates on both.}. Of course, all of this was carried out with $V$
    in a specific representation of $G$, but by starting from the defining
    representation of the structure group $G$, we can use the same connection
    to define parallel transport on vector bundles for any representation.

    \sk From its index structure and the way it transforms, $A^I_{Ji}$ looks
    very similar to a non-abelian gauge field. Let us show that this is exactly what it is.
    Suppose we have a vector field $V(x)$ on $\cM$, and we want to define its
    derivative in the direction $v$ in a coordinate independent way. We would
    like to write down
    \be
     \d_v V(x) = \lim_{\eps \to 0} \frac{V(\xhat) - V(x)}{\eps},
    \ee
    but here we are subtracting two vectors at different points in the numerator
    -- an operation which is clearly dependent on our choice of coordinates. Therefore, the right
    thing to do is to first parallel transport $V(x)$ to $\Vhat(\xhat)$
    and then subtract the two vectors. That is,
    \bea
     D_v V(x) & = & \lim_{\eps \to 0} \frac{V(\xhat) - \Vhat(\xhat)}{\eps} \ret
     & = & \lim_{\eps \to 0} \frac{V(\xhat) - V(x) - \eps A_i V(x)
     dx^i(v)}{\eps} \ret
     & = & \left( \d_i V(x) - A_i V(x) \right) \, v^i,
    \eea
    which is precisely the definition of a covariant derivative in terms of a
    gauge field $A$.

  \subsubsection{Chern classes}
   Inspired by the identification of a connection with a gauge field, let us
   consider the analogue of the non-abelian field strength:
   \be
    F = d A - A \wedge A,
   \ee
   where $A \wedge A$ is a shorthand for $A^I_{Ji} A^J_{Kj} \, dx^i \wedge
   dx^j$. (Note that $A \wedge A \neq 0$, contrary to what the notation might
   suggest!) A short calculation shows that on the overlap of two patches of
   $\cM$ (or equivalently, under a gauge transformation), this quantity
   transforms as
   \be
    F_{(b)} = \gL_{(ba)} F_{(a)} \gL_{(ba)}^{-1}.
   \ee
   This is exactly the transformation (\ref{eq:liealgpform}), and we see that
   $F$ can be viewed as a section of a true vector bundle of Lie algebra valued
   two-forms! In particular, we can take its trace and obtain a genuine
   two-form:
   \be
    c_1 = \frac{i}{2 \pi} \Tr(F),
   \ee
   where the prefactor is convention. It is easily seen that this two-form is
   closed:
   \be
    d (\Tr(F)) = \Tr(d F) = - \Tr(d (A \wedge A)) = - \Tr(d A \wedge A - A \wedge
    d A) = 0.
   \ee
   Therefore, we can take its cohomology class, for which we would like to argue
   that it is a topological invariant. It is clear that this
   construction is independent of the choice of coordinates on $\cM$. Moreover,
   it is independent of gauge transformations of the form
   (\ref{eq:gaugefieldtransform}). However, 
   on a general vector bundle there may of course be connections which cannot be
   reached in this way from a given connection. Changing to such a connection
   is called a ``large gauge transformation'', and from what we have said it is
   not clear a priori that the Chern classes do not depend on this choice of
   ``equivalence class'' of connections. However, with some work we can also
   prove this fact\footnote{Nevertheless, the reader should be aware of the fact
   that there may be similar constructions which are invariant under ordinary
   gauge transformations, but {\em not} under large ones. We will see an example
   of this when we discuss Chern-Simons theory in the next chapter.}. The
   invariant $[c_1]$ is called the {\em first Chern class}. In fact, it might be
   better to call it a ``relative topological invariant'': given a base manifold
   $\cM$ of fixed topology, we can topologically distinguish vector bundles over
   it by calculating the above cohomology class.

   \sk Of course, by taking the trace of $F$, we loose a lot of information.
   There turns out to be a lot more topological information in $F$, and it can
   be extracted by considering the expression
   \be
    c(F) = \det\left( 1 + \frac{i F}{2 \pi} \right),
   \ee
   where $1$ is the identity matrix of the same size as the elements of the Lie
   algebra of $G$. Again, it can be checked that this expression is invariant
   under a change of coordinates for $\cM$ and under a change of connection.
   Since the matrix components inside the determinant consist of the zero-form 1
   and the two-form $F$, expanding the determinant will lead to an expression
   consisting of forms of all even degrees. One writes this as 
   \be
    c(F) = c_0(F) + c_1(F) + c_2(F) + \ldots
   \ee
   The sum terminates either at the highest degree encountered in expanding the
   determinant, or at the highest allowed even form on $\cM$. Note that $c_0(F)
   = 1$, and $[c_1(F)]$ is exactly the first Chern class we defined above. The
   cohomology class of $c_n$ is called the {\em $n$th Chern class}.

   \sk As an almost trivial example, let us consider the case of a product bundle
   $\cM \times W$. As we have mentioned in the previous section, in this case
   there is a global section $g(x)$ of the principal bundle $\Etilde$, and we
   can use this to construct a connection $A = - g d g^{-1}$. Now we find that
   \bea
    F & = & - d g d g^{-1} - g d g^{-1} g d g^{-1} \ret
    & = & - d g d g^{-1} + d g d g^{-1} \ret
    & = & 0,
   \eea
   where in the second line we did a ``partial integration'' in the second
   term. Therefore, for a trivial bundle with this connection, $c_0 = 1$ and
   $c_n = 0$ for all $n>0$.

 \section{Topological field theories -- generalities}
  \label{sec:topfieldth}
  \subsection{General definition}
   There exists a mathematically rigorous, axiomatic definition of topological
   field theories due to M.~Atiyah. Instead of giving this definition, we will
   define topological field theory in a more physical, but as a result somewhat
   less rigorous way. The reader interested in the mathematical details is
   referred to \cite{Atiyah:1988tq}.

   \sk The output of a quantum field theory is given by its observables:
   correlation functions of products of operators,
   \be
    \langle \cO_1(x_1) \cdots \cO_n(x_n) \rangle_b.
    \label{eq:observables}
   \ee
   Here, the $\cO_i(x)$ are {\em physical} operators of the theory. What one
   calls ``physical'' is of course part of the definition of the theory, but it
   is important to realize that in general not all combinations of fields are
   viewed as physical operators. For example, in a gauge theory, we usually
   require the observables to arise from gauge-invariant operators. That is,
   $\Tr~ F$ would be one of the $\cO_i$, but $\Tr~ A$ or $A$ itself would not.

   \sk The subscript $b$ in the above formula serves as a reminder that the 
   correlation function is usually calculated in a certain background. That is, the
   definition of the theory may involve a choice of a manifold $\cM$ on which
   the theory lives, it may involve choosing a metric on $\cM$, it may involve
   choosing certain coupling constants, and so on.

   \sk The definition of a topological field theory is now as follows. Suppose
   that we have a quantum field theory where the background choices involve a
   choice of manifold $\cM$ and a choice of metric $h$ on $\cM$. Then the theory
   is called a {\em topological field theory} if the observables
   (\ref{eq:observables}) do {\em not} depend on the choice of metric $h$. Let
   us stress that it is part of the definition that $h$ is a background field --
   in particular, we do {\em not} integrate over $h$ in the path
   integral\footnote{Of course, one can try to avoid this restriction in the
   definition by first integrating over $h$ to obtain a metric-independent
   effective theory, and then putting an arbitrary metric $h'$ on $\cM$, on
   which the theory will clearly not depend. This will indeed result in a
   topological field theory of the ``Schwarz-type'' that we will discuss in the
   next section. However, in general it is of course practically impossible to
   do the path-integral over $h$ and arrive at such an effective theory of quantum
   gravity.}. One may of course wonder what happens if, once we have a
   topological field theory, we do make the metric $h$
   dynamical and integrate over it. This is exactly what we will do once we
   start considering topological {\em string} theories.

   \sk Note that the word ``topological'' in the definition may be somewhat of a
   misnomer. The reason is that the above definition does not strictly imply
   that the observables depend only on the topology of $\cM$ -- there may be
   other background choices hidden in $b$ on which they depend as well. For
   example, in the case of a complex manifold $\cM$, correlation functions will in
   general not only depend on the topology of $\cM$ and its metric, but also on
   our specific way of combining the $2d$ real coordinates on $\cM$ into $d$
   complex ones. This choice, called a {\em complex structure}, is part of the
   background of the quantum field theory, and correlation functions in a
   topological field theory will in general still depend on it.

   \sk If our quantum field theory has general coordinate invariance, as we will
   usually assume to be the case, then the above definition has an interesting
   consequence. The reason is that in such a case we can do an arbitrary
   general coordinate transformation, changing both the coordinates on $\cM$ and
   its metric, under which the correlation functions should be invariant. Then,
   using the topological invariance, we can transform back the metric to its old
   value. The combined effect is that we have only changed the $x_i$ in
   (\ref{eq:observables}). That is, in a generally coordinate invariant
   topological field theory, the observables do not depend on the insertion
   points of the operators.

  \subsection{Chern-Simons theory}
   \label{sec:chernsimons}
   The easiest way to construct a topological field theory is to construct a
   theory where both the action (or rather the ``quantum measure'' $e^{iS}$) and
   the fields do not include the metric at all. Such topological field theories
   are called ``Schwarz-type'' topological field theories. This may sound like a
   trivial solution to the problem, but nevertheless it can lead to quite
   interesting results! To see this, let us consider an example: Chern-Simons
   theory on a three-dimensional manifold $\cM$.

   \sk Chern-Simons theory is a gauge theory -- that is, in the language
   developed in the previous chapter, it is constructed from a vector bundle $E$
   over the base space $\cM$, with a structure group (gauge group) $G$ and a
   connection (gauge field) $A$. The Lagrangean of Chern-Simons theory is then
   given by
   \be
    L = \Tr \left( A \wedge d A - \frac{2}{3}A \wedge A \wedge A
    \right).
   \ee
   It is a straightforward exercise to check how this Lagrangean changes under
   the gauge transformation (\ref{eq:gaugefieldtransform}), and one finds
   \bea
    \Ltilde & \equiv & \Tr \left( \Atilde \wedge d \Atilde - \frac{2}{3} \Atilde
    \wedge \Atilde \wedge \Atilde \right) \\
    & = & \Tr \left( A \wedge d A - \frac{2}{3}A \wedge A \wedge A \right) - d \,
    \Tr(g A \wedge dg^{-1}) + \frac{1}{3} \Tr(g dg^{-1} \wedge d g \wedge dg^{-1}).
    \nonumber
   \eea
   The second term is a total derivative, so if $\cM$ does not have a boundary,
   the action, being the integral of $L$ over $\cM$, does not get a contribution
   from this term. The last term is not a total derivative, but its integral
   turns out to be a topological invariant of the map $g(x)$, which is
   quantized as
   \be
    \frac{1}{24 \pi^2} \int_{\cM} \Tr(g dg^{-1} \wedge d g \wedge dg^{-1}) = m \in
    \bZ .
   \ee
   From this, we see that if we define the action as
   \be
    S = \frac{k}{4 \pi} \int_\cM L,
   \ee
   with $k$ an integer, the action changes by $2 \pi k m$ under gauge
   transformations, and the quantum measure $e^{iS}$ is invariant.

   \sk So from this discussion we seem to arrive at the conclusion that the
   partition function
   \be
    Z = \int DA \, e^{iS[A]}
   \ee
   for a line bundle of a fixed topology $E$ is a topological invariant of
   $\cM$, as are the correlation functions of 
   gauge-invariant operators such as $\Tr~ F$. However, there is one more detail
   we have to worry about: there may be an anomaly in the quantum theory. That
   is, it may not be possible to define the path integral measure $DA$ in a
   gauge-invariant way.

   \sk One way to see what problems can arise is to note that to actually
   calculate the path integral, one has to pick a gauge condition on $A$. That
   is, we have to pick one representative of $A$ in each equivalence class under
   gauge transformations. To
   make such a choice will in general require a choice of metric. For example,
   from electromagnetism (where $E$ is a one-dimensional complex line bundle and
   $G = U(1)$) we know that a useful gauge is the Feynman gauge, in which the
   equation of motion for $A$ becomes
   \be
    \gD A = 0.
   \ee
   As we have seen before, the Laplacian $\gD$ is an operator which, through the
   Hodge star, depends on the metric, and hence the results we find will a
   priori be metric dependent. To show that the results are truly metric
   independent, one needs to show that the quantum results do not depend on our
   arbitrary choice of gauge.

   \sk We will not go into the details of this, but simply state that one can
   show that Chern-Simons theory on a compact three-manifold is anomaly-free, so
   our naive argument above was correct, and one can indeed calculate
   topological invariants of $\cM$ in this way.

   \sk Let us briefly discuss the kind of topological invariants that
   Chern-Simons theory can lead to. Recall that one can construct a Lie group
   element $g$ from a Lie algebra element $\cA$ as follows:
   \be
    g = e^{\cA}.
   \ee
   Now suppose we have a path $\gam(t)$ inside $\cM$. Suppose that we chop up
   $\gam$ into very small line elements given by tangent vectors
   \be
    \frac{d \gam}{dt} \gd t
   \ee
   Then we can insert this tangent vector into the connection one-form $A$, and
   obtain a Lie algebra element. As we have seen, it is precisely this Lie
   algebra element which transports vectors in $E$ along this small distance: we
   have to multiply these vectors by $1 + \cA$. Of course, this is a linear
   approximation to the finite transformation $e^{\cA}$. So if we transport a
   vector along the entire closed curve $\gam$, it will return multiplied by a
   group element
   \be
    g = \lim_{\gd t \to 0} \left[
    \exp \left( A \left( \frac{d \gam}{dt}(0) \right) \gd t \right)
    \exp \left( A \left( \frac{d \gam}{dt}(\gd t) \right) \gd t \right)
    \exp \left( A \left( \frac{d \gam}{dt}(2 \gd t) \right) \gd t \right)
    \cdots
    \right].
   \ee
   Now it is tempting to add all the exponents and write their sum in
   the limit as an integral, but this is not quite allowed since the different
   group elements may not commute, so $e^X e^Y \neq e^{X+Y}$. Therefore, one uses the following notation
   for this type of expression:
   \be
    g = P \exp \int_\gam A
   \ee
   where $P$ stands for path ordering. The element $g$ is called the {\em
   holonomy} of $A$ around the closed curve $\gam$. An interesting gauge and
   metric independent object turns out to be the trace of this group
   element. This trace is called the {\em Wilson loop} $W_\gam(A)$:
   \be
    W_\gam(A) = \Tr~ P \exp \int_\gam A.
   \ee
   The topological invariants we are interested in are now the correlation
   functions of such Wilson loops in Chern-Simons theory. Since these
   correlation functions are independent of the parametrization of
   $\cM$, we can equivalently say that they will be independent of the precise
   location of the loop $\gam$; we have in fact constructed a topological
   invariant of the embedding of $\gam$ inside $\cM$! This embedding takes the
   shape of a knot, so the
   invariants we have constructed are knot invariants. One can show that the
   invariants are actually polynomials in the variable $y = \exp{2 \pi i
   /(k+2)}$, where $k$ is the integer ``coupling constant'' of the Chern-Simons
   theory.

   \sk The above construction is due to E.~Witten, and was carried out
   in \cite{Witten:1988hf} in 1988. Before Witten's work, several polynomial
   invariants of knots were known, one of the simplest ones being the so-called
   Jones polynomial. It can be shown that many of these polynomials arise
   as special cases of the above construction, where one takes a certain 
   structure group $G$, $SU(2)$ for the Jones polynomial, and a certain
   vector bundle (representation) $E$, the fundamental representation
   for the Jones polynomial. That is, using this ``trivial'' topological field
   theory, Witten was able to reproduce a large number of the known knot
   invariants in a unified framework, and construct a huge number of new
   invariants as well!

  \subsection{Cohomological field theories}
   \label{sec:cohomologicalft}
   Of course, even though the above example leads to quite interesting
   topological invariants, the construction itself is somewhat trivial: given
   the absence of anomalies that we mentioned, the
   independence of the metric is completely manifest throughout the procedure.
   There exists a different way of constructing topological field theories in
   which the definition of the theory {\em does} use a metric, but one can still
   show that the partition function and the physical correlation functions of
   the theory are metric-independent. The theories constructed in this way are
   called topological theories of ``Witten-type'', or cohomological field
   theories.

   \sk Cohomological field theories are field theories that possess a very
   special type of symmetry. Recall that from Noether's theorem a global
   symmetry of a theory leads to a conserved charge $S$, and that after
   quantizing the theory, the symmetry is generated by the corresponding
   operator:
   \be
    \gd_\eps \cO_i = i \eps [S, \cO_i] \qquad \mbox{or} \qquad \gd_\eps \cO_i =
    i \eps \{S, \cO_i\},
   \ee
   depending on whether $S$ and $\cO_i$ are fermionic or bosonic. Furthermore,
   the symmetry-invariant states $|j \rangle$ satisfy
   \be
    S \, |j \rangle = 0.
    \label{eq:symmetricstate}
   \ee
   In particular, if the symmetry is not spontaneously broken, the vacuum of
   such a theory will be symmetric:
   \be
    S \, |0 \rangle = 0,
   \ee
   and expectation values of operators will be unchanged after a symmetry
   transformation:
   \bea
    \langle 0 | \cO_i + \gd \cO_i | 0 \rangle & = & \langle 0 | \cO_i  | 0
    \rangle + i \eps \langle 0 | S \cO_i \pm \cO_i S | 0 \rangle \ret
    & = & \langle 0 | \cO_i  | 0 \rangle,
   \eea
   since $S$ annihilates the vacuum. Note that, to linear order in $\eps$, the
   second term in the first line (with $|0 \rangle$ replaced by an arbitrary
   state $| \psi \rangle$) can also be obtained if instead of on the operators,
   we let the symmetry operator $S$ act on the {\em state} as
   \be
    |\psi \rangle \to |\psi \rangle + i \eps S |\psi \rangle.
    \label{eq:symmetrystate}
   \ee
   In case $S$ is the Hamiltonian, this is the infinitesimal version of the
   well-known transition between the Schr\"odinger and Heisenberg pictures.
   Note that the equations (\ref{eq:symmetricstate}) and
   (\ref{eq:symmetrystate}) already contain some flavor of cohomology.
   Cohomological field theories are theories where this analogy can be made
   exact.

   \sk With this in mind, the first property of a cohomological field theory
   will not come as a surprise: it should contain a fermionic symmetry operator
   $Q$ which squares to zero,
   \be
    Q^2 = 0.
   \ee
   This may seem like a strange requirement for a field theory, but symmetries
   of this type occur for example when we have a gauge symmetry and fix it by
   using the Faddeev-Popov procedure; the resulting theory will then have a
   global symmetry called {\em BRST-symmetry}, which satisfies precisely this
   constraint. Another example is found in supersymmetry, where one also
   encounters symmetry operators that square to zero, as we will see in detail
   later on.

   \sk The second property a cohomological field theory should have is really
   a definition: we define the physical operators in this theory to be
   the operators that are closed under the action of this
   $Q$-operator\footnote{From now on, we denote both the commutator and the
   anti-commutator by curly brackets, unless it is clear that one of the
   operators inside the brackets is bosonic.}:
   \be
    \{Q, \cO_i\} = 0.
    \label{eq:symmetricoperator}
   \ee
   Again, this may seem to be a strange requirement for a physical theory, but
   again it naturally appears in BRST quantization, and for example in conformal
   field theories, where we have a one-to-one correspondence between operators
   and states. In such theories, the symmetry requirement
   (\ref{eq:symmetricstate}) on the states translates into the requirement
   (\ref{eq:symmetricoperator}) on the operators.

   \sk Thirdly, we want to have a theory in which the $Q$-symmetry is not
   spontaneously broken, so the vacuum is symmetric. Note that this implies the
   equivalence
   \be
    \cO_i \sim \cO_i + \{ Q, \gL \}.
    \label{eq:operatorequivalence}
   \ee
   The reason for this is that the expectation value of an operator product
   involving a $Q$-exact operator $\{Q, \gL \}$ takes the form
   \be
    \langle 0 | \cO_{i_1} \cdots \cO_{i_j} \{Q, \gL \} \cO_{i_{j+1}} \cdots
    \cO_{i_n}| 0 \rangle = \langle 0 | \cO_{i_1} \cdots \cO_{i_j} (Q \gL - \gL
    Q) \cO_{i_{j+1}} \cdots \cO_{i_n}| 0 \rangle,
   \ee
   and each term vanishes separately, e.\ g.\
   \bea
    \langle 0 | \cO_{i_1} \cdots \cO_{i_j} Q \gL \cO_{i_{j+1}} \cdots
    \cO_{i_n}| 0 \rangle & = & \pm \langle 0 | \cO_{i_1} \cdots Q \cO_{i_j} \gL
    \cO_{i_{j+1}} \cdots \cO_{i_n}| 0 \rangle \ret
    & = & \pm \langle 0 | Q \cO_{i_1} \cdots \cO_{i_j} \gL \cO_{i_{j+1}} \cdots
    \cO_{i_n}| 0 \rangle \ret
    &=& 0,
    \label{eq:vanishing}
   \eea
   where we made repeated use of (\ref{eq:symmetricoperator}). Together,
   (\ref{eq:symmetricoperator}) and (\ref{eq:operatorequivalence}) mean that
   our physical operators are $Q$-cohomology classes.

   \sk The fourth and final requirement for a cohomological field theory is
   that the energy-momentum tensor is $Q$-exact:
   \be
    T_{\ga \gb} \equiv \frac{\gd S}{\gd h^{\ga \gb}} = \{ Q, G_{\ga \gb} \}
    \label{eq:exactem}
   \ee
   for some operator $G_{\ga \gb}$. The physical interpretation of this is the
   following. The integrals of the components $T_{0\ga}$ over a space-like
   hypersurface are conserved quantities. For example, the integral of $T_{00}$
   gives the Hamiltonian:
   \be
    H = \int_{space} T_{00}.
   \ee
   Similarly, $T_{0a}$ for $a \neq 0$ give the momentum charges. Certainly, the
   Hamiltonian $H$ should commute with all symmetry operators of the theory, and
   one usually takes the other space-time symmetries to commute with the
   internal symmetries as well. Of course, the choice of the first lower index 0
   here is related to a choice of Lorentz frame, so in general the integrals of
   all $T_{\ga \gb}$ will commute with the internal symmetries. In a local
   theory, it is then natural to assume that also the {\em densities}
   commute with $Q$. However, (\ref{eq:exactem}) is an even stronger
   requirement: the energy-momentum densities should not only commute with $Q$
   (that is, be $Q$-closed), but they have to do so in a trivial way (they
   should be $Q$-commutators, that is, $Q$-exact) for the theory to be
   cohomological.

   \sk This fourth requirement is the crucial one in showing the topological
   invariance of the theory. Let us consider the functional
   $h^{\ga \gb}$-derivative of an observable:
   \bea
    \frac{\gd}{\gd h^{\ga \gb}} \langle \cO_{i_1} \cdots \cO_{i_n} \rangle & = &
    \frac{\gd}{\gd h^{\ga \gb}} \left( \int D \phi \, \cO_{i_1} \cdots \cO_{i_n}
    e^{iS[\phi]} \right) \ret
    & = & i \int D \phi \, \cO_{i_1} \cdots \cO_{i_n} \frac{\gd S}{\gd h^{\ga
    \gb}} e^{iS[\phi]} \ret
    & = & i \langle \cO_{i_1} \cdots \cO_{i_n} \{ Q, G_{\ga \gb} \} \rangle \ret
    & = & 0,
   \eea
   where in the last line we used the same argument as in (\ref{eq:vanishing}).
   One might be worried about the operator ordering in going from the operator
   formalism to the path integral formalism and back. We should really have
   inserted a time ordering operator on the right hand side in the first step
   above. However, the result then shows us that we can arbitrarily
   change the metric -- and hence the time ordering of the operators, so
   with hindsight we may actually think of the operators as being arbitrarily
   ordered. Finally, we have of course assumed that our operators do not depend
   explicitly on the metric.

   \sk A very practical way to ensure (\ref{eq:exactem}) is to use a Lagrangean
   which itself is $Q$-exact:
   \be
    L = \{ Q, V \}
   \ee
   for some operator $V$. This choice has an extra virtue, which we can see if
   we explicitly include Planck's constant in our description: the quantum
   measure then reads
   \be
    \exp \frac{i}{\hbar} \left\{ Q, \int_\cM V \right\}.
   \ee
   Then, we can use exactly the same argument as before to show that
   \be
    \frac{d}{d \hbar} \langle \cO_{i_1} \cdots \cO_{i_n} \rangle = 0.
   \ee
   That is, the correlators we are interested in are independent of $\hbar$,
   and we can therefore calculate them {\em exactly} in the classical limit!
   This observation of course makes life a lot easier, and it will be of
   crucial importance to us.

   \subsubsection{Descent equations}
    \label{sec:descenteqns}
    An important property of cohomological field theories is that, given a
    scalar physical operator on $\cM$ -- where by ``scalar'' we mean an operator
    that does not transform under coordinate transformations of $\cM$, so in
    particular it has no $\ga$-indices -- we can construct further operators
    which behave like $p$-forms on $\cM$.  The basic observation is that we can
    integrate (\ref{eq:exactem}) over a spatial hypersurface to obtain a similar
    relation for the momentum operators:
    \be
     P_\ga = \{ Q, G_\ga \}
    \ee
    where $G_\ga$ is a fermionic operator. Now consider the operator
    \be
     \cO^{(1)}_\ga = i \{ G_\ga, \cO^{(0)} \},
    \ee
    where $\cO^{(0)}(x)$ is a scalar physical operator: $\{ Q, \cO^{(0)}(x) \} =
    0$. Let us calculate
    \bea
     \frac{d}{d x^\ga} \cO^{(0)} & = & i [P_\ga, \cO^{(0)}] \ret
     & = & i [ \{ Q, G_\ga \} , \cO^{(0)} ] \ret
     & = & \pm i \{ \{G_\ga, \cO^{(0)} \}, Q \} - i \{ \{ \cO^{(0)}, Q \}, G_\ga
     \}\ret
     & = & \{ Q, \cO^{(1)}_\ga \}.
    \eea
    In going from the second to the third line, we have used the Jacobi
    identity. The first sign in the third line depends on whether $\cO^{(0)}$ is
    bosonic or fermionic, but there is no sign ambiguity in the last line. 
    By defining the one-form operator
    \be
     \cO^{(1)} = \cO^{(1)}_\ga dx^\ga
    \ee
    we can write this result as
    \be
     d \cO^{(0)} = \{ Q, \cO^{(1)} \}.
    \ee
    Then, we can integrate this equation over a closed curve $\gam \subset \cM$
    to find
    \be
     \{ Q, \int_\gam \cO^{(1)} \} = 0.
    \ee
    That is, by constructing a $\int_\gam \cO^{(1)}$ for each $\cO^{(0)}$, we
    have found a whole class of new, non-local, physical operators!

    \sk The above procedure can now be repeated in exactly the same way starting
    from $\cO^{(1)}$, and doing this we find a whole tower of $p$-form
    operators:
    \bea
     \{ Q, \cO^{(0)} \} & = & 0 \ret
     \{ Q, \cO^{(1)} \} & = & d \cO^{(0)} \ret
     \{ Q, \cO^{(2)} \} & = & d \cO^{(1)} \ret
     & \cdots & \ret
     \{ Q, \cO^{(n)} \} & = & d \cO^{(n-1)} \ret
     0 & = & d \cO^{(n)}.
    \eea
    The last equation is of course trivial, since there are no $(n+1)$-forms on
    an $n$-dimensional manifold.

    \sk
    Following the same reasoning, the integral of $\cO^{(p)}$ over a
    $p$-dimensional submanifold of $\cM$ is now a physical operator. This gives
    us a large class of new physical operators, starting from the scalar ones.
    Note that these operators, being integrated over a submanifold, are
    inherently nonlocal. Nevertheless, they can have a very physical
    interpretation. Particularly important examples of this are the
    ``top-form'' operators $\cO^{(n)}$ that can be integrated over the whole
    manifold, leading to
    \be
     \left\{ Q, \int_\cM \cO^{(n)} \right\} = 0.
    \ee
    This implies that we can add terms $t^a \cO_a^{(n)}$, with $t^a$ arbitrary
    coupling constants, to our Lagrangean without spoiling the fact that the
    theory is cohomological. These deformations of the theory will be important
    to us later.

   \subsubsection{Two-dimensional cohomological field theories}
    Since string theories are two-dimensional field theories, we will in
    particular be interested in cohomological field theories in two dimensions.
    These theories have some extra properties which will be important in our
    discussion. Let us begin by reminding the reader of the relation between
    states in the operator formalism of quantum field theories, and boundary
    conditions in the path integral formalism. In its simplest form, this
    relation looks like
    \be
      \int_{BC1}^{BC2} D \phi \cdots e^{iS[\phi]} = \langle BC1
      | \, T(\cdots) \, | BC2 \rangle.
     \label{eq:relstatebc}
    \ee
    Here, we included a time-ordering operator for completeness, but as we have
    stated before, for the theories we are interested in this ordering is
    irrelevant\footnote{To avoid confusion, note that one of course still
    obtains a minus sign when changing the order of two fermionic
    fields on the left hand side or of the corresponding operators on the right
    hand side!}. The notation $|BCi \rangle$ indicates the state
    corresponding to the incoming or outgoing boundary condition. For example,
    if on the path integral side we prescribe all fields at a certain initial
    time, $\phi(t\nolinebreak=\nolinebreak t_1) = f(t_1)$ , on the operator side this corresponds to the
    incoming state satisfying
    \be
     \phi(t_1) | BC1 \rangle = f(t_1) | BC1 \rangle.
     \label{eq:statebc}
    \ee
    where on the left hand side we have an operator acting, but on the right
    hand side there is a simple scalar multiplication.
    More generally, in the operator formalism we can have linear combinations
    of states of the type (\ref{eq:statebc}). Therefore, we should allow for
    linear combinations on the left hand side of (\ref{eq:relstatebc}) as well.
    In other words, in the path integral formalism {\em a state is an operator
    which adds a number (a weight) to each possible boundary condition on the
    fields}. From this point of view, the states in (\ref{eq:statebc}) are like
    ``delta-functionals'': they assign weight 1 to the boundary condition
    $\phi(t_1) = f(t_1)$, and weight 0 to all other boundary
    conditions\footnote{Of course, for general states, one gets into serious
    mathematical problems if one requires that the integrated weight over
    the space of all boundary conditions is 1, since this would require a
    rigorous definition of a functional integral. As is usual in physics, we
    will assume that these problems can be overcome in a yet to be defined
    mathematical formalism.}.

    \begin{figure}[ht]
     \begin{center}
      \includegraphics[height=5cm]{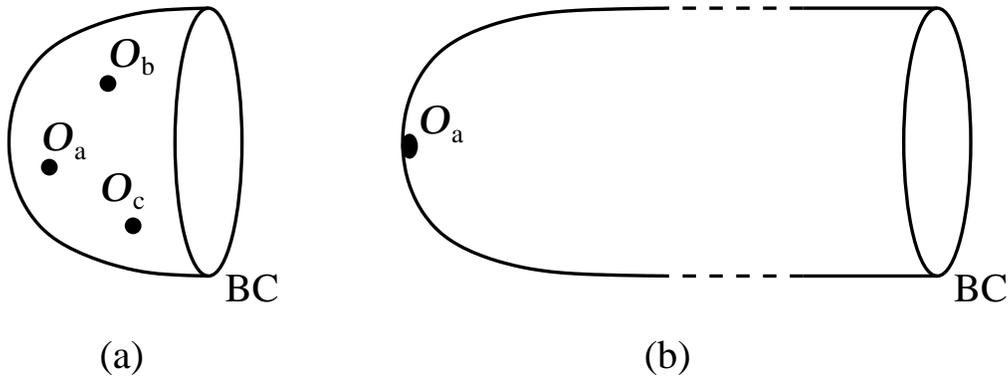}
     \end{center}
     \caption{(a) A state which assigns to each boundary condition BC on the
     circle the value $\langle \cdots \rangle_{BC}$ of a certain path integral,
     possibly with operator insertions, over a hemisphere which has the circle as
     a boundary . (b) An operator defines an ``asymptotic'' state by inserting
     it on an infinitely stretched hemisphere.}
     \label{fig:pistate}
    \end{figure}

\begin{figure}[ht]
     \begin{center}
      \includegraphics[height=2cm]{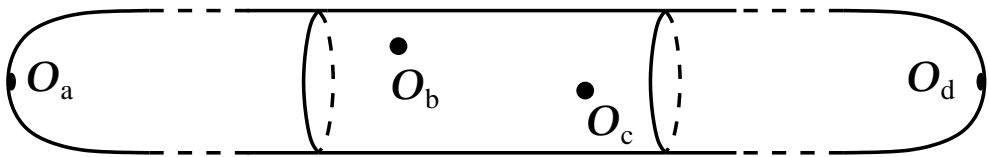}
     \end{center}
     \caption{A graphical representation of the correlation function
     (\ref{eq:correlator}).}
     \label{fig:correlator}
    \end{figure}

    \sk Now, let us specialize to two-dimensional field theories. Here, the
    boundary of a compact manifold is a set of circles. Let us for simplicity
    assume that the ``incoming'' boundary consists of a single circle. We can
    now define a state in the above sense by doing a path integral over a second
    surface with the topology of a hemisphere, as shown in figure
    \ref{fig:pistate}a. This path integral gives a number for every boundary
    condition of the fields on the circle, and this is exactly what a state in
    the path-integral formalism should do. In particular, one can use this
    procedure to define a state corresponding to every operator $\cO_a$ by
    inserting $\cO_a$ on the hemisphere and then stretching this hemisphere to
    infinite size, as indicated in figure \ref{fig:pistate}b. An expectation
    value in the operator formalism, such as
    \be
     \langle \cO_{a} |  \cO_{b}(x_2) \cO_{c}(x_3) | \cO_{d}
     \rangle_{cyl},
     \label{eq:correlator}
    \ee
    on a cylinder of finite length, can then schematically be drawn as in figure
    \ref{fig:correlator}. Here, instead of first doing the path integrals over
    the semi-infinite hemispheres and inserting the result in the path integral
    over the cylinder, one can of course just as well integrate over the whole
    surface at once. Of course, one could give another definition of the
    state related to an operator which does not involve the infinite stretching,
    but in general field theories, the reason for doing things in this way is
    that it is exactly the kind of thing one does when one defines asymptotic
    states and an S-matrix. However, note
    that in topological field theories, there is simply no {\em need} to do the
    stretching, since the path integral only depends on the topology of the
    surface, and hence not on the size of the hemisphere.

\begin{figure}[ht]
     \begin{center}
      \includegraphics[height=8cm]{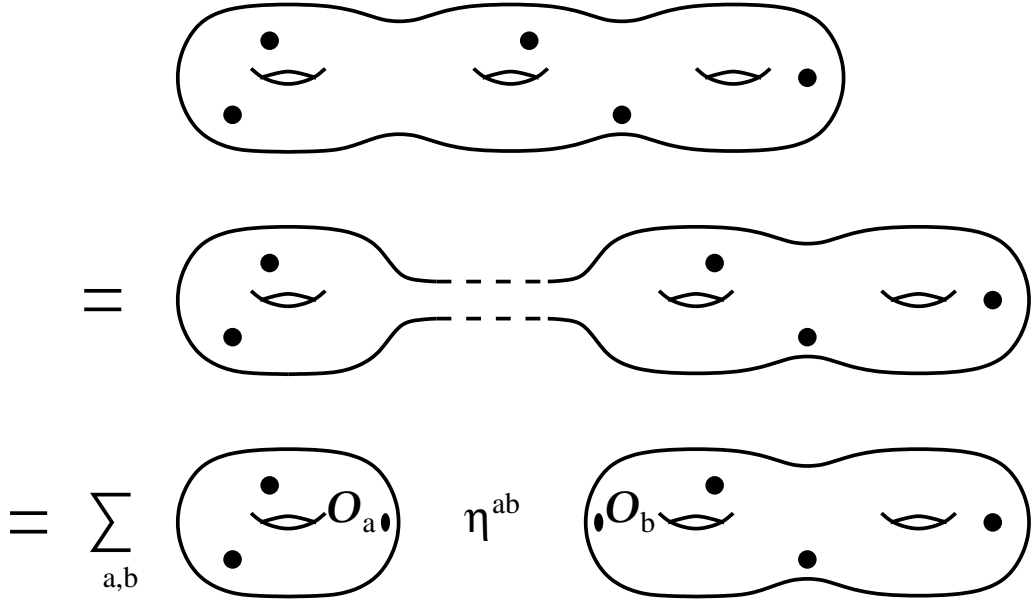}
     \end{center}
     \caption{A correlation function on a Riemann surface factorizes into
     correlation functions on two Riemann surfaces of lower genus.}
     \label{fig:factorization}
    \end{figure}

    \sk Until further notice, we will assume that all states that we are
    interested in are of the above form\footnote{This will change as soon as one
    adds D-branes to the theory, for example.}, which in particular means that
    we will integrate only over Riemann surfaces without boundary. Moreover, we
    assume these surfaces to be orientable. An important property of topological
    field theories in two dimensions is now
    that its correlation functions factorize in the following way:
    \be
     \langle \cO_1 \cdots \cO_n \rangle_\gS = \sum_{a,b} \langle \cO_1 \cdots
     \cO_i \cO_a \rangle_{\gS_1} \eta^{ab} \langle \cO_b \cO_{i+1} \cdots
     \cO_n \rangle_{\gS_2},
     \label{eq:factorization}
    \ee
    where the genus of $\gS$ is the sum of the genera of $\gS_1$ and $\gS_2$.
    This statement is explained in figure \ref{fig:factorization}. By using the
    topological invariance, we can deform a Riemann surface $\gS$ with a set of
    operator insertions in such a way that it develops a long tube. From general
    quantum field theory, we know that if we stretch this tube long enough, only
    the asymptotic states -- that is, the states in the physical part of the
    Hilbert space -- will propagate. But as we have just argued, instead of
    inserting these asymptotic states, we may just as well insert the
    corresponding operators at a finite distance. However, to conclude that this
    leads to (\ref{eq:factorization}), we have to show that this definition of
    ``physical states'' -- being the ones that need to be inserted as asymptotic
    states -- agrees with our previous definition in terms of $Q$-cohomology.
    Let us argue that it does by first writing the factorization as
    \be
     \langle \cO_1 \cdots \cO_n \rangle_\gS = \sum_{A,B} \langle \cO_1 \cdots
     \cO_i | \cO_A \rangle_{\gS_1} \eta^{AB} \langle \cO_B | \cO_{i+1} \cdots
     \cO_n \rangle_{\gS_2}.
     \label{eq:factorization2}
    \ee
    where the $\cO_A$ with capital index now correspond to a {\em complete}
    basis of asymptotic states in
    the Hilbert space. The reader may be more used to this type of expressions
    in the case where $\eta_{AB} = \gd_{AB}$, but since we have not shown that
    with our definitions $\langle \cO_A | \cO_B \rangle = \gd_{AB}$, we have to
    work with this more general form of the identity operator, where $\eta$ is
    a metric that we will determine in a moment.
    
    \sk
    Now, we can write the Hilbert space as a direct sum, $\cH =
    \cH_0 \oplus \cH_1$, where $\cH_0$ consists of states $| \psi \rangle$ for
    which $Q | \psi \rangle = 0$ and $\cH_1$ is its orthogonal complement. Since
    \be
     Q \left( \cO_1 \ldots \cO_i | 0 \rangle \right) = 0,
    \ee
    the states in $\cH_1$ are in particular orthogonal to states of the form
    $\cO_1 \ldots \cO_i | 0 \rangle$, and hence the states in $\cH_1$ do not 
    contribute to the sum in (\ref{eq:factorization2}). Moreover, changing
    $\cO_A$ to $\cO_A + \{ Q, \gL \}$ does not change the result in
    (\ref{eq:factorization2}), so we only need to sum over a basis of $\cH_0 /
    \Im(\{ Q, \cdot \})$, which is exactly the space $\cH_{phys}$ of
    ``topologically physical'' states! (In fact, we also have to multiply by
    the number of states in $\Im (\{Q, \cdot \})$ as well of course, but we can
    absorb this factor in the definition of $\eta^{ab}$.)

    \sk Finally, let us determine the metric $\eta^{ab}$. We can easily deduce
    its form by factorizing the two-point function
    \be
     C_{ab} = \langle \cO_a \cO_b \rangle
    \ee
    in the above way, resulting in
    \be
     C_{ab} = C_{ac} \eta^{cd} C_{db}.
    \ee
    In other words, we find that the metric $\eta^{ab}$ is simply the matrix
    inverse of the two-point function $C_{ab}$, which for this reason we will
    write as $\eta_{ab}$ from now on.

    \begin{figure}[ht]
     \begin{center}
      \includegraphics[height=8cm]{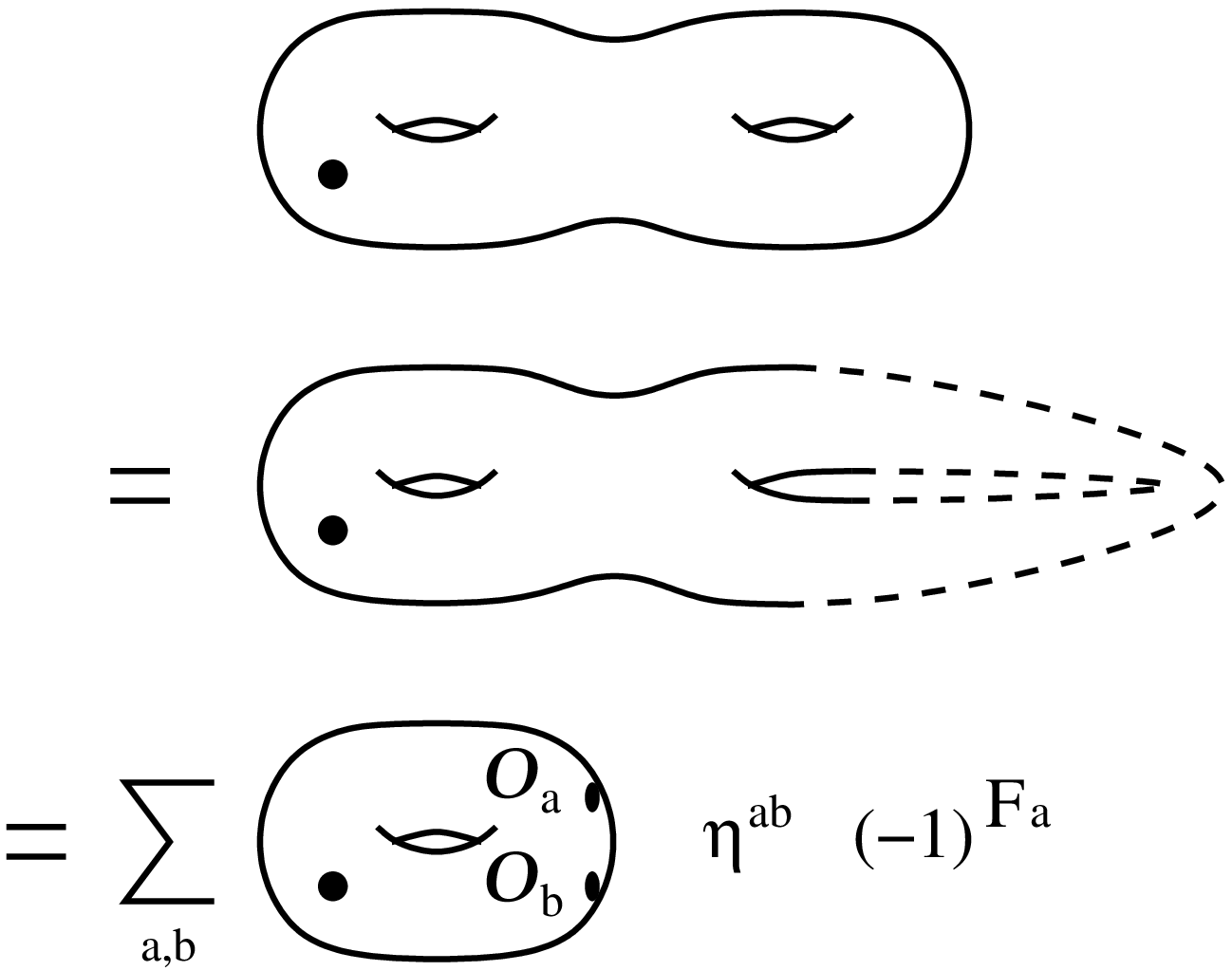}
     \end{center}
     \caption{A correlation function on a Riemann surface factorizes into
     correlation functions on a Riemann surface of lower genus by
     stretching an internal tube to infinite length.}
     \label{fig:factorinternal}
    \end{figure}

    \sk One can apply a similar procedure to ``cut open'' internal loops in a
    Riemann surface $\gS$. This is illustrated in figure
    \ref{fig:factorinternal}, from which we see that
    \be
     \langle \cO_1 \cdots \cO_n \rangle_\gS = (-1)^{F_a} \eta^{ab} \sum_{a,b}
     \langle \cO_a \cO_b \cO_1 \cdots \cO_n \rangle_{\gS'},
     \label{eq:factorinternal}
    \ee
    where the genus of $\gS'$ is one less than the genus of $\gS$.
    The factor $(-1)^{F_a}$ multiplies the expression on the right hand side by
    $-1$ if the inserted operator $\cO_a$ (and hence also $\cO_b$) is fermionic.
    Proving that it needs to be included is not straightforward, but one can
    think of it as the ``stringy version'' of the well-known statement from
    quantum field theory that fermion loops add an extra minus sign to a Feynman
    diagram.

    \sk The reader should convince himself that together, the equations
    (\ref{eq:factorization}) and (\ref{eq:factorinternal}) imply that we can
    reduce all $n$-point correlation functions to products of three-point
    functions on the sphere. We denote these important quantities by
    \be
     c_{abc} \equiv \langle \cO_a \cO_b \cO_c \rangle_0,
    \ee
    where the label 0 denotes the genus of the sphere. By using
    (\ref{eq:factorization}) to separate two insertion points on a sphere, we
    see that
    \bea
     \langle \cdots \cO_a \cO_b \cdots \rangle_\gS & = & \sum_{c,d} \langle
     \cdots \cO_c \cdots \rangle_\gS \, \eta^{cd} c_{abd} \ret
     & = & \sum_{c} \langle \cdots \cO_c \cdots \rangle_\gS \, {c_{ab}}^c,
    \eea
    where we raised an index of $c_{abc}$ with the metric $\eta$. We can view
    the above result as the definition of an operator algebra with structure
    constants ${c_{ab}}^c$:
    \be
     \cO_a \cO_b = \sum_c {c_{ab}}^c \cO_c.
    \ee
    From the metric $\eta_{ab}$ and the structure constants ${c_{ab}}^c$, we can
    now calculate any correlation function in the cohomological field theory.

    \sk As a final remark, note that when we turn on deformations $t^a
    \cO_a^{(2)}$ in the Lagrangean, $\eta_{ab}$ and $c_{abc}$ will become
    $t$-dependent. On can calculate the $t^d$-derivative of $c_{abc}$, for
    example, as
    \be
     \frac{d c_{abc}}{d t^d} = \langle \cO_a \cO_b \cO_c \int \cO_d^{(2)}
     \rangle,
    \ee
    as is clear from writing this expression in the path integral language.
    After we have introduced topological string theory, we will further study
    this $t$-dependence of the three-point functions. An important point will be
    that once we introduce integrated operators in our correlation functions, we
    can no longer naively perform the simple genus-reducing operations of figure
    \ref{fig:factorization} and \ref{fig:factorinternal}, since the operator
    $\cO^{(2)}$ has to be integrated over the infinite tube as well, and moreover
    it may have contact terms when its position equals that of one of the other
    $\cO$.

 \section{Calabi-Yau manifolds and their moduli spaces}
  \label{sec:calabiyau}
  In this section we will introduce Calabi-Yau manifolds and their moduli
  spaces. Later, we will see that Calabi-Yau spaces are the most natural target
  spaces for topological strings to propagate in. The dependence of the
  topological string results on the moduli of the Calabi-Yau will be crucial in
  our story.

  \subsection{Calabi-Yau manifolds}
  Let us begin with a bit of notation. The tangent bundle of a complex
  $m$-dimensional manifold has basis vectors $\d / \d z^i$ and $\d / \d
  \zbar^i$ for $i=1, \ldots, m$. A general element of the tangent vector space
  at a certain point can be written as
  \be
   v^i \frac{\d}{\d z^i} + \vbar^i \frac{\d}{\d \zbar^i}.
  \ee
  Recall that a complex $m$-dimensional manifold is automatically a real
  $2m$-dimensional manifold. This means that if we write $z = x + iy$, the
  prefactors of $\d / \d x^i$ and $\d / \d y^i$ must be real. From this, it
  follows that $v^i$ and $\vbar^i$ are complex conjugates. Let us write this out
  in detail:
  \bea
   \frac{\d}{\d x^i} & = & \frac{\d}{\d z^i} + \frac{\d}{\d \zbar^i} \ret
   \frac{\d}{\d y^i} & = & i \frac{\d}{\d z^i} - i \frac{\d}{\d \zbar^i},
  \eea
  so a vector
  \be
   a^i \frac{\d}{\d x^i} + b^i \frac{\d}{\d y^i},
  \ee
  with $a^i$ and $b^i$ real, can be written in terms of complex coordinates as
  \bea
   (a^i + i b^i) \frac{\d}{\d z^i} + (a^i - i b^i) \frac{\d}{\d z^i},
  \eea
  where indeed the coefficients are complex conjugates. An important operator on
  the tangent bundle is now the operator induced by multiplication of the
  complex coordinates $z$ by $i$:
  \be
   z' = i z, \qquad \zbar' = - i \zbar
  \ee
  This operator is called the {\em complex structure} operator, since its
  eigenvectors are the holomorphic and the anti-holomorphic coordinates. Note
  that it can also be written as
  \be
   x' = - y \qquad y' = x
  \ee
  This map extends to the tangent vector spaces in a natural way:
  \be
   a' = - b \qquad b' = a
  \ee
  which can be written as an action on the complex components $v$:
  \be
   (v')^j = i v^j \qquad (\vbar')^j = - i \vbar^j.
  \ee
  We write this as a linear map $J$: $v' = J v$, so $J$ has diagonal
  components $\pm i$.

  \sk
  Now, as discussed in section \ref{sec:vbexamples}, we can go to the {\em
  complexified} 
  tangent space, where we allow the coefficients $a$ and $b$ to be complex.
  As a result the $v^i$ and $\vbar^i$ are independent complex numbers. Of
  course, the notation $v^i, \vbar^i$ is then somewhat confusing, so we change
  it to
  \be
   v^i \frac{\d}{\d z^i} + v^{\ibar} \frac{\d}{\d \zbar^i}.
   \label{eq:vhol}
  \ee
  Here, one should think of $\ibar$ as $i+m$, and of $\zbar^i$ as $z^{i+m}$.
  (Sometimes the notation $\zbar^{\ibar}$ is also used, but this feels a
  bit like ``double counting''.) As in (\ref{eq:vhol}), if we do not explicitly
  state otherwise, the summation convention will always imply a sum over $m$
  indices, either $i$ or $\ibar$, but not both. Of course, the
  complex structure $J$ also straightforwardly extends to a linear operator on
  the complexified tangent space. However, note that here the operator no
  longer has the simple interpretation of ``multiplying complex conjugate
  components by $\pm i$''.

  \sk Usually, a metric $g$ on the real tangent space of $\cM$ is taken to be
  real itself, meaning that $g(v,w) \in \bR$ for any (real) tangent vectors $v$
  and $w$. It is a straightforward exercise to check that in complex coordinates,
  this implies that
  \bea
   g_{\ibar \jbar} & = & (g_{ij})^* \ret
   g_{\ibar j} & = & (g_{i \jbar})^*.
  \eea
  If we have a metric on the complexified tangent space with these same
  properties, this is also called a ``real metric''. Moreover, we can now
  consider metrics that are compatible with the complex
  structure, meaning that
  \be
   g(v,w) = g(Jv, Jw).
  \ee
  Writing out the definitions, one finds that such a metric satisfies
  \be
   g_{ij} = g_{\ibar \jbar} = 0.
  \ee
  A real metric of this type is called {\em Hermitean}.

  \sk A {\em local K\"ahler metric} is now a Hermitean metric on a certain patch
  which we can write as a second derivative of some function:
  \be
   g_{i \jbar} = \frac{\d K(z, \zbar)}{\d z^i \d \zbar^j}.
  \ee
  Since the patch has the topology of an $m$-dimensional ball, the famous
  Poincar\'e lemma states that this condition is equivalent to the
  ``integrability condition''
  \be
   \frac{\d g_{i \jbar}}{\d z^k} = \frac{\d g_{k \jbar}}{\d z^i},
  \ee
  and the similar condition for the anti-holomorphic derivatives. Note that we
  can write this in form notation as
  \be
   d \go = 0,
   \label{eq:Kahlercondition}
  \ee
  where
  \be
   \go \equiv 2 i \d \delbar K, \qquad \mbox{i.\ e.} \qquad \go_{i \jbar} = -
   \go_{\ibar j} = i g_{i \jbar},
  \ee
  and the other components of $\go$ vanish. The conventional factor of $i$ is
  inserted in this definition to guarantee that for a real metric $g$, the
  two-form $\go$ is also real. (Again, this means that $\go(v,w) \in \bR$
  for real vectors.) Note that, in this way, $\go$ can also be {\em globally}
  defined in terms of the metric, so we define a (global) {\em K\"ahler
  manifold} to be a complex manifold with a Hermitean metric satisfying
  (\ref{eq:Kahlercondition}). As we have shown, 
  locally this means that one can write such a metric in terms of a function
  $K$, the {\em K\"ahler potential}. The globally defined two-form $\go$ is called
  the {\em K\"ahler form}, and its cohomology class the {\em K\"ahler class}.

  \sk From the metric $g$, we can now calculate the Riemann and Ricci tensors.
  It turns out that for a K\"ahler metric, many components of these tensors
  vanish identically. In particular, for the Ricci tensor the components
  $R_{ij}$ and $R_{\ibar \jbar}$ vanish. A {\em Calabi-Yau manifold} is now
  defined to be a Ricci-flat K\"ahler manifold, i.\ e.\ one for which also
  \be
   R_{i \jbar} = 0.
  \ee
  One reason this condition is interesting is of course that Ricci-flat
  manifolds are solutions to the vacuum Einstein equations. (However, this does
  of course not require the K\"ahler condition, or even the Hermitean one.) A
  second reason is that in string theory compactifications, both for technical
  and for phenomenological reasons\footnote{Of course, in our real, low-energy
  world we do not see supersymmetry. However, it is likely that if
  supersymmetry is present in nature at all, the breaking of the last
  supersymmetry will take place at a much larger distance scale than the usual
  scale of string compactification, so the compactification procedure would
  need to preserve at least $N=1$ supersymmetry.}, we are interested
  in preserving some supersymmetry after the compactification. If we denote the
  supersymmetry generator by $Q$, then a supersymmetry transformation is
  performed by acting on the fields with $\eps_{(10)} Q$, where $\eps_{(10)}$
  is some ten-dimensional spinor. To get four-dimensional supersymmetry, we want
  to write $\eps_{(10)} = \eps_{(4)} \otimes \eps_{(6)}$, but to properly
  compactify the theory we have to be able to pull $\eps_{(6)}$ through all the
  covariant derivatives -- i.\ e.\ it must be covariantly constant\footnote{The
  reader who is interested in a precise version of this handwaving argument
  is referred to the seminal paper \cite{Candelas:1985en} by P.~Candelas,
  G.~Horowitz, A.~Strominger and E.~Witten.}. Therefore,
  supersymmetry is only preserved after compactification if the internal
  manifold $\cM$ admits a covariantly constant spinor. One can now show that
  this condition is equivalent to the conditions that $\cM$ be complex,
  K\"ahler and Ricci-flat\footnote{Often, Ricci-flatness is replaced by the
  condition of $SU(d)$-holonomy, but one can show that given the other two
  conditions, these two are equivalent.}.

  \sk It turns out to be extremely hard to construct six-dimensional compact
  complex manifolds with Ricci-flat K\"ahler metrics; in fact, not a single
  nontrivial example is known! However, a very useful fact is that the
  Ricci-flatness implies that the first Chern class of the tangent bundle (also
  simply referred to as the first Chern class of the manifold itself,
  $c_1(\cM)$) vanishes. One can
  now ask the converse question: given a K\"ahler manifold with vanishing first
  Chern class, is it Ricci-flat? The answer is ``no'', since
  infinitesimal changes of the metric do not change the first Chern class, but
  they can change the Ricci-flatness without destroying the K\"ahler condition.
  A better question to ask is therefore: given a K\"ahler manifold with
  vanishing first Chern class, can we deform the K\"ahler metric in such a way
  that one obtains a Ricci-flat metric? In 1957, E.~Calabi showed in
  \cite{Calabi:1957zz} that if this is possible, then the solution is unique:
  for a given K\"ahler class  there is at most one Ricci-flat metric such that
  its K\"ahler form is in this class. In 1978, S.-T.~Yau proved
  in \cite{Yau:1978zz} that such a metric indeed
  always exists. Therefore, given a complex manifold with vanishing first Chern
  class, there is precisely one Ricci-flat K\"ahler metric in each K\"ahler
  class. In general, this existence and uniqueness is enough to derive global
  physical results; we can get very far without actually being able to {\em
  construct} such a metric. For this reason, K\"ahler manifolds with vanishing
  first Chern class (but not necessarily the Calabi-Yau metric) are also
  usually called ``Calabi-Yau manifolds''.

  \subsection{Moduli spaces}
   \label{sec:modulispaces}
   The most important consequence of Yau's theorem is the following. We know
   from the above remarks that given a certain topological space and a complex
   structure and a K\"ahler class on it, we have a unique Calabi-Yau metric.
   Therefore, the space of {\em all} Calabi-Yau manifolds of a given topology is
   exactly the space of all possible K\"ahler classes and complex structures on
   $\cM$. This space is called the {\em moduli space} of the Calabi-Yau. In this
   section, we will study this moduli space in some detail. (The reader
   interested in more details should consult the lecture notes
   \cite{Greene:1996cy} by B.~Greene.) A 
   global description of the moduli space is very hard to give, so we will content ourselves with giving
   a local description. That is, given a certain ``background'' complex
   structure and K\"ahler class, we will discuss how each of these can be
   deformed.
   
   \subsubsection{The K\"ahler moduli}
   The ``K\"ahler part'' of the moduli space is the easiest one to describe, and
   we can even describe it globally.
   We have seen that the K\"ahler form is a $(1,1)$-form, so given a complex structure on
   $\cM$, we can choose a K\"ahler class by choosing a cohomology class of
   degree $(1,1)$ -- that is, an element of $H^{1,1}(\cM)$.
   Two different cohomology classes of $(1,1)$-forms will lead to two
   inequivalent Calabi-Yau manifolds (simply because any K\"ahler metric gives
   rise to a unique K\"ahler class), but not all classes of $(1,1)$-forms can
   appear as a K\"ahler class. For example, we want to choose this cohomology
   class in such a way that all $p$-cycles of the manifold have
   positive\footnote{The case of zero volume is thought of as ``pinching the
   cycle'': a certain cycle is shrunk to zero size, and in the limit of zero
   volume the topology of the manifold changes. As a lower-dimensional example,
   when pinching a one-cycle on a two-torus, one obtains a two-sphere. There is
   a long and interesting story about whether one can use these topology changes
   to glue together the boundaries of all Calabi-Yau moduli spaces for different
   topologies into a single large moduli space, but except for some remarks in
   chapter \ref{sec:applications} we will not discuss this here. Again, the
   interested reader is referred to the lecture notes \cite{Greene:1996cy} by
   B.~Greene.} volume.
   For this, it turns out to be sufficient to require that
   \be
    \int_{S_{(2p)}} \go \wedge \cdots \wedge \go > 0,
   \ee
   for all $2p$-cycles $S_{(2p)}$,
   where of course we inserted $p$ factors of the K\"ahler form $\go$.
   This leads to a finite number of inequalities on the cohomology class, which
   results in the fact that the allowed K\"ahler classes form a cone.
   (A cone is defined to be a submanifold of a vector space which contains the
   origin, and for which any positive linear combination of its elements is
   again an element of the cone.) One can argue that these positivity
   conditions are actually the only ones one needs to impose, and that, given a
   complex structure, all cohomology classes in this so-called {\em K\"ahler
   cone} lead to a well-defined Calabi-Yau manifold.
   
   \sk For local investigations, however, the inequalities defining the K\"ahler
   cone do not bother us much: locally any allowed K\"ahler metric (assuming it
   is not at the boundary of the cone) can be deformed by adding an arbitrary
   $(1,1)$-form -- that is, given a complex structure, locally the K\"ahler
   moduli space of a Calabi-Yau is 
   \be
    M_K \sim H^{1,1}(\cM),
   \ee
   where we stressed in the notation that this is only a {\em local}
   description. More precisely, the {\em tangent space} -- which can be viewed
   as the space of infinitesimal deformations -- to the K\"ahler part of
   the moduli space at a non-boundary point is isomorphic to $H^{1,1}(\cM)$.

   \subsubsection{The complex structure moduli}
   Conversely, suppose we now fix a K\"ahler class, and we want to know
   what the allowed complex structures are. In contrast to the K\"ahler cone,
   globally it is very hard to describe this space. However, locally this is
   much easier. Choosing a complex structure is equivalent to splitting up the
   coordinates of $\cM$ into three holomorphic
   coordinates $z^i$ and their complex conjugates $\zbar^i$. This information
   can be neatly encoded by prescribing a nowhere vanishing holomorphic
   three-form $\gO^{(3,0)}$, which on a certain coordinate patch can be written
   as
   \be
    \gO(z) = f(z) \, dz^1 \wedge dz^2 \wedge dz^3,
   \ee
   with $f$ purely holomorphic. Note that such a form is harmonic: $\d \gO =
   \delbar \gO = 0$. It therefore represents a cohomology class in
   $H^{3,0}(\cM)$. An important result about Calabi-Yau manifolds is now that
   such a $(3,0)$-form can always be found, and that its cohomology class is
   {\em unique} up to multiplication by a constant -- that is, $h^{3,0}=1$. So
   given a complex structure on $\cM$, we have a unique one-dimensional linear
   subspace of $H^3(\cM)$ that we can identify as $H^{3,0}(\cM)$. From the
   above formula, it is intuitively clear, and not very hard to prove, that the
   converse is also true: once we know which linear subspace of $H^3(\cM)$ is
   $H^{3,0}(\cM)$, we can reconstruct the complex structure from this.
   
   \sk This of course immediately raises the same question as in the K\"ahler
   moduli case: which linear subspaces of $H^3(\cM)$ lead to a well-defined
   complex structure? We need to answer this question to have a workable
   description of our complex structure moduli space. The equivalent question
   for K\"ahler classes resulted in a number of inequalities and a nice and
   simple global description in terms of a cone, but here such a description is
   not straightforward. Therefore, let us again study the question
   locally: let us assume that a certain ``background'' holomorphic $(3,0)$-form
   $\gO$ is given. In what directions can we now slightly change $\gO$? We know
   that locally, $\gO$ can be written as $f(z) \, dz^1 \wedge dz^2 \wedge dz^3$,
   and changing the complex structure infinitesimally (say with some small
   parameter $\eps$) changes the coordinates to
   \be
    z^i \to {a^i}_j z^j + {b^i}_{\jbar} \zbar^j,
   \ee
   where the coefficients can be Taylor expanded in $\eps$ as
   \bea
    {a^i}_j & = & \gd^i_j + \eps \, {\ga^i}_j + O(\eps^2) \ret
    {b^i}_{\jbar} & = & \eps \, {\gb^i}_{\jbar} + O(\eps^2).
   \eea
   Note that for this result, we do not need to give an exact definition of
   $\eps$; the only input is the correct limit as $\eps \to 0$, so any
   infinitesimal parameter will suffice. Inserting the above result
   in the expression for $\gO$,
   we see that to order $\eps$, the change of $\gO$ is a $(3,0)$-form plus a
   $(2,1)$-form; all other forms only appear with higher powers of $\eps$.
   
   \sk Of course, the deformed $\gO$ will in general no longer be a closed,
   holomorphic form; we would have to pick a very specific transformation of the
   coordinates for that. However, what the argument above shows us is that in
   the new complex structure, the old $(3,0)$-form is now a $(3,0) +
   (2,1)$-form, or by symmetry, that any $(3,0)$-form in the new complex
   structure will be a $(3,0) + (2,1)$-form in the old complex structure.
   In other words, to first order the cohomology class of $\gO$ only changes by
   a $(3,0)$-cohomology class plus a $(2,1)$-cohomology class if we change the
   complex structure. Since $H^{3,0}$ is one-dimensional and scalings do not
   influence the complex structure, the only changes in the complex structure
   can to first order in $\eps$ correspond to the addition of cohomology classes
   of degree $(2,1)$ to the cohomology class of $\gO$.
   
   \sk A more detailed investigation that we will not carry out here shows that
   in fact all of these additions correspond in a one-to-one manner to changes
   in the complex  structure, and hence we find the result that the tangent
   space of infinitesimal deformations of the complex structure is isomorphic to
   $H^{2,1}(\cM)$:
   \be
    M_C \sim H^{2,1}(\cM),
   \ee
   In particular, this implies that the moduli space of complex structures of
   $\cM$ is of dimension $h^{2,1}$.

   \sk Let us now define a local set of variables on this moduli space. To this
   end, it is useful to have a basis of the homology group $H_3(\cM)$.
   Note that this group has dimension $2h^{2,1} + 2$, since $h^{3,0} = 1$, we
   saw in section \ref{sec:relationshodgenumbers} that $h^{p,q} = h^{q,p} =
   h^{3-p,3-q}$, and we know that $H^3(\cM)$ and $H_3(\cM)$ have the same
   dimension. It is well-known (and can
   be proven by putting the symplectic form $\int A \wedge B$ on $H^3(\cM_6)$
   in the canonical form and using Poincar\'e duality) that similarly to the
   middle homology group on Riemann surfaces, one can always choose a so-called
   {\em canonical basis} for this group, i.\ e.\ a set of $2h^{2,1} + 2$
   three-cycles $A^I, B_J$ ($I, J = 0, \ldots, h^{2,1}$) such that the
   intersection numbers of these cycles are
   \bea
    A^I \cap A^J & = & 0 \ret
    A^I \cap B_J & = & \gd^I_J \ret
    B_I \cap B_J & = & 0.
    \label{eq:canonicalbasis}
   \eea
   We will not give a precise definition of these intersection numbers (in fact,
   the easiest definition is in terms of the Poincar\'e dual cohomology
   classes), but one should think of $A \cap B$ as counting intersection points with signs
   depending on the orientation of the two cycles. It can be shown that such
   intersection numbers are invariant under deformations of the two cycles
   involved: under deformations, extra intersections may appear, but they will
   always come in pairs with signs that cancel. Moreover, the dependence on the
   orientation implies that $A \cap B = - B \cap A$.

   \sk As we have seen in section \ref{sec:homology}, a three-form cohomology
   class is completely specified by its integrals over a certain basis of
   three-cycles. Hence, we can define $2 h^{2,1} + 2$ coordinates for the
   cohomology class of $\gO$ as 
   \bea
    x^I & = & \int_{A^I} \gO \ret
    F_I & = & \int_{B_I} \gO.
    \label{eq:canonicalcoordinates}
   \eea
   Defining a dual basis $\ga_I, \gb^J$ of $H^3(\cM_6)$ (where $\int_{A^I}
   \ga_J = \gd^I_J$, etc.) we see that we can also write this as
   \be
    \gO = x^I \ga_I + F_I \gb^I.
   \ee
   Of course, the $2h^{2,1} + 2$ coordinates $(x^I, F_I)$ form a highly
   redundant set if we only want to describe the moduli space of complex
   structures, since we know that it has dimension $h^{2,1}$. In fact, it can be shown
   that if $\gO$ defines a complex structure, one can always choose a canonical
   basis of three-cycles for which the $F_I$ can be determined in terms of the
   $x^I$. To see how exactly the $F_I$ 
   depend on $x^I$, we make use of the fact explained above that the first order
   change of $\gO$ with respect to $x^I$ (and hence its derivative with respect
   to this coordinate) can be at most a $(2,1)$-form. From this we see
   that the wedge product of $\gO$ with its $x^I$-derivative is zero, and hence 
   that
   \bea
    0 & = & \int_\cM \gO \wedge \frac{\d \gO}{\d x^I} \ret
    & = & \int_\cM (x^J \ga_J + F_J \gb^J)
    \wedge (\ga_I + \d_I F_K \gb^K).
   \eea
   To evaluate this integral, we use the so-called Riemann bilinear relation
   (which can be proven using Poincar\'e duality):
   \be
    \int_{\cM} \gL \wedge \Psi = \sum_I \left( \int_{A^I} \gL \int_{B_I} \Psi -
    \int_{B_I} \gL \int_{A^I} \Psi \right),
   \ee
   where $\gL$ and $\Psi$ are arbitrary three-forms, which gives us that
   \be
    0 = F_I - x^J \d_I F_{J}.
   \ee
   By pulling $x^J$ through the derivative, this leads us to the result that
   \be
    F_I = \half \d_I (x^J F_J).
   \ee
   First of all, this shows us that the $F_I(x)$ can be written as partial
   derivatives of a function $F(x)$; secondly, from the
   resulting equation $2 F = x^I F_I$ for this function we see that it is
   homogeneous of degree two in the $x^I$. This in its turn implies that
   rescaling all $x^I$ leads to the same rescaling of $F_I$ and hence of the
   whole $\gO$, so multiplying all $x^I$ by an overall factor does not change
   the complex structure. In the language of projective spaces, the $h^{2,1} +
   1$ coordinates $x^I$ are {\em homogeneous} coordinates on the space of
   complex structures. The function $F$ is called the {\em prepotential} of the
   Calabi-Yau manifold.

 \section{Twisting supersymmetric field theories}
  \label{sec:twisting}
  \subsection{$N=(2,2)$ supersymmetry in two dimensions}
   The discussion so far has been very general, but the abstract concept of a
   cohomological field theory is of course not very useful until we manage to
   construct concrete examples. There is a
   large class of such examples which can be obtained by ``twisting'' $N=(2,2)$
   supersymmetric field theories in two dimensions. Before explaining the
   twisting procedure, we review some details of $N=(2,2)$ supersymmetric
   theories on $\bC$. We will see that the structure of such
   theories is already very similar to that of cohomological theories. In
   section \ref{sec:twistingss}, we will generalize this structure by replacing
   $\bC$ by an arbitrary curved Riemann surface, and see how the procedure of
   twisting and the resulting truly cohomological field theories naturally
   arise.

   \subsubsection{Superspace}
   Supersymmetry generators usually transform as spin $\half$ fermions under the
   Lorentz group. An $N=1$ supersymmetric theory has one such spinor of
   supercharges, an $N=2$ theory has two, and so on. However, as we have seen
   in section \ref{sec:vectorbundles}, in two dimensions the Lorentz group is
   simply $SO(2) = U(1)$, and the spin $\half$ representation is reducible
   since each ``fundamental'' spinor can be split up into two components which
   transform with opposite charge under this $U(1)$. For this
   reason, one uses the notation $N=(p,q)$ when there are $p$ irreducible
   spinor supercharges with positive $U(1)$-charge and $q$ with negative
   $U(1)$-charge. The theories we will be interested in will have two
   ``fundamental'' spinors as their supercharges -- that is, they will have
   $N=(2,2)$ supersymmetry.
   
   \sk The easiest way to describe $N=(2,2)$ supersymmetric theories in two
   dimensions is to use
   so-called {\em superspace}. This is a space which is obtained by adding four
   fermionic coordinates $\gt^\pm, \gtbar^\pm$ to the original coordinates $z,
   \zbar$ of $\bC$. Under complex conjugation, $\gt^+$ is mapped to $\gtbar^-$
   and $\gt^-$ to $\gtbar^+$. The reason for this seemingly strange notation is that
   we want the $\pm$-indices to reflect the transformation properties of the
   coordinates under the Lorentz group: if $z \mapsto e^{i \ga
   z}$, then
   \be
    \gt^{\pm} \mapsto e^{\pm i \ga/2} \, \gt^{\pm}, \qquad
    \gtbar^{\pm} \mapsto e^{\pm i \ga/2} \, \gtbar^{\pm}
   \ee
   Note the factor of $1/2$ in the transformations, denoting the fermionic
   nature of the variables. We will often use a similar notation for $z$-derivatives,
   denoting $\d_z$ by $\d_+$ and $\d_{\zbar}$ by $\d_-$. (An even more precise
   notation would have been something like $\d_{++}$ and $\d_{--}$, indicating
   that the charge of these operators is twice the charge of the fermionic ones,
   but we will refrain from doing this.)

   \sk A superfield is now a ``function'' on this enlarged space of coordinates.
   By this, we mean the following: in a way similar to how one Taylor expands
   an ordinary function in powers of $z$ and $\zbar$, one can write down a
   finite ``fermionic Taylor expansion'' (finite, since the non-commuting
   variables square to zero) for a superfield $\Phi$ as
   \be
    \Phi(z, \zbar, \gt^{\pm}, \gtbar^{\pm}) = \phi(z, \zbar) + \psi_+(z, \zbar)
    \, \gt^+ + \psi_-(z, \zbar) \, \gt^- + \ldots
   \ee
   which includes a total of $2^4 = 16$ ordinary fields. In this expression, the
   prefactors $\phi, \psi_+, \psi_-$, etc., can themselves be ordinary functions
   or Grassmann-valued ``functions'' on $\bC$. Usually, one considers superfields
   of fixed statistics, so if $\phi$ is an ordinary function then $\psi_\pm$
   are Grassmann-valued functions, and conversely.

   \sk Just as on ordinary space one considers the Poincar\'e group,
   consisting of translations and rotations (we will be working in the Euclidean
   setting), we are now interested in the linear coordinate transformations
   which result in a symmetry of superspace. By ``resulting in a symmetry'',
   we mean that such a coordinate transformation should leave the measure of
   integration,
   \be
    dz \, d \zbar \, d \gt^+ d \gt^- d \gtbar^+ d \gtbar^-,
    \label{eq:measure}
   \ee
   invariant. As a consequence, integrals of the type
   \be
    S_D = \int d^2 z d^4 \gt \, K(\Phi^i, \Phibar^i).
    \label{eq:Dterm}
   \ee
   are then automatically invariant under such transformations if $K$ is
   a scalar function of a set of fields $\Phi^i$. Terms in the action of this
   form (where usually one requires $K$ to be a real function 
   of its complex arguments) are called {\em $D$-terms}.

   \sk Which symmetries acting linearly on the coordinates can we find? To
   begin with, the ordinary Poincar\'e group is also a symmetry group of
   superspace. Let us write out the generators of this group as they would act
   on scalar superfields. We write $z = x^1 + i x^0$ and think of $ix^0$ as
   Euclidean time, so the Hamiltonian and momentum operators are
   \bea
    H & = & -i \frac{d}{d(ix^0)} = - i (\d_+ - \d_-) \ret
    P & = & -i \frac{d}{d x^1} = - i (\d_+ + \d_-).
   \eea
   Moreover, there is the single $U(1)$ ``Lorentz'' rotation generator which, as
   we mentioned before, also naturally acts on the fermionic coordinates, and
   which can be written as
   \bea
    M & = & 2 z \d_+ - 2 \zbar \d_- + \gt^+ \frac{d}{d \gt^+} - \gt^- \frac{d}{d
    \gt^-} + \gtbar^+ \frac{d}{d \gtbar^+} - \gtbar^- \frac{d}{d \gtbar^-},
    \label{eq:mgenerator}
   \eea
   where we normalized $M$ such that $e^{2 \pi i M}$ rotates the Grassmann
   variables once and the ordinary complex variables twice. The operators
   satisfy the familiar algebra:
   \bea
    {[}M, H] & = & - 2 P \ret
    {[}M, P] & = & - 2 H.
   \eea
   Here, and in everything that follows, commutators which are not mentioned
   are vanishing.
   
   \sk Next, there are several fermionic operators which do not change
   the measure. First, there are bosonic shifts in the fermionic coordinates:
   $\gt \to \gt + c$, which we can apply to each type of $\gt$ separately. These
   shifts are generated by four operators of the form $\d / \d \gt$. Similarly,
   there are fermionic shifts in the bosonic coordinates: $z \to z + c \gt$ for
   each type of $\gt$. Since this maps $dz \to dz + c d \gt$, it also does not
   change the measure (\ref{eq:measure}). These shifts are generated by eight
   operators of the form $\gt \d_\pm$; we will only be interested in the four
   that have $U(1)$-charges $\pm \half$. It turns out to be useful to group
   these four with the previous four operators into four complex combinations
   and their conjugates: 
   \bea
    \cQ_{\pm} & = & {\phantom -} \frac{\d}{\d \gt^{\pm}} + i \gtbar^{\pm} \d_{\pm} \ret
    \cQbar_{\pm} & = & - \frac{\d}{\d \gtbar^{\pm}} - i \gt^{\pm} \d_{\pm} \ret
    D_{\pm} & = & {\phantom -} \frac{\d}{\d \gt^{\pm}} - i \gtbar^{\pm} \d_{\pm} \ret
    \Dbar_{\pm} & = & - \frac{\d}{\d \gtbar^{\pm}} + i \gt^{\pm} \d_{\pm}.
   \eea
   Note that it is natural to view $- \d / \d \gtbar^-$ as the complex conjugate
   of $\d / \d \gt^+$, since when acting on the real and Grassmann-even quantity
   $\gt^+ \gtbar^-$ (we use a convention where complex conjugation also
   reverses the order of Grassmann quantities) these operators give complex
   conjugate results. With this conjugation property in mind, one can view
   $\cQbar_-$ as the complex conjugate of $\cQ_+$, etc. There are four
   nonzero anti-commutators between these operators:
   \bea
    \{ \cQ_\pm, \cQbar_\pm \} & = & -2i \d_\pm = {\phantom -}P \pm H \ret
    \{ D_\pm, \Dbar_\pm \} & = & {\phantom -} 2i \d_\pm = -P \mp H.
    \label{eq:Qcomm}
   \eea
   Moreover, these operators have nonzero commutators with $M$:
   \bea
    {[}M, \cQ_\pm] & = & \mp \cQ_\pm \ret
    {[}M, \cQbar_\pm] & = & \mp \cQbar_\pm \ret
    {[}M, D_\pm] & = & \mp D_\pm \ret
    {[}M, \Dbar_\pm] & = & \mp \Dbar_\pm.
   \eea
   These commutators are as expected: our notation is such that an object with
   a lower $+$-index transforms with the opposite charge from an object with an
   upper $+$-index, so we can multiply these into a scalar.
   
   \sk So far, the construction has been completely symmetric in $D$ and $\cQ$, so
   it might seem strange to give these sets of operators such different names.
   A related remark is that the $D$-terms we have constructed so far still
   satisfy something which we could call $N=(4,4)$--supersymmetry, meaning that
   these terms are still invariant under
   all of the transformations associated to the $\cQ$'s and $D$'s. However, now
   let us introduce some constrained fields. The constraints will be of the
   following form. We define a {\em chiral superfield} to be a superfield
   $\Phi$ satisfying the two conditions
   \be
    \Dbar_{\pm} \Phi = 0.
   \ee
   Each of these conditions can be viewed as a relation between the components
   of the superfield $\Phi$, and hence one expects this to reduce the number of
   degrees of freedom of a superfield from sixteen to four. For example,
   $\Dbar_+$ acting on $\phi(z)$ gives a single term, $i \gt^+ \d_+ \phi$. There
   is another term at this order in the $\gt$-expansion of $\Dbar_+ \Phi$,
   coming from the $- \d / \d \gtbar^+$ operator acting on the prefactor of $\gt^+
   \gtbar^+$, which we will denote by $\Phi_{+ \overline{+}}$. We define all of
   our fermionic derivatives to be left derivatives, so this operator gives us
   $+\Phi_{+ \overline{+}}$.  One can easily check that these are the only
   terms at this order in the $\gt$-expansion, so we find the relation that
   \be
    \Phi_{+ \overline{+}} = - i \d_+ \phi.
   \ee
   Continuing like this, one finds a whole set of relations between the
   component fields. In the end, one finds that the four components $\phi(z),
   \psi_\pm(z)$ and $F(z)$ (the latter being the prefactor of $\gt^+ \gt^-$) can
   be arbitrary functions; there are five components which depend on these,
   \bea
    \Phi_{+ \overline{+}} & = & - i \d_+ \phi \ret
    \Phi_{- \overline{-}} & = & - i \d_- \phi \ret
    \Phi_{+ - \overline{+}} & = & {\phantom-} i \d_+ \psi_- \ret
    \Phi_{+ - \overline{-}} & = & - i \d_- \psi_+ \ret
    \Phi_{+ - \overline{+} \overline{-}} & = & {\phantom {-i}} \d_+ \d_- \phi
    \label{eq:chiralcomponents}
   \eea
   and all other components have to vanish identically. These conditions can
   formally be written as
   \be
    \Phi = \phi(z', \zbar') + \psi_+(z', \zbar') \gt^+ +
    \psi_-(z', \zbar') \gt^- + F(z', \zbar') \gt^+ \gt^-,
   \ee
   where
   \bea
    z' & = & z - i \gt^+ \gtbar^+ \ret
    \zbar' & = & \zbar - i \gt^- \gtbar^-,
   \eea
   and the formal change of variables should be viewed as defined by a Taylor
   series in the fermionic variables $\gt$.

   \sk In a similar way, one defines {\em anti-chiral superfields} satisfying
   $D_{\pm} \Phi = 0$, and two types of ``mixed chiral superfields'' (also called
   ``twisted chiral'', but we don't use this terminology because we will use the
   word ``twisting'' with a different meaning) which satisfy one barred and one
   unbarred condition.

   \sk An important observation is now that since the $\cQ$- and the
   $D$-operators anti-commute, a $\cQ$-transformed chiral superfield is still
   chiral:
   \be
    \Dbar_{\pm} (\gd_{\eps^+} \Phi) =  \Dbar_{\pm}(\eps^+ \cQ_+ \Phi) =
     \eps^+ \cQ_+ \Dbar_{\pm} \Phi = 0,
   \ee
   and similarly for all other $\cQ$-operators. As a result, we can exponentiate
   the infinitesimal supersymmetry transformations to get finite ones, and the
   resulting fields will still be chiral. In other words, we can view the
   action of the generators $\cQ$ on a chiral superfield $\Phi$ as an action on
   its four ``fundamental'' component fields $\phi, \psi_\pm$ and $F$, and the
   other twelve component fields will automatically transform in the correct
   way if we express them in terms of the fundamental components using
   (\ref{eq:chiralcomponents}). We therefore reach the conclusion that we can
   start with a $D$-term as in (\ref{eq:Dterm}) involving $n$ chiral
   superfields, write it out in terms of $4n$ ordinary fields, and thus
   construct an action  which is invariant under an $N=(2,2)$ supersymmetry
   generated by the $\cQ$-operators, under which the $4n$ fields transform into
   each other.
   
   \sk It is instructive to write out part of such a $D$-term in terms of
   component fields. In particular, let us focus on the terms which contain only
   the fields $\phi^i(z)$ and their complex conjugates. (We will write out the
   full component expression for a $D$-term in the next subsection.) Since the
   ordinary action is obtained by integrating $K(\Phi^i, \Phibar^i)$ over the
   Grassmann
   coordinates, the only terms which appear are the ones multiplying
   $\gt^+ \gt^- \gtbar^+ \gtbar^-$ in the expansion of $K$. The fields $\phi^i$
   themselves appear only without factors of $\gt$, but we have seen in
   (\ref{eq:chiralcomponents}) that their derivatives do appear with factors of
   $\gt$. Let us first consider terms which only depend on $\phi^i$ but not on
   $\phibar^i$. We get a factor of $\gt^+ \gt^- \gtbar^+ \gtbar^-$ in the
   expansion term linear in $\Phi_{+ - \overline{+} \overline{-}}$ and one in
   the term bilinear in $\Phi_{+ \overline{+}}$ and $\Phi_{- \overline{-}}$;
   these terms give
   \be
    \frac{d K}{d \phi^i} \d_+ \d_- \phi^i + \frac{d^2 K}{d \phi^i d \phi^j}
    \d_+ \phi^i \d_- \phi^j  = - \frac{d^2 K}{d \phi^i d \phibar^j}
    \d_+ \phibar^j \d_- \phi^i,
   \ee
   where in the last expression we subtracted a total $z$-derivative of
   $\frac{dK}{d \phi^i} \d_- \phi^i$, so the equality is only true under the integral sign
   of the action, assuming that the fields decay rapidly enough at infinity. From the
   $\phibar$-dependent terms, one finds this same term with the $+$ and $-$
   indices interchanged. Finally, from the mixed $\phi, \phibar$ terms in the
   expansion, we find exactly the same terms again. Therefore, the total $(\phi,
   \phibar)$-dependent part of the action is
   \bea
    S_\phi & = & - \int d^2 z \, g_{i \jbar} \, \eta^{\ga \gb} \d_\ga \phi^i \d_\gb \phibar^j,
    \label{eq:bosonicstring}
   \eea
   where we introduced the ``worldsheet'' metric
   \be
    \eta^{+-} = \eta^{-+} = 2, \qquad \eta^{++} = \eta^{--} = 0,
   \ee
   and the ``space-time'' metric
   \be
    g_{i \jbar}(\phi, \phibar) = \frac{d^2 K}{d \phi^i d \phibar^j}.
    \label{eq:kahlermetric}
   \ee
   Note that (\ref{eq:bosonicstring}) is precisely the action for a bosonic
   string with a gauge-fixed worldsheet metric $\eta_{\ga \gb}$ on a space-time
   manifold with metric $g_{i \jbar}$. If we add all 
   other terms in the expansion of $K$, we therefore find an $N=(2,2)$
   supersymmetric string action! Moreover, from (\ref{eq:kahlermetric}) we see
   that the target space is in fact a K\"ahler manifold with K\"ahler potential
   $K$.

   \sk Before continuing, we should make some remarks here to prevent
   confusion. The reader who studied some string theory from, for example, the
   book by M.~Green, J.~Schwarz and E.~Witten \cite{Green:1987sp} might be
   worried about the large number of supersymmetries that we have here. It is
   known that as a result of the so-called conformal anomaly, a string theory
   with $N=1$ worldsheet supersymmetry must have ten target space dimensions,
   whereas for $N=2$ we need to consider the rather unphysical case of a
   four-dimensional target space with an even number of timelike directions.
   Therefore, should we not expect to run into
   trouble once we take the conformal anomaly into account? The answer is
   ``no'', as we will see explicitly in what follows\footnote{The remarks here
   apply to the untwisted theories. As we will see, after we have twisted, the conformal
   anomaly will no longer play any role at all, even though there will still be
   a ``critical dimension'' in a different sense.}. The reason for this is that the analysis
   leading to the number of space-time dimensions counts {\em local}
   supersymmetries (that is, supersymmetries whose parameter $\eps$ may depend
   on $z$ and $\zbar$), whereas in our story we will only need {\em global}
   supersymmetries. Of course, one may still be surprised by the fact that here
   one finds some ``extra'' global supersymmetries that do 
   not correspond to the well-known local ones. The reason for this is that we
   are studying particularly symmetric target spaces -- K\"ahler manifolds. The
   extra symmetry that these spaces have gives rise to the extra
   supersymmetries. Or, vice versa, we have seen that by requiring $N=(2,2)$
   global supersymmetry, the target space automatically will be a K\"ahler
   manifold.
   
   \sk Finally, let us mention that using the chiral superfields, one can
   construct more supersymmetry invariant functionals as follows:
   \be
    S_F = \int d^2 z d^2 \gt \,  W(\Phi^i) \Big|_{\gtbar^\pm = 0},
   \ee
   where $W$ is a {\em holomorphic} function of the $\Phi^i$ -- that is, it does
   not depend on $\Phibar^i$. Note that since we set $\gtbar^\pm = 0$, we only
   integrate over this slice of superspace, with coordinates $\gt^\pm$. This
   action is clearly $\cQ^\pm$-invariant, since these operators consist of a total
   $\gt$- plus a total $z$-derivative. To show invariance under $\cQbar_\pm$ we
   need a slightly different argument, since $\d / \d \gtbar^\pm$ does not
   remove a coordinate over which we subsequently integrate. However, note that
   $\cQbar_\pm = \Dbar_\pm - 2 i \gt^\pm \d_\pm$, so we find that
   \bea
    \eps \int d^2 z d^2 \gt \, \Qbar_\pm W(\Phi_i) & = & \eps \int d^2 z d^2 \gt
    (\Dbar_\pm - 2 i \gt^\pm \d_\pm) W(\Phi_i) \ret
    & = & 0
   \eea
   since the first term vanishes by the chirality of the fields, and the second
   term vanishes if the fields fall off fast enough at infinity. Terms
   constructed in this way are called {\em $F$-terms}. They will not play an
   important role in these notes, but they are crucial in applications of
   topological string theory to mirror symmetry, for example.

   \subsubsection{R-symmetry}
    We have not finished our list of linear symmetry operators on superspace
    yet. Note that the $U(1)$ Lorentz symmetry acts on the variables $z, \zbar,
    \gt^\pm$ and $\gtbar^\pm$ simultaneously. However, one can construct many
    more $U(1)$-groups which act only on a subset of these variables and leave
    the measure invariant. Of these operators, we are only
    interested in the ones which at worst multiply the $D$-operators by a
    constant, since we want the rotated chiral fields to remain chiral. There
    are two independent rotations\footnote{We do not consider rotations which
    rotate one $z$- and one $\gt$-coordinate, since these will destroy the
    reality of the K\"ahler potential.} which satisfy this:
    \bea
     R_V(\ga): & & (\gt^+, \gtbar^+) \to (e^{-i \ga} \gt^+, e^{i \ga}
     \gtbar^+), \qquad (\gt^-, \gtbar^-) \to (e^{-i \ga} \gt^-, e^{i \ga}
     \gtbar^-) \ret
     R_A(\ga): & & (\gt^+, \gtbar^+) \to (e^{-i \ga} \gt^+, e^{i \ga}
     \gtbar^+), \qquad (\gt^-, \gtbar^-) \to (e^{i \ga} \gt^-, e^{-i \ga}
     \gtbar^-).
     \label{eq:rsymmetry}
    \eea
    Here, the $V$ stands for ``vector'' and $A$ stands for ``axial'', in (not
    too perfect) analogy with the way spinors in four dimensions transform
    under axial and vector symmetries. The corresponding $U(1)$-groups are
    called the vector and axial {\em R-symmetry groups}. The symmetries are
    generated by the following operators:
    \bea
     F_V & = & - \gt^+ \frac{d}{d \gt^+} - \gt^- \frac{d}{d \gt^-} + \gtbar^+
     \frac{d}{d \gtbar^+} + \gtbar^- \frac{d}{d \gtbar^-} \ret
     F_A & = & - \gt^+ \frac{d}{d \gt^+} + \gt^- \frac{d}{d \gt^-} + \gtbar^+
     \frac{d}{d \gtbar^+} - \gtbar^- \frac{d}{d \gtbar^-}.
     \label{eq:rgenerator}
    \eea
    The nonzero commutators with the operators we have already constructed are
    easily found to be
    \bea
     {[}F_V, \cQ_\pm ] & = & {\phantom -} \cQ_\pm \ret
     {[}F_V, \cQbar_\pm ] & = & - \cQbar_\pm \ret
     {[}F_A, \cQ_\pm ] & = & \pm \cQ_\pm \ret
     {[}F_A, \cQbar_\pm ] & = & \mp \cQbar_\pm.
    \eea
    Note that we use the standard ``passive'' notation for coordinate
    transformations; for example, for an invariant field under the vector
    rotation, the above notation really means that
    \be
     \Phi_{new} (z, e^{-i \ga} \, \gt^\pm, e^{i \ga} \, \gtbar^\pm) = \Phi_{old}
     (z, \gt^\pm, \gtbar^\pm),
    \ee
    so we can also view this transformation as an ``active'' transformation on
    the component fields, e.\ g.\
    \be
     \psi_{+, new} = e^{i \ga} \psi_{+, old}
     \label{eq:fieldtransform}
    \ee
    Of course, a superfield may also have charges $(q_V, q_A)$ under these
    operators, in which case it transforms as
    \be
     \Phi_{new} (z, e^{-i \ga} \, \gt^\pm, e^{i \ga} \, \gtbar^\pm) = e^{i \ga
     q_V} \Phi_{old} (z, \gt^\pm, \gtbar^\pm),
    \ee
    leading for example to
    \be
     \psi_{+, new} = e^{i \ga (1 + q_V)} \psi_{+, old},
     \label{eq:fieldtransform2}
    \ee
    and similarly for the axial operator.

    \sk By construction, if we set all field charges $(q_A, q_V)$ to zero, the
    $D$-terms we have constructed will be invariant under the two $R$-symmetry
    groups. Moreover, if the $D$-term is constructed from a real function $K$
    (as it must be to identify $K$ with a K\"ahler potential), we can actually
    make any choice for the charges of $\Phi^i$, as long as we require that the
    complex conjugate fields $\Phibar^i$ have opposite charges.

    \sk Now, we have to ask ourselves an important question: are the symmetries
    $R_V$ and $R_A$ also symmetries of the quantum theory? To find this out, we
    have to figure out whether the path integral measure is invariant under
    these operations. This measure can be written as
    \be
     \prod_i D \phi^i \, D \psi_+^i \, D \psi_-^i \, DF^i \times c.c.
    \ee
    Since the $\phi^i$ do not multiply any $\gt$ in the expansion of the
    superfield, they do not transform under the $R$-symmetries, so their
    measure is also invariant. (Or, if the fields have explicit $R$-charges,
    these will be cancelled by those of the complex conjugates.) Moreover, note
    from (\ref{eq:chiralcomponents}) that $F$ appears without any derivatives,
    and recall that it multiplies two $\gt$-variables, so it will appear at
    most quadratically in the action. This means we can easily integrate it out
    of our path integral and replace every appearance
    of $F$ with its value according to its equation of motion. Doing this, the
    somewhat tedious calculation of writing down the full $D$-term Lagrangean
    in terms of component fields actually results, after subtracting some total
    derivatives, in a rather simple expression:
    \be
     L = - g_{i \jbar} \d^\ga \phi^i \d_\ga {\phibar}^{j} - 2 i g_{i \jbar}
     {\psibar}_-^{j} \gD_+ \psi^i_- - 2 i g_{i \jbar} {\psibar}_+^{j} \gD_-
     \psi^i_+ - R_{i \jbar k \lbar} \psi^i_+ \psi^k_- {\psibar}_+^{j}
     {\psibar}_-^{l}
     \label{eq:fullDterm}
    \ee
    Here, we have inserted the expressions
    \bea
     R_{i \jbar k \lbar} & = & g^{m \nbar} g_{m \jbar, \lbar} g_{\nbar i, k} -
     g_{i \jbar, k \lbar} \ret
     \gD_{\pm} \psi^i & = & \d_{\pm} \psi^i + \Gam^{i}_{jk} \d_\pm \phi^j
     \psi^k \ret
     \Gam^i_{jk} & = & g^{i \lbar} g_{\lbar j, k}
    \eea
    for the Riemann tensor, (pullback) covariant derivatives and Christoffel
    symbols of a K\"ahler metric. The notation $g_{i \jbar, k}$ etc.\ is the
    standard general relativity notation for the derivative of $g_{i \jbar}$
    with respect to $\phi^k$.

    \sk Since the $\phi^i$ measure does not transform under $R$-symmetry and we
    have integrated out $F^i$, the crucial thing to check is whether the fermion
    measure is invariant under the $R$-symmetries. To see whether this is the
    case, let us begin by ignoring the four-fermion term in (\ref{eq:fullDterm}).
    Moreover, we will suppress the $i$-indices for notational convenience. The
    path integral for $\psi_-$ then has the form
    \be
     \int D \psi_- D \psibar_- \exp \left( \psibar_-,  \gD_+ \psi_- \right)
     \label{eq:psiminpi}
    \ee    
    where $\gD_+$ is a linear operator, and we wrote the contraction of the
    indices and the integration over the complex plane as an inner product.
    The fields $\psi_-$ and $\psibar_-$ live in an infinite-dimensional vector
    space, so let us expand them in terms of a basis:
    \be
     \psi_- = \sum_a \psi_-^{(a)} \qquad \psibar_- = \sum_b \psibar_-^{(b)}.
    \ee
    Two remarks are in order here. First of all, since $a$ and $b$ run over an
    infinite number of values, it does not really make sense to say that we
    have ``the same number'' of basis vectors in both spaces. In particular, we
    will see that it is useful not to require any relation between
    $\psibar_-^{(a)}$ and $\psi_-^{(a)}$. Secondly, the reader
    might have expected an expression of the form $\psi_- = \sum_a c_a \,
    \psi_-^{(a)}$, with $\psi_-^{(a)}$ a set of fixed basis vectors and $c_a$
    a set of complex numbers. The reason we do not write an expression of
    this form is that it would imply integrating over the values of $c_a$,
    which is an ordinary integration. As we will see, it will be crucial that
    the path integral over $\psi_-$ consists of {\em Grassmann} integrations, so
    we will think of the $\psi_-^{(a)}$ as living in a one-dimensional subspace
    of the whole $\psi_-$ space, and we write the path integral measure as
    \be
     \int \prod_a d \psi_-^{(a)} \prod_b d \psibar_-^{(b)}
     \label{eq:productmeasure}
    \ee
    Let us now for a moment pretend that $\psi_-$ and $\psibar_-$ are elements
    of a {\em finite dimensional} vector space. Following our remark above, we
    will not assume that these vector spaces are necessarily of the same
    dimension; let us for concreteness assume that the dimension of the
    $\psi_-$ space is larger than that of the $\psibar_-$ space. In general, we
    will then be able to find bases for the two spaces such that the exponent
    in (\ref{eq:psiminpi}) takes the form
    \be
     \left (\psibar_-, \gD_+ \psi_- \right) =
     \left( \psibar_-^{(1)} \cdots \psibar_-^{(n)} \right)
     \left( \begin{array}{ccc|c}
      \gl_1 & & 0 &  \\
      & \ddots & & \hspace{1 em} 0 \hspace{.5 em} \\
      0 & & \gl_n & 
     \end{array} \right)
     \left( \begin{array}{c}
      \psi_-^{(1)} \\
      \vdots \\
      \psi_-^{(n)} \\
      \hline
      \psi_-^{(n+1)} \\
      \vdots \\
      \psi_-^{(n+k)}      
     \end{array} \right).
     \label{eq:zeromodeex}
    \ee
    An important role is now played by the so-called ``zero modes'': the
    eigenvectors of the $\gD_+$-operator with zero eigenvalues. As we can see
    in the above equation, there are at least $k$ of those, but there may of
    course be ``accidental'' zero modes when some of the $\gl _i$ are zero. Note
    that by transposing the above equation, we can also write it as
    \be
     ( \psi_-, \gD^\dagger_+ \psibar_-)
    \ee
    which defines the adjoint operator $\gD^\dagger_+$, which in this case
    is simply the transpose of $\gD_+$. Note that the number of ``accidental''
    zero modes above is precisely the number of zero modes of this adjoint
    operator.
    
    \sk If the inner product $(\cdot , \cdot)$ and the linear operator $\gD_+$
    depend on a set of background parameters, the eigenvalues $\gl_i$ will vary
    continuously as a function of these
    parameters. At special values of the parameters, one or more of the $\gl_i$
    may become zero, and hence the number of zero modes of the
    $\gD_+$-operator is not constant. On the other hand, from the reasoning
    above we see that the difference between this number and the number of
    $\gD^\dagger_+$ zero modes {\em is} constant under deformations. We denote
    this number (which equals $k$ in (\ref{eq:zeromodeex})) by $k_-$, and write
    it as 
    \be
     k_- \equiv \dim \Ker \gD_+ - \dim \Ker \gD^\dagger_+.
     \label{eq:defindex}
    \ee
    The number $k_-$ is called the {\em index} of the $\gD_+$-operator.
    
    \sk Of course, the assumption that $\psi_-$ lives in a finite-dimensional
    space is not true in our problem. However, one can show that in
    general\footnote{At least
    in the case where the worldsheet is compact, which will be the case of our
    true interest. So far, our discussion has only used the flat metric of
    $\bC$, so one could replace it with a complex flat torus, to avoid the
    complications of a noncompact worldsheet.} it is still true that $\gD_+$
    and $\gD_+^\dagger$ have a 
    finite-dimensional space of zero modes, and that the index $k_-$ defined as
    in (\ref{eq:defindex}) does not change under a change of parameters. Note
    that the parameters here are encoded in the metric $g_{i \jbar}$ of the
    target space and in the map $\phi^i$, so what we are really saying is that
    $k_-$ is a topological invariant of the embedding of the worldsheet
    into the target space! There is a a very famous (and nontrivial)
    result called the ``Atiyah-Singer index theorem'' which gives an explicit
    formula for such an index invariant. In this simple case, it can be written
    as
    \be
     k_- = \int_{\phi(\gS)} c_1(\cM)
     \label{eq:atiyahsinger}
    \ee
    where $\phi(\gS)$ is the image of the two-dimensional worldsheet inside
    the target space. So far, we have been assuming that $\gS=\bC$ (which in
    fact is a problematic case since it is non-compact), but the formula is
    true for any worldsheet topology.

    \sk Just as in our finite-dimensional example, one can show that except in
    special cases, one of the two dimensions appearing in (\ref{eq:defindex})
    will be zero. We will always assume that we are in this generic situation.
    Moreover, using complex conjugation one finds that for the $\psi_+$-part of
    the action,
    \be
     k_+ \equiv \dim \Ker \gD_- - \dim \Ker \gD^\dagger_- = - k_-.
    \ee    
    For definiteness, let us assume furthermore that $k_- \geq 0$, and relabel
    this positive integer as $k$. That is, we will have $k$ zero modes
    $\psi_-^{(a)}$, $k$ zero modes $\psibar_+^{(a)}$, and no zero modes for
    $\psibar_-$ and $\psi_+$.
    
    \sk Why do the zero modes of the $\gD$-operators interest us so much? To see
    this, let us rewrite (\ref{eq:psiminpi}) as
    \be
     \prod_{a,b} \left( \int d \psi_-^{(a)} d \psibar_-^{(b)} \exp \left(
     \psibar_-^{(b)},  \gD_+ \psi_-^{(a)} \right) \right)
    \ee
    If $\psi_-^{(a)}$ is a zero mode (or if $\psibar_-^{(b)}$ is, but we have
    chosen a situation where this does not happen), the exponent will
    simply be $1$, and the Grassmann integral will give a vanishing result.
    Therefore, we find that the path integral, defined in our naive way, is
    always zero if there are fermionic zero modes! One way to resolve this
    problem is to remove the integrations over the zero modes from the
    definition of the path integral by hand. This seems rather ugly, since
    which mode is a zero mode depends on the parameters of the problem. By the
    rules of Grassmann integration, removing the integral over a zero mode is
    equivalent to inserting an extra copy of the zero mode into the path
    integral. This of course has the same ugliness to it, but we can now remove
    this problem by inserting the whole $\psi_-(z) = \sum \psi_-^{(a)} (z)$
    into the path integral. Of course, we have to insert $k$ copies of this
    field to make sure that there is a term which contains all of the
    zero modes exactly once. This procedure 
    is known as the ``absorption of zero modes''. Similarly, we have to insert
    $k$ copies of $\psibar_+(z)$ to cancel the zero modes of $\gD_-^\dagger$.
    Putting everything together, we find that the naive path integral gives
    zero, but that the path integral with the zero modes absorbed,
    \be
     \int D \psi_+ D \psi_- D \psibar_+ D \psibar_- \, \left\{ g_{i_1 \jbar_1}
     \psi_-^{i_1} (z_1) \psibar_+^{j_1} (z_1) \dots g_{i_k \jbar_k}
     \psi_-^{i_k} (z_k) \psibar_+^{j_k} (z_k) \right\} \, e^{-S}
    \ee
    is a nonvanishing quantity. 
    
    \sk Now, we can readdress the issue of $R$-symmetry. By symmetry arguments,
    it seems natural that the naive path integral measure
    (\ref{eq:productmeasure}), when defined as an appropriate limit of ordinary
    Grassmann integrals, is $R$-symmetry invariant. However, we have seen
    that we need to remove some integrations from this path integral in an
    asymmetric way -- or what is equivalent from an $R$-symmetry point of view,
    insert the above operators. From (\ref{eq:rsymmetry}), we see that this
    product of operators is invariant under the $R_V$-symmetry (if we take all
    the explicit $R$-charges to vanish), since there is an equal number of
    $\psi_-$ and $\psibar_+$ insertions and these transform oppositely under
    the vector $R$-symmetry. On the other hand, the insertions transform in
    the {\em same} way under the axial $R$-symmetry, so the path integral is not
    invariant under this symmetry unless $k=0$. From (\ref{eq:atiyahsinger}),
    we see that there is one case in which this certainly is true for any
    worldsheet: if $c_1(\cM) = 0$. This is precisely the case of a Calabi-Yau
    target space! Therefore, our
    long discussion can be summarized in two brief points:
    \begin{itemize}
     \item
      The $R_V$ vector $R$-symmetry is present in the quantum theory for any
      K\"ahler target space if we take the explicit $R$-charges of the fields
      to vanish.
     \item
      The $R_A$ axial $R$-symmetry is only present in the quantum theory for a
      Calabi-Yau target space.
    \end{itemize}
    These results will be crucial in our discussion of the twisting of these
    $N=(2,2)$ sigma-models.
    
    \sk As a final remark: in all of the above we have ignored the
    $\psi^4$-terms in the Lagrangean (\ref{eq:fullDterm}). In the limit where
    the target space is large compared to the generic size of the worldsheet
    (the ``string length''), the Riemann curvature will be small and this term
    can therefore be viewed as a small perturbation. By the topological nature
    of our argument, we do not expect small perturbations to affect the integer
    number of zero mode insertions that is needed. Therefore, for a large
    enough target space, all of our discussion above will be true. The results
    might break down when the target space reaches a certain critical size at
    which the number of required insertions jumps, but one expects this to
    happen at a size of the order of the string scale. In all that follows, we
    will assume that our target spaces are ``large enough'' in the above sense.

  \subsection{Twisted $N=2$ theories}
   \label{sec:twistingss}
   The $N=(2,2)$ supersymmetric theories we have constructed are already very
   similar to cohomological field theories. In particular, note from
   (\ref{eq:Qcomm}) that
   \bea
    \{ \cQbar_+ + \cQ_-, \cQ_+ - \cQbar_- \} & = & 2H \ret
    \{ \cQbar_+ + \cQ_-, \cQ_+ + \cQbar_- \} & = & 2P, \ret \ret
    \{ \cQbar_+ + \cQbar_-, \cQ_+ - \cQ_- \} & = & 2H \ret
    \{ \cQbar_+ + \cQbar_-, \cQ_+ + \cQ_- \} & = & 2P, 
   \eea
   so if we define
   \bea
    Q_A & = & \cQbar_+ + \cQ_- \ret
    Q_B & = & \cQbar_+ + \cQbar_-
   \eea
   then we find two\footnote{We could have constructed more of these
   operators by adding some minus signs and switching complex conjugates, but
   all of these choices are related to the above two by discrete automorphisms
   of the $N=(2,2)$ supersymmetry algebra.} operators which square to zero, and
   such that the Hamiltonian and the momentum are $Q_{A/B}$-exact. Of course,
   this does not prove yet that the whole energy-momentum tensor is
   $Q_{A/B}$-exact, but in all reasonable examples this actually turns out to
   be the case.

   \sk Does this mean we have already constructed our first examples of
   cohomological field theories? The answer is ``no'', for the following reason.
   The supersymmetric field theories we have constructed so far have lived on
   {\em flat} worldsheets $\bC$ or $T^2$. However, to claim that these
   theories are metric-independent, we should be able to define them for
   arbitrary worldsheet metrics. This is no problem insofar as writing down the
   Lagrangean is concerned: one simply replaces all worldsheet derivatives with
   their covariant versions. In other words, the vector bundles in which the
   fields live now obtain nontrivial connections. On the other hand, a much
   more serious problem is the definition of the supersymmetries corresponding
   to $\cQ_\pm$. We would like to write down expressions of the form
   \be
    \gd \Phi^i = \eps^+ \cQ_+ \Phi^i.
   \ee
   To define a global symmetry, $\eps^+$ should be a ``constant'' spinor. Here,
   ``constant'' actually means {\em covariantly constant}, since only when the
   spinor is covariantly constant can we pull it outside covariant derivatives
   to prove the invariance of the action. But for a general worldsheet metric,
   there simply are no covariantly constant spinor fields! The reason is that the
   values of a covariantly constant field at two arbitrary points should be
   related by parallel transport. In particular, after parallel transport of
   such a spinor around a closed curve, it should come back to itself. But of
   course, if on a general surface we would parallel transport a spinor
   $\eps^+(z)$ around an arbitrary closed  curve, it would come back to itself up
   to a rotation. The requirement that there exists a spinor at every point on
   the surface such that for any curve this rotation is the identity is a very
   special requirement on the metric. (This is exactly the same reasoning as
   the one which leads to the interest in Calabi-Yau manifolds in three complex
   dimensions. However, in the current situation we have nothing to choose:
   since we want to define a path integral over all possible metrics we would
   need a covariantly constant spinor for {\em any} metric, which is not
   possible.)
   
   \sk One should contrast this problem with the case of ordinary bosonic
   symmetries, where the infinitesimal parameter is simply a number. Such a
   number can be viewed to be an element of the trivial bundle $\gS \times
   \bC$ over the worldsheet $\gS$, and it can therefore always be chosen to be
   constant. (In other words,
   a covariantly constant scalar is simply a constant one.) One can therefore
   solve the above problem by constructing a theory where the supercharge we
   are interested in lives in a trivial bundle. Note that the type of bundle in
   which an object lives is determined by its Lorentz charge (i.\ e.\ its
   spin): the scalar $\phi(z)$ takes values in the trivial bundle $\gS \times
   \bC$, objects with a $+$ index such as $\psi_+, \psibar_+, \cQ_+$ and
   $\eps_+$ take values in $S_+$, and objects with a $-$ index take values in
   $S_-$. Therefore, to construct a $\cQ$-symmetric theory, we have to
   construct a theory in which there is a different Lorentz group under which
   some of the $\cQ$-operators transform with spin zero.

   \sk The way to construct such a theory follows from the $N=(2,2)$ algebra
   itself: note that the operators $M, F_A$ and $F_V$ have exactly the same
   types of commutators with the other operators, and have vanishing commutators
   among themselves. Therefore, we could define a new operator\footnote{The
   reader should remember that the $A$ in $F_A$ stands for ``axial'', and is
   unrelated to the label on $M_{A/B}$. The somewhat confusing notation has
   grown historically from the fact that the $A$- and $B$-models that we will
   define are related to the type IIA and IIB strings respectively.}
   \be
    M_A = M - F_V \qquad \mbox{or} \qquad M_B = M - F_A,
   \ee
   and simply {\em declare} this to be the Lorentz symmetry generator. That is,
   when we now go from the ``flat'' theory to the theory with arbitrary metric,
   we replace the ordinary derivatives with covariant derivatives {\em with
   respect to the new Lorentz quantum numbers.} Let us now look at the
   following commutators:
   \bea
    {[}M_A, \cQ_+] = - 2 \cQ_+ && [M_B, \cQ_+] = -2 \cQ_+ \ret
    {[}M_A, \cQ_-] = {\phantom -} 0 {\phantom \cQ_{\phantom -}} && [M_B, \cQ_-] = 2 \cQ_- \ret
    {[}M_A, \cQbar_+] = {\phantom -} 0 {\phantom \cQ}_{\phantom -} && [M_B, \cQbar_+] = 0 \ret
    {[}M_A, \cQbar_-] = {\phantom -} 2 \cQbar_- && [M_B, \cQbar_-] = 0.
   \eea
   We see that for $M_A$, the operator $Q_A$ has become a scalar, and for
   $M_B$ this happens to the operator $Q_B$. Therefore, the corresponding
   symmetry operations can now be defined {\em on an arbitrary curved
   worldsheet}! This construction is called {\em twisting}, and we 
   reach the conclusion that these twisted theories can truly be viewed as
   cohomological field theories. (What we have not shown yet is that the
   energy-momentum tensor of these theories is $Q$-exact, but we will do so when we
   discuss the $A$- and $B$-models in more detail below.)

   \sk Recall from the previous section that the vector $R$-symmetry is always
   a quantum symmetry, whereas the axial $R$-symmetry is broken unless
   $c_1(\cM) = 0$. This means that the $A$-twisting can be performed for any
   K\"ahler target space, but the $B$-twisting can only be done if our target
   space is Calabi-Yau.

   \sk Thus, we have now constructed a large class of topological field
   theories. The next question is, of course: what kind of topological
   information about $\cM$ do these theories compute? The answer to this
   question turns out to depend crucially on the type of twisting, and we will
   discuss it in sections \ref{sec:amodel} and \ref{sec:bmodel}. However, before
   doing so we have to say a few words about virtual dimensions.

   \subsubsection{Virtual dimensions}
    \label{sec:virtualdim}
    Loosely speaking, the virtual dimension (also called ``formal dimension'')
    of a space is the dimension it should have by a naive counting of degrees of
    freedom and equations. The concept in particular applies to mathematical
    problems involving moduli spaces. We have seen that the dimension of the
    moduli space of a Calabi-Yau manifold can easily be calculated, but in
    general the dimensions of moduli spaces are much harder to calculate
    exactly. On the other hand, it is often quite easy to make an ``educated
    guess'' of what the dimension of a moduli space should be. However, this
    virtual dimension need not always be the actual dimension 
    of the space! The virtual dimension may even turn out to be a negative number.
    Nevertheless, in sufficiently generic situations, the virtual and the
    actual dimensions of a space will coincide, and one can even give an
    interpretation to the negative dimensions. Since there are quite a few
    technical details involved in rigorously defining what the virtual dimension
    of a moduli space is, and how it is related to its actual dimension, we
    will refrain from doing this. Instead, we will discuss a ``toy model''
    example of these concepts to get some intuition.

    \sk Our example is the following: suppose we have $n$-dimensional Euclidean
    space, $\bR^n$, parameterized by coordinates $x^1, \ldots, x^n$. Next,
    suppose we have a $p$-dimensional subspace $X_1^p$ and a $q$-dimensional
    subspace $X_2^q$. What is the dimension of the intersection of $X_1^p$ and
    $X_2^q$?

    \begin{figure}[ht]
     \begin{center}
      \includegraphics[height=4cm]{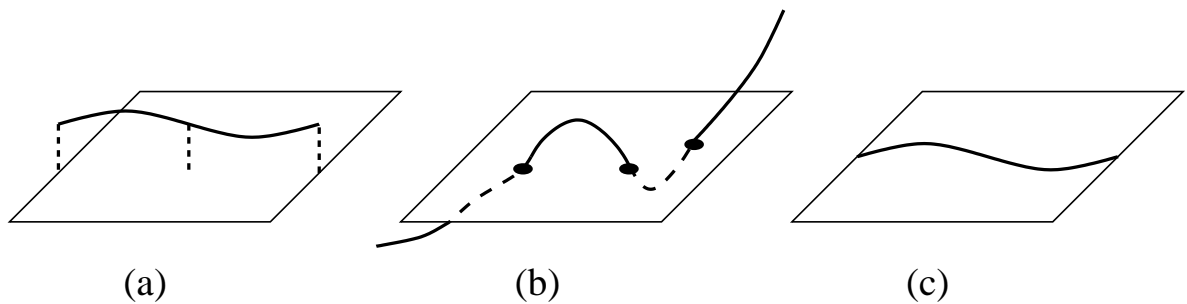}
     \end{center}
     \caption{Intersections of a curve and a plane in $\bR^3$. (a) A
     non-generic case with empty intersection. (b) The generic case with a
     $0$-dimensional intersection. (c) A non-generic case with a
     $1$-dimensional intersection.}
     \label{fig:virtualdim}
    \end{figure}

    \sk Of course, the answer to this question cannot be given from this
    information alone. For example, in figure \ref{fig:virtualdim}, we have
    drawn three examples where $n=3, p=2$ and $q=1$, and the intersections of
    $X_1^p$ and $X_2^q$ are the empty set (which does not really have a
    well-defined dimension)
    and a $0$- and a $1$-dimensional set respectively. However, note that the
    situations in figure \ref{fig:virtualdim}a and \ref{fig:virtualdim}c are
    ``fine-tuned'' by choosing the curve to be parallel to the plane. In the
    ``generic'' situation, of which \ref{fig:virtualdim}b is an example, the
    dimension of the intersection will be $0$ -- that is, the intersection
    consists of a number of points.

\begin{figure}[ht]
     \begin{center}
      \includegraphics[height=5cm]{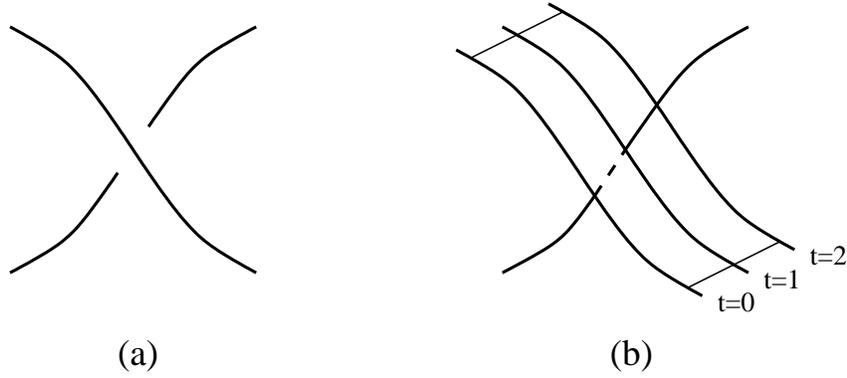}
     \end{center}
     \caption{(a) Two curves in $\bR^3$ in general have an empty intersection.
     (b) If one of the curves is part of a one-parameter family of curves, then
     the intersection of this family with the second curve will generically be
     a $0$-dimensional set.}
     \label{fig:negativedim}
    \end{figure}

    \sk One can explain this result as follows. Locally, a $p$-dimensional
    subset of $\bR^n$ can be viewed as the solution to $n-p$ equations
    \be
     f^a(x^i) = 0, \qquad a=1, \ldots, n-p.
    \ee
    In other words, we can locally choose coordinates $f^i$ with $i=1, \ldots,
    n$, such that $X_1^p$ is the subset obtained by setting the first $n-p$
    coordinates $f^a$ to zero. Similarly, $X_2^q$ is defined by $n-q$ equations
    $g^b = 0$. Therefore, generically their intersection is defined by $2n-p-q$
    equations, so we expect its dimension to be
    \be
     n - (2n-p-q) = p + q - n.
     \label{eq:virtualdim}
    \ee
    This dimension, obtained by counting equations, is the virtual dimension of
    the intersection of $X_1^p$ and $X_2^q$. Note that we indeed find 0 in the
    example of figure \ref{fig:virtualdim}. Of course in specific situations,
    several things may go wrong in this naive reasoning:
    \begin{itemize}
     \item
      The sets of equations $\{ f^a \}$ and $\{ g^b \}$ may not be independent.
      For example, if one of the $f^a$ equals one of the $g^b$, the dimension of
      the intersection will be higher than the predicted one. This is the
      situation in figure \ref{fig:virtualdim}c.
     \item
      There may be incompatibilities between the sets $\{ f^a \}$ and $\{ g^b
      \}$. For example, if $f^a = g^b + c$ for some $a$ and $b$ and for some
      nonzero constant $c$, there will be no solution to the combined set of
      equations at all. This is the case in figure \ref{fig:virtualdim}a. The
      incompatibilities may also be less severe, in which case there is still an
      intersection left, but the dimension is lower then the virtual one.
      (Example: two two-spheres in $\bR^3$ touching at a single point.)
    \end{itemize}
    These are two ways in which our specific situation can be non-generic.
    However, there is another way in which (\ref{eq:virtualdim}) can fail which
    is much more interesting to us:
    \begin{itemize}
     \item
      The outcome $p + q - n$ may be negative! An example is the case of $n=3$
      and $p=q=1$; two curves inside $\bR^3$. In general, two such curves
      will not intersect, see figure \ref{fig:negativedim}a, so we can interpret
      a negative virtual dimension to mean that the generic intersection is the
      empty set. However, there is more information contained in the result $p
      + q - n=-1$ than just this fact. The $-1$ is in a sense the ``missing
      number of dimensions''. For example, if we would now take a one-parameter
      family of curves $X^p_1[t]$, then for certain discrete values of $t$ we
      would find a finite number of intersection points. This is illustrated in
      figure \ref{fig:negativedim}b. In other words, by ``adding one dimension
      to the   moduli space'', we can get back to a situation where the
      intersection (in this case of $X^p_1[t]$ and $X^q_2$) is a zero-dimensional set
      of points\footnote{Of course, one can even add more parameters and obtain
      intersection spaces of positive dimension as well. Here, one may even run
      into the opposite problem, where the virtual dimension will be larger
      than $d$ and we have to enlarge our ``target space'' to get back to a
      reasonable interpretation of the answer.}.
    \end{itemize}
    Everything we said above will translate to our case of interest: the
    dimensions of moduli spaces. These moduli spaces usually have a virtual
    dimension which is easy to calculate. In non-generic cases this virtual
    dimension may be too high or too low, but in what follows we will mostly
    ignore these non-generic cases. (Even though, from a mathematical point of
    view, these are often the truly interesting ones!) More importantly,
    however, we will find negative virtual dimensions of moduli spaces, meaning
    that we will somehow have to extend the dimension of our moduli space to get
    nonzero answers.

    \sk After this intermezzo, let us now study the contents of the A- and
    B-twisted $N=(2,2)$ sigma models.

   \subsubsection{The A-model}
    \label{sec:amodel}
    From equations (\ref{eq:mgenerator}) and (\ref{eq:rgenerator}), we can
    easily read off in which bundles over the worldsheet the fermion fields
    live after the $A$-twist:
    \bea
     \psi_+^i \equiv \psi^i_{z} & \in & \gO^{1,0} \otimes \phi^*(T^{(1,0)} \cM) \ret
     \psi_-^i \equiv \chi^i & \in & \phi^*(T^{(1,0)} \cM) \ret
     \psibar_+^{i} \equiv \chi^{\ibar} & \in & \phi^*(T^{(0,1)} \cM) \ret
     \psibar_-^{i} \equiv \psi^{\ibar}_{\zbar} & \in & \gO^{0,1} \otimes
     \phi^*(T^{(0,1)} \cM),
    \eea
    where ``$\in$'' really means ``is a section of''. For example, $\psi_+^i$
    becomes a worldsheet $(1,0)$-form after the twisting, and its
    unchanged $i$-index denotes that it takes values in the holomorphic tangent
    bundle on $\cM$ -- or more precisely, as is denoted by the
    $\phi^*(\cdot)$-notation: it takes values in the pullback of this
    bundle to the worldsheet $\gS$, where we need to pull back using the map
    $\phi$ which embeds $\gS$ in $\cM$. In the above list, we gave the fields 
    new names which better represent their new transformation properties. In
    particular, note that $\psi_-^i$ and $\psibar^{i}_+$ naturally combine
    into a section of the pull-back of the full complexified tangent bundle to
    $\cM$.

    \sk In terms of these new fields, the Lagrangean density
    (\ref{eq:fullDterm}) can be written as
    \bea
     L & =  -2t & \Big( g_{i \jbar} \d_z \phi^i \d_{\zbar} {\phibar}^j +
     g_{i \jbar} \d_{\zbar} \phi^i \d_{z} {\phibar}^j \ret
     && + i g_{i \jbar} \psi_{z}^i \gD_{\zbar} \chi^{\jbar} 
     + i g_{i \jbar} \psi_{\zbar}^{\jbar} \gD_{z} \chi^i_- 
     + \half R_{i \jbar k \lbar} \psi^i_{z} \psi^{\jbar}_{\zbar} \chi^k \chi^{\lbar} \Big)
     \label{eq:amodellagr}
    \eea
    where we did a partial integration to write the worldsheet scalars $\chi$
    to the right of the covariant derivatives and rewrote $\gD_+ = \gD_z, \gD_-
    = \gD_{\zbar}$. Moreover, we multiplied the expression by a coupling (or
    inverse Planck) constant $t$. Note that all terms in the above Lagrangean come
    with one $z$ and one $\zbar$-index. This implies that we can write the whole
    Lagrangean as $\eta^{\ga \gb} L_{\ga \gb}$, and using this form we can
    straightforwardly covariantize it to an action on any curved worldsheet with
    metric $h^{\ga \gb}$. By our twisting procedure, we have achieved that this
    covariantized action is still invariant under transformations generated by
    $\eps Q_A$, where $\eps$ is now a constant spin zero (but anticommuting)
    parameter. For simplicity, in what follows we will keep working with the
    case of a flat worldsheet metric, but the generalization to curved
    worldsheets will always be conceptually (though not necessarily
    calculationally) straightforward.

    \sk First of all, we would like to see if we can write the Lagrangean in
    the form $\{Q_A, V\}$ so that, using the arguments of section
    \ref{sec:cohomologicalft}, we can immediately conclude that our theory is
    topological and independent of $t$. It turns out that this is almost
    possible: one can write
    \be
     L' = - i t \{ Q_A, V \}
     \label{eq:lprime}
    \ee
    with
    \be
     V = g_{i \jbar} \left( \psi^{i}_z \d_{\zbar} \phibar^j + \d_z
     \phi^{i} \psi^{\jbar}_{\zbar} \right).
     \label{eq:amodelv}
    \ee
    It is a rather tedious exercise to write out (\ref{eq:lprime}), but doing
    so one finds an expression which quite closely resembles the Lagrangean
    $L$. In fact, by adding multiples of the $\psi^i_z$- and
    $\psi^{\ibar}_{\zbar}$-equations of motion, one can write $L'$ in a
    form where all the terms involving fermions equal those in $L$. To be precise,
    one finds
    \be
     L' = L - 2 t g_{i \jbar} \Big( \d_z \phi^i \d_{\zbar} {\phibar}^j -
     \d_{\zbar} \phi^i \d_z \phibar^j \Big).
     \label{eq:ldiffamodel}
    \ee
    The difference between the two Lagrangeans has a simple geometric
    interpretation in terms of the pullback to the worldsheet of the target
    space K\"ahler form $\go = 2 i g_{i \jbar} dz^i \wedge d \zbar^j$:
    \bea
     S - S' & = & 2 t \int_\gS d^2 z g_{i \jbar}  \Big( \d_z \phi^i
     \d_{\zbar} {\phibar}^j - \d_z \phi^i \d_{\zbar}
     {\phibar}^j \Big) \ret
     & = & t \int_\gS \phi^*(\go) \ret
     & = & t \int_{\phi(\gS)} \go
    \eea
    Now, it is crucial that $\go$ is a closed form -- that is, its integral only
    depends on the homology class of $\phi(\gS)$! This homology class is usually
    denoted by $\gb \in H_2(\cM)$, and the integral of the K\"ahler form over
    it is written as $\go \cdot \gb$. Using this notation, we have that
    \be
     S - S' = t \go \cdot \gb.
    \ee
    For a fixed homology class $\gb$ of $\phi(\gS)$, this term will simply
    contribute a prefactor $\exp(-t \go \nolinebreak \cdot \nolinebreak \gb)$.
    This number does not depend
    on the metric $h^{\ga \gb}$ on the worldsheet, and the rest of the quantum
    measure, $e^{-S'}$, is $Q_A$-exact. This explicitly shows, using the
    arguments of section \ref{sec:cohomologicalft} for $Q$-exact actions, that
    our theory is topological with respect to the worldsheet metric. Similarly,
    since $d S' / d t$ is $Q_A$-exact, we see that after extracting the simple
    $t$-dependent prefactor the rest of the path integral becomes
    $t$-independent, so we can calculate it exactly by taking the classical $t
    \to \infty$ limit!

    \sk Of course, this reasoning can be extended to the statement that the
    theory does not depend on {\em any} structure which appears only in $V$,
    but not in $S - S'$. In particular, $S - S'$ does not depend on the complex
    structure of $\cM$, so our model is
    independent of a choice of complex structure. However, through $\go$ it
    clearly depends on a choice of K\"ahler class on 
    $\cM$. We see that the theory is ``half-topological'' with respect to the
    {\em target space} manifold $\cM$: it depends on ``half'' of the moduli of
    this Calabi-Yau manifold.

    \sk The reader might be worried about the fact that in deriving the crucial
    equation (\ref{eq:ldiffamodel}) we have used the equations of motion.
    Our goal is to calculate exact path integrals, which involve integrals
    over {\em all} field configurations, and not just those that satisfy the
    equations of motion. Therefore, we should really work off-shell and
    not invoke the equations of motion at all. The reason that our procedure is
    still valid is the following. Instead of using the $\psi$-equations of
    motion in the Lagrangean, we could equivalently define a new operator
    $\Qtilde_A$ whose action on the $\psi$-fields equals that of $Q_A$ up to a
    term proportional to the
    $\psi$-equations of motion, and whose action on the other fields equals
    that of $Q_A$. With such an operator, we can achieve $\{\Qtilde_A, V \} =
    L'$ without further use of the equations of motion. Of course, since we
    added multiples of the equations of motion, $\Qtilde_A^2 = 0$ only
    on-shell. Now however, one can reintroduce the auxiliary fields $F$ to find
    a theory where $\Qtilde_A^2 = 0$ off-shell as well. This therefore gives us
    a new topological theory, which is only marginally different from the
    previous one. In particular, changing the $Q_A$-action does not change the
    path integrals. However, what {\em does} change is the set of physical
    operators, which is now given by the $\Qtilde_A$-cohomology, and hence the
    set of correlation functions that are topologically invariant. In
    what follows, however, we will only be interested in correlation functions
    of operators that do not involve $\psi$, and since $Q_A$ and $\Qtilde_A$
    only differ in their action on $\psi$, it is not hard to show that this
    class of physical correlation functions does not change at all, and we are
    allowed to work with the $Q_A$-theory instead.

    \sk Let us now discuss what topological information of the target space the
    $A$-model observables calculate. First of all, many of the observables one
    could write down are zero, by the same argument we have used before: one
    should write down correlators which are (quantum-) invariant under the
    $R_V$ and the $R_A$ symmetry. However, we cannot simply copy the results
    from the untwisted $N=(2,2)$-theory, since the fermions now live in
    different bundles, and may have different numbers of zero modes!

    \sk So, let us study the expectation value of a general local operator
    \be
     \cO_B = B_{i_1 \cdots i_p \jbar_1 \cdots \jbar_q} (\phi) \chi^{i_1} \cdots
     \chi^{i_p}\chi^{\jbar_1} \cdots \chi^{\jbar_q}.
     \label{eq:amodeloperator}
    \ee
    The reason for not including factors of $\psi$ in this construction is
    that they carry $z$-indices, and we can only remove these in a covariant
    way by using the worldsheet metric, which of course we are not allowed to
    do if we want to obtain topological quantities. For the same reason, we
    cannot use derivatives of $\chi$ and $\phi$. Of course, one could take such
    quantities and integrate them to remove the $z$-indices, but this will no
    longer result in {\em local} operators. Therefore, the above construction
    gives all invariant local operators one can construct from our fields.

    \sk Note that using the identification
    \bea
     \chi^i & \leftrightarrow & d \phi^i \ret
     \chi^{\ibar} & \leftrightarrow & d \phi^{\ibar},
    \eea
    the above operator has all the transformation properties of a $(p,q)$-form on
    $\cM$, and in fact that is what we will identify it with. Actually, writing
    out the necessary terms in $\{ Q_A, \Phi^i \}$ for a chiral superfield
    $\Phi^i$ (and doing the same calculation for $\Phibar^j$) one finds the
    component field transformations
    \bea
     \{ Q_A, \phi^i \} & = & - \chi^i \ret
     \{ Q_A, \phibar^i \} & = & - \chi^{\ibar} \ret
     \{ Q_A, \chi^i \} & = & 0 \ret
     \{ Q_A, \chi^{\ibar} \} & = & 0.
    \eea
    These show that one can identify $Q_A$ with (minus) the de Rham cohomology
    operator:
    \be
     \{ Q_A, \cO_B \} = - \cO_{dB}.
    \ee
    As discussed in section \ref{sec:cohomologicalft}, the physical operators
    of a cohomological field theory are by definition given by its
    $Q$-cohomology classes. We therefore find an important result: {\em 
    the local physical operators in the $A$-model are in one-to-one
    correspondence with de Rham cohomology elements on $\cM$}! 

    \sk Now, as in the untwisted case, one can show by using an
    index theorem that the differential operators in the action acting on
    the fermions have zero modes. Even though the $A$-model can be defined for
    any K\"ahler target space, the results take the simplest form if we choose
    a Calabi-Yau target, so let us restrict to this case. Then the number of
    $\gD_z$-zero modes is given by the simple expression
    \be
     k = m(1-g).
     \label{eq:numberofzm}
    \ee
    By complex conjugation, the number of $\gD_{\zbar}$-zero modes is the
    same. Just as before, $k$ is really the {\em difference} between the
    number of $\gD_z$ and $\gD_z^\dagger$-zero modes. Again, we will assume
    that we are in the generic situation where only one of these two operators
    has zero modes, so if $k>0$ there will be $\chi^i$ and $\chi^{\ibar}$ zero
    modes, and if $k<0$ there will be $\psi^i_z$- and
    $\psi^{\ibar}_{\zbar}$-zero modes.
    
    \sk We see that to have a chance of getting nonzero results, we
    either have to take $g=1$, in which case the partition function itself is
    nonzero, or $g=0$, in which case the nonzero expectation values are the
    operators which have $m$ insertions of $\chi^i$ and $\chi^{\ibar}$ -- i.e.\
    those corresponding to degree $(m,m)$ ``top-forms''. For $g>1$, we would
    have to absorb the $\psi^{\ibar}_z$ zero modes, but as we have seen this is
    not possible if we insist on using local and metric-independent
    operators only.

    \sk What do these correlation functions calculate? Recall that up to the
    factor of $\exp(-t \go \cdot \gb)$, the expectation value can be calculated
    in the classical large $t$ limit. In this limit, $\phi$ has to satisfy the
    equations of motion following from $L'$. In $L'$, only $\d_{\zbar} \phi^i$
    and $\d_z \phi^{\ibar}$ appear, and we find the equations of motion
    \be
     \d_{\zbar} \phi^i = \d_z \phi^{\ibar} = 0,
    \ee
    that is, $\phi$ should be a {\em holomorphic} map from $\gS$ to $\cM$. In
    general, the space of such maps is finite-dimensional, and hence the path
    integral reduces to a finite-dimensional integral over this space! This
    space, which depends on the genus $g$ of the worldsheet $\gS$, on the target
    space $\cM$, and on the homology class (also called the ``degree'') $\gb$
    of $\phi(\gS)$, is called ``the moduli space of holomorphic instantons of
    degree $\gb$'' and denoted by $M_g(\cM, \gb)$. Its virtual complex
    dimension can be calculated, and turns out to be $m(1-g)$. That this equals
    the number $k$ of zero modes we found before is not a coincidence: one can
    show that each $\chi^i$-zero mode in fact corresponds to a possible
    holomorphic deformation of the embedding of $\gS$ in $\cM$. Intuitively,
    this may seem natural from our identification of $\chi^i$ with $dz^i$. One
    should also compare this result to the worldvolume theories on D-branes,
    where the massless scalar fields also correspond to the possible
    deformations of the embedding.

    \sk The $Q_A$-cohomology classes of the operators $\cO_B$ in
    (\ref{eq:amodeloperator}), for the moment without restriction on the
    degree of $B$, span a vector space which as we have shown is naturally
    isomorphic to the de Rham cohomology of $\cM$. A useful basis to take for
    this vector space is $\{ \cO_S \}$, where the $S$ are the Poincar\'e dual
    cohomology classes to a set of cycles $\{ \cS \}$ which form a basis for the
    homology of $\cM$. Recall
    from section \ref{sec:relationshodgenumbers} that for a cycle $\cS_{(p)}$
    of real dimension $p$, the Poincar\'e dual class $S^{(2m-p)}$ has degree
    $2m-p$, and satisfies
    \be
     \int_{\cS_{(p)}} A^{(p)} = \int_{\cM} A^{(p)} \wedge S^{(2m-p)}
    \ee
    for all $p$-forms $A^{(p)}$. In particular, one can take a
    ``delta function'' representative of $S^{(2m-p)}$, which has support only
    on $\cS$. By choosing a basis $\cS^i_{(p)}$ for the $p$th homology of
    $\cM$, one can in this way construct a basis $S_i^{(2m-p)}$ for 
    the $(2m-p)$th cohomology of $\cM$.

    \sk Let us study a general correlator
    \be
     \langle \cO_{S_1}(P_1) \cdots \cO_{S_r}(P_r) \rangle,
     \label{eq:Acorrelator}
    \ee
    where the $P_i$ are certain marked points on the worldsheet\footnote{We
    assume here for simplicity that the worldsheet with these marked points
    has no remaining symmetries.}. As we have seen, for this expression to be
    nonzero, the sum of the degrees
    of the forms $S_i$ must be $2m(1-g)$. Moreover, if we take the delta-function
    representatives of $S_i$, the point $\phi(P_i)$ must be on $\cS^i$ to lead
    to a nonzero result. Note
    that if the dimension of $\cS^i$ is $p$, this sets $2m-p$ constraints on
    the map $\phi$, which is exactly the degree of the form $S_i$. Since the
    sum of these degrees should be $2m(1-g)$ to give a nonzero result, there is a
    total of $2m(1-g)$ constraints on $\phi$. But $\phi$ itself lives in a
    space of real dimension $2m(1-g)$, so we are really integrating a
    ``delta-function'' over a zero-dimensional space -- that is, we are
    counting a number of points!

    \sk Note that in this reasoning we have not fixed $g$. Therefore formally,
    our argument applies to any $g$. Of course in practice, we can only use it
    for $g=0, 1$, since for higher genus the virtual dimension of the moduli
    space is negative -- that is, generically there will be no complex maps
    from the curve $\gS$ of genus $g>1$ to an arbitrary manifold $\cM$. As we
    have seen in the simple example in the previous section, the solution to
    this is to add extra parameters to our moduli space. This is exactly what
    topological string theory (as opposed to topological field theory) will do
    for us, and the cancellation for arbitrary $g$ above gives us some hope of
    finding nonzero correlators at any genus. We will see exactly how this
    comes about in the next chapter.

    \sk So, to summarize, the correlator (\ref{eq:Acorrelator}) gives the
    result
    \be
     \langle \cO_{S_1}(P^1) \cdots \cO_{S_r}(P^r) \rangle_{\gb} =e^{-t \go
     \cdot \gb} \cdot \# M_{g,r}(\cM, \gb, \cS^i),
     \label{eq:almostgromovwitten}
    \ee
    where $M_{g,r}(\cM, \gb, \cS^i)$ is the moduli space of holomorphic maps of
    a Riemann surface $\gS$ of degree $g$ with $r$ labelled points $P_i$ on it,
    into a certain homology class $\gb \in H_2 (\cM)$, in such a way that
    $\phi(P_i)$ is mapped into $\cS^i$. Of course in quantum field theory, we
    would sum this expression over all homology classes $\gb$ to obtain the
    full correlation function.
    
    \sk Note that if we really were to take the large $t$-limit also in the
    first factor of the above expression, then only the holomorphic maps for
    which $\go \cdot \gb=0$ would contribute. These are the holomorphic maps
    for which the ``K\"ahler volume'' of the image vanishes, and one can show
    that this only happens for constant maps. Of course, the moduli space of such 
    maps is simply $\cM$ itself. Moreover, a constant map which intersects all
    of the $\cS^i$ is simply an intersection point of all the $\cS^i$, so the
    correlator then simply reduces to
    \be
     \langle \cO_{S_1}(P^1) \cdots \cO_{S_r}(P^r) \rangle_{t \to \infty} =
     \# (\cS^1 \cap \cdots \cap \cS^r).
    \ee
    One can think of (\ref{eq:almostgromovwitten}) as ``quantum intersection
    numbers'' which generalize this.

    \sk A remark may be in place here. Note that in the path integral, we are
    integrating over a moduli space $M_{g,r}$. However, the operators $\cO$
    seem to correspond to differential forms on $\cM$. How can these concepts
    be matched and how can they lead to topological invariants? The answer is
    that one can always ``pull back'' differential forms on the manifold to
    differential forms on moduli space using the distinguished points $P_i$.
    That is, every point $Y$ in moduli space stands for a map $\phi_Y$ from
    $\gS$ to $\cM$. In particular, by choosing an $i \in \{1, \ldots, r \}$ we
    can define a map $ev_i$ (``evaluation in point $i$'') from $M_{g,r}$ into
    $\cM$ by $ev_i: Y \mapsto \phi_Y(P_i)$. Using these maps, one can now
    pull back a $p$-form $B$ on $\cM$ to a $p$-form $ev_i^*(B)$ on $M_{g,r}$.
    In this way, all of the $\cO_{S_i}$ can be pulled back to differential
    forms on the moduli space by using the appropriate evaluation map. It is
    a well-known and easy to prove result that the pull-backs of forms in the
    same cohomology class also end up in the same cohomology class, so it does
    not matter which representative inside a particular cohomology class we choose for this. We
    have seen that the degrees of these forms precisely add up to the (virtual) 
    dimension of the moduli space, and therefore we can indeed naturally
    integrate the resulting form over the whole moduli space.

    \sk So far, we have fixed the positions of the points $P_i$. It may
    actually seem more natural to add these positions as extra moduli to our
    problem. In fact, as soon as we start considering topological {\em string}
    theory in the next section, we will want to integrate over worldsheet
    metrics, so the notion of a ``fixed'' point does not make sense anymore and
    we are forced to add the positions of $P_i$ as moduli. The appearance of
    these $r$ extra moduli means that we have to make
    the dimensions of the $\cS^i$ smaller as well to end up with a simple counting
    problem. (Pictorially, this can be seen from the fact that now the
    points can move around on the surface, so we can for example make each
    $\cS^i$ one dimension smaller since the point on the surface has one
    complex degree of freedom itself to ``find'' $\cS^i$.) The famous
    invariants one obtains by 
    this generalization are called {\em Gromov-Witten invariants}.

   \subsubsection{The B-model}
    \label{sec:bmodel}
    We begin our study of the $B$-model in exactly the same way as for the
    $A$-model: by listing in which bundles the fermions live after the
    $B$-twist.
    \bea
     \psi_+^i & \in & \gO^{1,0} \otimes \phi^*(T^{(1,0)} \cM) \ret
     \psi_-^i & \in & \gO^{0,1} \otimes \phi^*(T^{(1,0)} \cM) \ret
     \psibar_+^{i} & \in & \phi^*(T^{(0,1)} \cM) \ret
     \psibar_-^{i} & \in & \phi^*(T^{(0,1)} \cM).
    \eea
    The difference from the $A$-model seems minimal. Now the two worldsheet
    scalar fields have the {\em same} type of tangent space index -- i.e.\ both
    of them are space-time $(0,1)$-forms, whereas in the $A$-model case we had
    one $(1,0)$-form and one $(0,1)$-form. Similarly, the worldsheet one-forms
    now both are space-time $(1,0)$-forms. We will see, however, that the
    consequences of these small differences turn out to be quite substantial.
    Again, it is convenient to relabel the fields in a way which makes their 
    worldsheet spin more manifest. We define
    \bea
     \eta^{\ibar} & = & \psibar^{i}_+ + \psibar^{i}_- \ret
     \gt_i & = & g_{i \jbar} (\psibar^{j}_+ - \psibar^{j}_-) \ret
     \rho_z^i & = & \psi^i_+ \ret
     \rho_{\zbar}^i & = & \psi^i_-.
    \eea
    Of course, we could just as well have defined $\gt^{\ibar}$ instead of
    $\gt_i$, but it turns out that the form introduced above leads to the
    simplest expressions, in particular for the $Q_B$-transformations. For
    example, one finds that $\{Q_B, \gt_i \} = 0$, whereas $\{ Q_B, \gt^{\ibar}
    \} = - 2 \Gam^{\ibar}_{\jbar \kbar} \eta^{\jbar} \gt^{\kbar}$.
    The Lagrangean density in terms of the above fields can be written as
    \bea
     L & = - t & \Big( g_{i \jbar} \eta^{\ga \gb} \d_\ga \phi^i \d_{\gb}
     \phibar^j
     + i g_{i \jbar} \eta^{\jbar} (\gD_{\zbar} \rho^i_{z} + \gD_{z}
     \rho^i_{\zbar}) 
     + i \gt_i (\gD_{\zbar} \rho^i_z - \gD_z \rho^i_{\zbar}) \ret
     && + \half {R_{i \jbar k}}^l \rho^i_z \rho^k_{\zbar} \eta^{\jbar} \gt_l
     \Big),
    \eea
    where we again introduced a coupling constant $t$. Note that this
    formula can also be straightforwardly covariantized -- compare our remarks
    in the $A$-model case. 
    
    \sk Just like for the $A$-model, to facilitate our study of the topological nature
    of the model we would now like to find a Lagrangean $L'$ which is as
    similar to $L$ as possible and which can be written as
    \be
     L' = - i t \{ Q_B, V \}.
    \ee
    Here, however, we run into a problem that we did not have with the
    $A$-model. We can write down a similar expression to the one in
    (\ref{eq:amodelv}):
    \be
     V = g_{i \jbar} \left( \rho_z^i \d_{\zbar} \phibar^j + \rho_{\zbar}^i \d_z
     \phibar^j \right),
    \ee
    for which we find
    \be
     L - L' = -t \left( i \gt_i (\gD_{\zbar} \rho^i_z - \gD_z \rho^i_{\zbar}) 
     + \half {R_{i \jbar k}}^l \rho^i_z \rho^k_{\zbar} \eta^{\jbar} \gt_l
     \right).
     \label{eq:ldiffbmodel}
    \ee
    In deriving this expression, we have not yet used the equations of motion.
    At first sight, it seems that we can now use the $\gt$-equation of motion
    to set this whole expression to zero. This, however, is not allowed, since
    we really need to work off-shell to calculate exact path integrals. In the
    $A$-model case, we could argue our way out of this problem by using the
    fact that the operators we wanted to work with did not involve the
    $\psi$-fields. Here, however, we will be interested in operators containing
    $\gt$, so we will have to live with the above expression as it is. (Some
    experimenting shows that other choices for $V$ do not help us either.)
    
    \sk Fortunately, things are not quite as bad as they seem. Since the right
    hand side of (\ref{eq:ldiffbmodel}) is antisymmetric in the exchange of the
    $z$- and $\zbar$-indices, we can write it as a differential $(1,1)$-form. The
    integral of such a form over a two-dimensional manifold is independent of
    the metric, so the only metric dependence of $L$ is in the $\{ Q_B, V \}$
    term, and we find by the standard argument that the theory is topological
    with respect to the worldsheet metric.
    
    \sk Moreover, note that the
    right-hand side of equation (\ref{eq:ldiffbmodel}) is linear in $\gt$, and
    that $V$ does not contain $\gt$ at all. This means that we can remove the
    $t$-dependence of (\ref{eq:ldiffbmodel}) by absorbing $t$ in a redefinition of
    $\gt$. As long as we will study correlation functions which are homogeneous
    in $\gt$ -- and this will be the case of our interest -- the path integral
    changes simply by an overall factor of $t$ to some power. This means that
    to calculate these correlation functions, we can again take the large $t$
    limit and obtain exact results. Note the big difference from the $A$-model:
    there, we found a different $t$-dependent prefactor for each homology class
    in the target space, so the final summed results still have a quite
    nontrivial $t$-dependence. Here, the overall prefactor is ``universal'', so
    even after summing over the different homologies of the image of the
    worldsheet, the correlation functions will still have a very simple
    power-law $t$-dependence.
    
    \sk Other results, however, are not so easily obtained. In particular, it
    is not easy to see on what {\em target space} structures the theory
    depends. A calculation of the relevant terms in $\{ Q_B, \Phi^i \}$ and $\{
    Q_B, \Phibar^i \}$ shows that $\{ Q_B, \phi^i \} = 0$ but $\{ Q_B,
    \phibar^i \} = - \eta^{\ibar}$. From this asymmetry it is clear that the
    theory depends on the choice of target space complex structure. What about
    the K\"ahler structure? Without doing the actual calculation, let us simply
    remark that one can show that the variation of the action with respect
    to the cohomology class of the K\"ahler form $\go$ is $Q_B$-exact, so 
    the theory does not depend on the K\"ahler moduli of the Calabi-Yau.
    One should compare this statement to the ``mirror'' result in the
    $A$-twisted case, where we found independence of the complex structure
    moduli but dependence on the K\"ahler ones.

    \sk Just like in the $A$-twisted case, let us now write down
    metric-independent local operators. They take the form
    \be
     \cO_B = {B_{\ibar_1 \cdots \ibar_p}}^{j_1 \cdots j_q} (\phi, \phibar)
     \eta^{\ibar_1} \cdots \eta^{\ibar_p} \gt_{j_1} \cdots \gt_{j_q}.
    \ee
    Our notation is again suggestive: we want to identify $\eta^{\ibar}$ with
    $d \phi^{\ibar}$ and $\gt_i$ with $\d / \d \phi^i$, so $\cO_B$ corresponds to a
    $(0,p)$-form with values in the antisymmetrized product of $q$ holomorphic
    tangent spaces, which we denote as $\bigwedge^q T^{(1,0)} \cM$. The reason
    for this identification, as well as for the specific linear combinations of
    fields that we used to define $\eta$ and $\gt$, is the form of the
    $Q_B$-transformations, most of which we have already mentioned before:
    \bea
      \{ Q_B, \phi^i \} & = & 0 \ret
      \{ Q_B, \phibar^i \} & = & - \eta^{\ibar} \ret
      \{ Q_B, \gt_i \} & = & 0 \ret
      \{ Q_B, \eta^{\ibar} \} & = & 0,
    \eea
    from which we find that
    \be
     \{ Q_B, \cO_B \} = - \cO_{\delbar B}.
    \ee
    (Again, we have a clash of conventional notations here: of course the
    $B$-index on $Q_B$ is not related to the subscript $B$ on $\cO_B$.)
    Thus, we see that $Q_B$ can be viewed as (minus) the Dolbeault exterior
    derivative $\delbar$ acting on differential forms with values in
    $\bigwedge^q T^{(1,0)} \cM$, and the physical operators in the theory are
    in one-to-one correspondence to the $\delbar$-cohomology classes of such
    forms.

    \sk Using the appropriate index theorems, we find that the
    correlation functions are nonzero if and only if
    \be
     p = q = m(1-g).
    \ee
    Again, this means that at genus $1$ the partition function is nonzero, and
    at genus $0$ we should study the expectation values of $(0,m)$-forms with
    values in $\bigwedge^m T^{(1,0)} \cM$.

    \sk We have already encountered several big differences between the $A$-
    and $B$-models, but perhaps the most important one is the following. In the
    $A$-model case, only the term $g_{i \jbar} \d_z \phi^i \d_{\zbar}
    \phi^{\jbar}$ appeared in the Lagrangean $L'$, which led to the conclusion
    that the path integral only got contributions from the holomorphic maps
    $\phi: \gS \to \cM$. Here, $L'$ also contains a term $g_{i \jbar}
    \d_{\zbar} \phi^i \d_{z} \phi^{\jbar}$, and as a result the equations of
    motion for $\phi$ and $\phibar$ are
    \be
     \d_z \phi^i = \d_{\zbar} \phi^i = \d_z \phi^{\ibar} = \d_{\zbar} \phi^{\ibar} =
     0,
    \ee
    which only have the constant maps as their solutions. Since the
    ``classical'' $t \to \infty$ limit, as we have argued above, up to an
    overall power of $t$ actually gives the exact correlation functions for any
    $t$, we thus find that the ``moduli space of curves'' over which we have to
    integrate is the space of constant maps -- which is simply the manifold
    $\cM$ itself. The evaluation of correlation functions therefore seems to be
    simply given by integrating $(0,m)$-forms with values in $\bigwedge^m
    T^{(1,0)} \cM$ over $\cM$.

    \sk However, here we encounter a puzzle. The objects which we know
    can be invariantly integrated over $\cM$ are $(m,m)$-forms. The question is
    therefore: how are the objects we have found related to $(m,m)$-forms?
    There seems to be only one topologically invariant way to achieve this, and
    this is by contracting the holomorphic indices of $B$ with the holomorphic
    indices of the $(m,0)$-form $\gO$ on $\cM$, and then multiplying the resulting
    $(0,m)$-form by this same $(m,0)$-form:
    \be
     {B_{\ibar_1 \cdots \ibar_m}}^{j_1 \cdots j_m} \mapsto {B_{\ibar_1 \cdots
     \ibar_m}}^{j_1 \cdots j_m} \gO_{j_1 \cdots j_m} \gO_{k_1 \cdots k_m}.
    \ee
    By carefully considering the definition of the path integral, one can
    indeed show (though the author admits
    he has never seen a proof he understood) that this is the procedure one
    needs to follow. (Note that we now even more explicitly see the dependence
    of the theory on the complex structure of $\cM$.) We thus arrive at the
    result that the observables of the $B$-model are nothing but integrals of
    wedge products of forms over the target space $\cM$! In this sense, the
    $B$-model is much simpler than the $A$-model, where we have to integrate
    over a moduli space, but on the other hand the results of the $A$-model at
    first sight seem mathematically much more interesting. Actually, in the
    next chapter we will briefly comment on how, using mirror symmetry, the
    virtues of the two different models can be combined.

 \section{Topological strings}
  \label{sec:topologicalstrings}
  The two-dimensional field theories we have constructed are already very
  similar to string theories. However, one ingredient from string theory is
  missing: in string theory, the worldsheet theory does not only involve
  a path integral over the maps $\phi^i$ to the target space and their
  fermionic partners, but also a path integral over the worldsheet metric
  $h_{\ga \gb}$. So far, we have set this metric to a fixed background value.

  \sk We have also encountered a drawback of our construction. Even though the
  theories we have found can give us some interesting ``semi-topological''
  information about the target spaces, one would like to be able to define
  general nonzero $n$-point functions at genus $g$ instead of just the
  partition function at genus one and the particular correlation functions we
  calculated at genus zero.
  
  \sk It turns out that these two remarks are intimately related. In this
  section we will go from topological field theory to topological string theory
  by introducing integrals over all metrics, and in doing so we will find
  interesting nonzero correlation functions at any genus.

  \subsection{Coupling to topological gravity}
   In coupling an ordinary field theory to gravity, one has to perform three
   steps.
   \begin{itemize}
    \item
     First of all, one rewrites the Lagrangean of the theory in a
     covariant way by replacing all the flat metrics by the dynamical ones,
     introducing covariant derivatives and multiplying the measure by a factor of
     $\sqrt{\det h}$.
    \item
     Secondly, one introduces an Einstein-Hilbert term as the
     ``kinetic'' term for the metric field, plus possibly extra terms and
     fields to preserve the symmetries of the original Lagrangean.
    \item
     Finally, one has to integrate the resulting theory over the space of all
     metrics.
   \end{itemize}

   \sk In these notes, I will not discuss the first two steps in
   this procedure. As we have seen in our discussion of topological field
   theories, the precise form of the Lagrangean only plays a comparatively minor
   role in determining the properties of the theory, and we can derive many
   results without actually considering a Lagrangean. Therefore, let us just
   state that it is possible to carry out the analog of the first two steps
   mentioned above, and construct a Lagrangean with a ``dynamical'' metric which
   still possesses the topological $Q$-symmetry we have constructed. The reader
   who is interested in the details of this construction is referred to the
   paper \cite{Witten:1989ig} by E.~Witten and to the lecture notes
   \cite{Dijkgraaf:1990qw} by R.~Dijkgraaf, E.~Verlinde and H.~Verlinde.

   \sk The third step, integrating over the space of all metrics, is the one we
   will be most interested in here. Naively, by the metric independence of our
   theories, integrating their partition functions over the
   space of all metrics, and then dividing the results by the volume of the
   topological ``gauge group'', would be equivalent to multiplication by a
   factor of 1:
   \be
    \frac{1}{G_{top}} \int Dh \, Z[h] \buildrel{?}\over{=} Z[h_0]
    \label{eq:naivetop}
   \ee
   for any arbitrary background metric $h_0$.
   There are several reasons why this naive reasoning might go wrong:
   \begin{itemize}
    \item
     There may be metric configurations which cannot be reached from a given
     metric by continuous changes.
    \item
     There may be anomalies in the topological symmetry at the quantum level
     preventing the conclusion that all gauge fixed configurations are
     equivalent.
    \item
     The volume of $G_{top}$ is infinite, so even if we could rigorously define
     a path integral the above multiplication and division would not be
     mathematically well-defined.
   \end{itemize}
   For these reasons, we should really be more careful and precisely define
   what we mean by the ``integral over the space of all metrics''. Let us
   note the important fact that just like in ordinary string theory (and even
   before twisting), the two-dimensional sigma models become conformal field
   theories when we include the metric in the 
   Lagrangean. This means that we can borrow the technology from string
   theory to integrate over all conformally equivalent metrics. As is well
   known, and as we will discuss in more detail later, the conformal symmetry
   group is a huge group, and integrating over conformally equivalent metrics
   leaves only a finite-dimensional integral over a set of worldsheet moduli.
   Therefore, our strategy will be to use the analogy to ordinary string theory
   to first do this integral over all conformally equivalent metrics, and then
   perform the integral over the remaining finite-dimensional moduli space.

   \sk In integrating over conformally equivalent metrics, one
   usually has to worry about conformal anomalies.
   However, here a very important fact comes to our help. To understand this
   fact, it is useful to rewrite our twisting procedure in a somewhat different
   language. 
   
   \sk Let us consider the energy-momentum tensor $T_{\ga \gb}$, which is the
   conserved Noether current with respect to global translations on $\bC$.
   From conformal field theory, it is known that $T_{z \zbar} = T_{\zbar z} =
   0$, and the fact that $T$ is a conserved current, $\d_\ga {T^\ga}_\gb = 0$,
   means that $T_{zz} \equiv T(z)$ and $T_{\zbar \zbar} \equiv \Tbar (\zbar)$
   are (anti-)holomorphic in $z$. One can now expand $T(z)$ in Laurent
   modes\footnote{The extra power of $-2$, just like the $-1$ in
   (\ref{eq:jmodes}), is conventional in conformal field theory.},
   \be
    T(z) = \sum L_m z^{-m-2}.
    \label{eq:lmodes}
   \ee
   The $L_m$ are called the ``Virasoro generators'', and it is a well-known
   result from conformal field theory that in the quantum theory their
   commutation relations are
   \be
    {[} L_m, L_n ] = (m-n) L_{m+n} + \frac{c}{12} m(m^2 - 1) \gd_{m+n}.
   \ee
   The number $c$ depends on the details of the theory under consideration, and
   it is called the central charge. When this central charge is nonzero, one
   runs into a technical problem. The reason for this is that the equation of
   motion for the metric field is simply
   \be
    \frac{\gd S}{\gd h^{\ga \gb}} = T_{\ga \gb} = 0.
   \ee
   In conformal field theory, one imposes this equation as a constraint in the
   quantum theory. That is, one requires that for physical states
   $| \psi \rangle$,
   \be
    L_m | \psi \rangle = 0 \qquad \forall m \in \bZ
   \ee
   However, this is clearly incompatible with the above commutation relation
   unless $c=0$. In string theory, this value for $c$ can be achieved by taking
   the target space of the theory to be ten-dimensional. If $c \neq 0$ the
   quantum theory is problematic to define, and we speak of a ``conformal
   anomaly''.

   \sk Of course, the whole above story repeats itself for $\Tbar(\zbar)$ and
   its modes $\Lbar_m$. At this point there is a crucial difference between
   open and closed strings. On an open string, left-moving and right-moving
   vibrations are related in such a way
   that they combine into standing waves. In our complex notation,
   ``left-moving'' translates into ``$z$-dependent'' (i.e.\ holomorphic), and
   ``right-moving'' into ``$\zbar$-dependent'' (i.e.\ anti-holomorphic). Thus,
   on an open string all holomorphic quantities are related to their
   anti-holomorphic counterparts. In particular, $T(z)$ and $\Tbar(\zbar)$, and
   their modes $L_m$ and $\Lbar_m$, turn out to be complex conjugates. There is
   therefore only one independent algebra of Virasoro generators $L_m$.

   \sk On a {\em closed} string on the other hand, which is the situation we
   have been studying so far, left- and right-moving waves are completely
   independent. This means that all holomorphic and anti-holomorphic
   quantities, and in particular $T(z)$ and $\Tbar(\zbar)$, are independent. One
   therefore has two sets of Virasoro generators, $L_m$ and $\Lbar_m$.

   \sk Let us now analyze the problem of central charge in the twisted
   theories. To twist the theory, we have used the $U(1)$ $R$-symmetries. Any
   global $U(1)$-symmetry of our theory has a conserved current $J_\ga$. The
   fact that it is conserved again means that $J_z \equiv J(z)$ is holomorphic
   and $J_{\zbar} \equiv \Jbar(\zbar)$ is anti-holomorphic. Once again, on an
   open string $J$ and $\Jbar$ will be related, but in the closed string theory we
   are studying they will be independent functions. In particular, this means
   that we can view a global $U(1)$-symmetry as really consisting of two
   independent, left- and right-moving, $U(1)$-symmetries, with generators
   $F_L$ and $F_R$.
   
   \sk Note from equation (\ref{eq:rgenerator}) that the sum of $R$-symmetries
   $F_V + F_A$ only acts on objects with a + index. That is, it acts purely
   on left-moving quantities. (Admittedly, the notation with the bars is
   somewhat confusing here: $z$, $\gt^+$ and $\gtbar^+$ are the
   left-moving and hence ``holomorphic'' quantities.) Similarly, $F_V - F_A$
   acts purely on right-moving quantities. From our discussion above, it is
   therefore natural to identify these two symmetries with the two components
   of a single global $U(1)$ symmetry:
   \be
    F_V = \half (F_L + F_R) \qquad F_A = \half (F_L - F_R).
   \ee
   A more detailed construction shows that this can indeed be done. 
   
   \sk Let us expand the left-moving conserved $U(1)$-current into Laurent
   modes:
   \be
    J(z) = \sum J_m z^{-m-1}.
    \label{eq:jmodes}
   \ee
   The commutation relations of these modes with one another and with the
   Virasoro modes can be calculated, either by writing down all of the modes in
   terms of the fields of the theory, or by using more abstract knowledge from
   the theory of superconformal symmetry algebras. In either case, one finds
   \bea
    {[} L_m, L_n ] & = & (m-n) L_{m+n} + \frac{c}{12} m(m^2 - 1) \gd_{m+n} \ret
    {[} L_m, J_n ] & = & -n J_{m+n}\ret
    {[} J_m, J_n ] & = & \frac{c}{3} m \gd_{m+n}.
    \label{eq:scalgebra}
   \eea
   Note that the same central charge $c$ appears in the $J$- and in the
   $L$-commutators. This turns out to be crucial.
   
   \sk Following the standard Noether procedure, we can now construct a
   conserved charge by integrating the conserved current $J(z)$ over a
   space-like slice of the $z$-plane. In string theory, the physical time
   direction is the radial direction in the $z$-plane, so a spacelike slice is
   just a curve around the origin. The integral is therefore simply calculated
   using Cauchy's theorem:
   \be
    F_L = \oint_{z=0} J(z) dz = 2 \pi i J_0.
   \ee
   In the quantum theory, it will be this operator that generates the
   $U(1)_L$-symmetry. Now recall that to twist the theory we want to introduce
   new Lorentz rotation generators,
   \bea
    M_A & = & M - F_V = M - \half (F_L + F_R) \ret
    M_B & = & M - F_A = M - \half (F_L - F_R).
   \eea
   A well-known result from string theory\footnote{This can be shown by
   representing the $c=0$ Virasoro algebra by operators of the form $z^p \,
   d/dz$, for example. Note that $L_0$ is not the conserved charge
   corresponding to $T(z)$ because of the $-2$ in equation (\ref{eq:lmodes}).
   Of course, this should not be the case, since we constructed $T(z)$ from the
   conserved current related to translations (instead of rotations and radial
   scalings) in the $z$-plane.} is that the generator of Lorentz rotations is
   $M = 2 \pi i (L_0 - \Lbar_0)$. Therefore, we find that the twisting
   procedure in this new language amounts to
   \bea
    A: && L_{0, A} = L_0 - \half J_0 \qquad \Lbar_{0,A} = \Lbar_0 + \half
    \Jbar_0 \ret
    B: && L_{0, B} = L_0 - \half J_0 \qquad \Lbar_{0,B} = \Lbar_0 - \half
    \Jbar_0.
   \eea
   Let us now focus on the left-moving sector; we see that for both twistings
   the new Lorentz rotation generator is the difference of $L_0$ and $\half
   J_0$. Of course, the new Lorentz generator should also correspond to a
   conserved two-tensor, and from (\ref{eq:lmodes}) and (\ref{eq:jmodes}) there
   is a very natural way to obtain such a current:
   \be
    \Ttilde(z) = T(z) + \half \d J(z),
    \label{eq:twistcurrent1}
   \ee
   which clearly satisfies $\delbar \Ttilde = 0$ and
   \be
    \Ltilde_m = L_m - \half (m+1) J_m,
    \label{eq:twistcurrent2}
   \ee
   so in particular we find that $\Ltilde_0$ can serve as $L_{0,A}$ or
   $L_{0,B}$. Of course, we should apply the same procedure (with a minus sign
   in the $A$-model case) in the right-moving
   sector. Equations (\ref{eq:twistcurrent1}) and (\ref{eq:twistcurrent2}) tell
   us how to implement the twisting procedure not only on the conserved
   charges, but on the whole $N=2$ superconformal algebra -- or at least on the
   part consisting of the $J$- and $L$-modes, but a further investigation shows
   that this is the only part that changes. We have motivated, but not
   rigorously derived (\ref{eq:twistcurrent1}); for a complete justification the
   reader is referred to the original papers by W.~Lerche, C.~Vafa and
   N.~Warner \cite{Lerche:1989uy} and by S.~Cecotti and C.~Vafa
   \cite{Cecotti:1991me}.

   \sk Now, we come to the crucial point. The algebra that the new modes
   $\Ltilde_m$ satisfy can be straightforwardly calculated from
   (\ref{eq:twistcurrent1}) and (\ref{eq:scalgebra}), and we find
   \be
    {[} \Ltilde_m, \Ltilde_n ] = (m-n) \Ltilde_{m+n}.
   \ee
   That is, there is no central charge left! This means that we do not have any
   restriction on the dimension of the theory, and topological strings will
   actually be well-defined in target spaces of any dimension.

\begin{figure}[ht]
    \begin{center}
     \includegraphics[height=5cm]{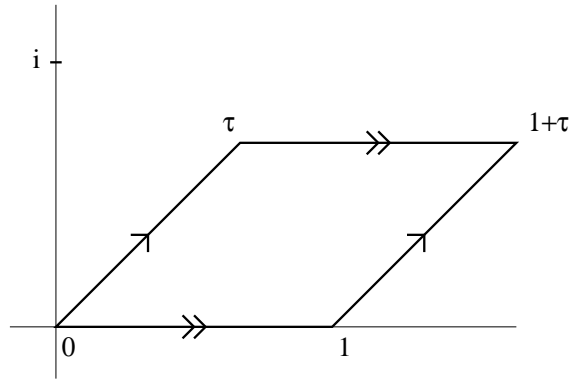}
    \end{center}
    \caption{The modulus $\tau$ of a torus.}
    \label{fig:tau}
   \end{figure}

   \sk From this result, we see that we can integrate our partition function
   over conformally equivalent metrics without having to worry about the
   conformal anomaly represented by the nonzero central charge. After having
   integrated over this large part of the space of all metrics, it turns out
   that there is a finite-dimensional integral left to do. In 
   particular, it is known that one can always find a conformal transformation
   which in the neighborhood of a chosen point puts the metric in the form
   \be
    h_{\ga \gb} = \eta_{\ga \gb},
   \ee
   with $\eta$ the usual flat metric with diagonal entries $\pm 1$. (Or $+1$ in
   the Euclidean setting.) On the other hand, when one considers the global
   situation, it turns out that one cannot always enforce this gauge condition
   everywhere. For example, if the worldsheet is a torus, there is a
   left-over complex parameter $\tau$ that cannot be gauged away. The easiest
   way to visualize this parameter\footnote{A confusing point about this way of
   presenting things is that this torus does seem to have the flat metric
   $\eta_{ij}$. However, the coordinate choice giving this metric is not
   periodic, so if we start from a geometry where the coordinates are periodic
   we cannot reach it. See the book by Polchinski\cite{Polchinski:1998rq},
   chapter 5.1, for a more thorough explanation of these issues.} is by drawing the
   resulting torus in the complex plane and rescaling it in such a way that one
   of its edges runs from $0$ to $1$; the other edge then runs from $0$ to
   $\tau$, see figure \ref{fig:tau}. It seems intuitively clear that a
   conformal transformation -- which should leave all angles fixed -- will
   never deform $\tau$, and even though intuition often fails when considering
   conformal mappings, in this case this can indeed be proven. Thus, $\tau$ is
   really a modular parameter which we need to integrate over. Another fairly
   intuitive result is that any locally flat torus can, after a rescaling, be
   drawn in this form, so $\tau$ indeed is the only modulus of the torus.
   
   \sk More generally, one can show that a Riemann surface of genus $g$ has
   \be
    m_g = 3(g-1)
   \ee
   complex modular parameters. As usual, this is the {\em virtual} dimension of the
   moduli space. If $g>1$, one can show that this virtual dimension equals the
   actual dimension. For $g=0$, the sphere, we have a negative virtual
   dimension $m_g = -3$, but the actual dimension is 0: there is always a flat
   metric on a surface which is topologically a sphere (just consider the
   sphere as a plane with a point added at infinity), and after having chosen
   this metric there are no remaining parameters such as $\tau$ in the torus
   case. For $g=1$, the virtual dimension is $m_g = 0$, but as we have seen the
   actual dimension is 1.
   
   \sk We can explain these discrepancies using the fact that, after we have
   used the conformal invariance to fix the metric to be flat, the sphere and
   the torus have leftover symmetries. In the case of the sphere, it is
   well known in string theory that one can use these extra symmetries to fix
   the positions of three labelled points. In the case of the torus, after
   fixing the metric to be flat we still have rigid translations of the torus
   left, which we can use to fix the position of a single labelled point.
   To see how this leads to a difference between the virtual and the actual
   dimensions, let us for example consider tori with $n$ labelled points on
   them. Since the virtual dimension of the moduli space of tori without
   labelled points is 0, the virtual dimension of the moduli space of tori with
   $n$ labelled points is $n$. One may expect that at some point (and in fact,
   this happens already when $n=1$), one reaches a sufficiently generic
   situation where the virtual dimension really is the actual dimension.
   However, even in this case we can fix one of the positions using the
   remaining conformal (translational) symmetry, so the positions of the points
   only represent $n-1$ moduli. Hence, there must be an $n$th modulus of a
   different kind, which is exactly the shape parameter $\tau$ that we have
   encountered above. In the limiting case where $n=0$, this parameter
   survives, thus causing the difference between the virtual and the real
   dimension of the moduli space.
   
   \sk For the sphere, the reasoning is somewhat more formal: we
   analogously expect to have three ``extra'' moduli when $n=0$. In fact,
   three extra parameters are present, but they do not show up as moduli. They
   must be viewed as the three parameters which need to be added to the problem
   (see section \ref{sec:virtualdim}) to find a zero-dimensional moduli space.
   (This counting of moduli is always a great source of confusion. It is
   helpful to realize that in the case of the sphere, two unrelated things
   happen at the same time: the virtual dimension is negative {\em and} the
   virtual dimension does not equal the actual dimension. If it were not for
   this last fact, we would have needed to add three parameters by hand to find
   a finite-dimensional moduli space. Now, however, the discrepancy between
   virtual and actual dimensions caused by the leftover symmetries of the flat
   sphere already gives us these parameters ``for free''.)
   
\begin{figure}[ht]
    \begin{center}
     \includegraphics[height=5cm]{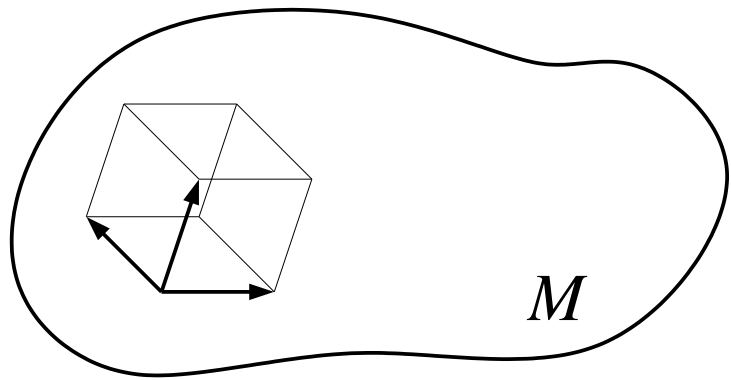}
    \end{center}
    \caption{A measure on the moduli space $M$ assigns a number to every set of
    three tangent vectors. This number is interpreted as the volume of the
    element spanned by these vectors.}
    \label{fig:measure}
   \end{figure}

   \sk Since the cases $g=0, 1$ are thus somewhat special, let us begin by
   studying the theory on a Riemann surface with $g>1$. To arrive at
   the  topological string correlation functions, after gauge fixing we have to
   integrate over the remaining moduli space of complex dimension $3(g-1)$. To
   do this, we need to fix a measure on this moduli space. That is, given a set
   of $6(g-1)$ tangent vectors to the moduli space, we want to produce a number
   which represents the size of the volume element spanned by these vectors,
   see figure \ref{fig:measure}. We should of course do this in a way which is
   invariant under coordinate redefinitions of both the moduli space and the
   worldsheet. Is there a ``natural'' way to do this?

   \sk To answer this question, let us first ask how we can describe the
   tangent vectors to the moduli space. In two dimensions, conformal
   transformations are equivalent to holomorphic transformations:
   \be
    z \mapsto f(z).
   \ee
   It thus seems natural to assume that the moduli space we have left labels
   different complex structures on $\gS$, and indeed this can be shown to
   be the case. Therefore, a tangent vector to the moduli space is  an
   infinitesimal change of complex structure, and these changes can be
   parameterized by holomorphic one-forms with anti-holomorphic vector indices:
   \be
    dz \mapsto dz + \eps \mu^z_{\zbar}(z) d \zbar.
   \ee
   The dimension counting above tells us that there are $3(g-1)$ independent
   $(\mu_i)^z_{\zbar}$, plus of course their $3(g-1)$ complex conjugates which
   change $d \zbar$. So the tangent space is spanned by these $\mu_i(z, \zbar),
   \mubar_i (z, \zbar)$. How do we get a number out of a set of these objects?
   Since $\mu_i$ has a $z$ and a
   $\zbar$ index, it seems natural to integrate it over $\gS$. However, the
   $z$-index is an upper index, so we need to lower it first with some tensor
   with two $z$-indices. It turns out that a good choice is to use the
   $Q$-partner $G_{zz}$ of the energy-momentum tensor component $T_{zz}$, and
   thus to define the integration over moduli space as
   \be
    \int_{M_g} \prod_{i=1}^{3g=3} \left( dm^i d \mbar^i \int_{\gS} G_{zz}
    (\mu_i)^z_{\zbar} \int_{\gS} G_{\zbar \zbar} (\mubar_i)^{\zbar}_z \right).
    \label{eq:modulimeasure}
   \ee
   Note that by construction, this integral is also invariant under a change of
   basis of the moduli space.
   There are several reasons why using $G_{zz}$ is a natural choice. First of
   all, this choice is analogous to what one does in bosonic string theory.
   There, one integrates over the moduli space using exactly the same formula,
   but with $G$ replaced by the conformal ghost $b$. This ghost is the
   BRST-partner of the energy-momentum tensor in exactly the same way as $G$ is
   the $Q$-partner of $T$. Secondly, one can make the not unrelated observation
   that since
   \be
    \{ Q, G \} = T,
   \ee
   we can still use the standard arguments to show independence of the theory
   of the parameters in a Lagrangean of the form $L = \{Q, V \}$. The only
   difference is that now we also have to commute $Q$ through $G$ to make it
   act on the vacuum, but since $T_{\ga \gb}$ itself is the derivative of the
   action with respect to the metric $h_{\ga \gb}$, the terms we obtain in this
   way amount to integrating a total derivative over the moduli space.
   Therefore, apart from possible 
   boundary terms these contributions vanish. We will work out this somewhat
   sketchy argument in more detail in section \ref{sec:holanom}, where we will
   also see that the boundary terms actually play an important role. For now,
   let us note that this reasoning also gives us an argument for using
   $G_{zz}$ instead of $T_{zz}$ (which is more or less the only other
   reasonable option) in (\ref{eq:modulimeasure}): if we had chosen $T_{zz}$
   then all path integrals would have been over total derivatives on the moduli
   space, and apart from boundary contributions the whole theory would have
   become trivial.

   \sk If we consider the vector and axial charges of the full path
   integral measure, including the new path integral over the worldsheet metric
   $h$, we find a surprising result. Since the worldsheet metric does not
   transform under $R$-symmetry, naively one might expect that its measure does
   not either. However, this is clearly not correct since one should also take into
   account the explicit $G$-insertions in (\ref{eq:modulimeasure}) that do
   transform under $R$-symmetry. From the $N=2$ superconformal algebra (or,
   more down-to-earth, from expressing the operators in terms of the fields),
   it follows that the product of $G$ and $\Gbar$ has vector charge zero and axial charge $2$.
   Therefore, the total vector charge of the measure remains zero, and the
   axial charge gets an extra contribution of $6(g-1)$. Added to the fermion
   zero-mode contributions that we found in the previous chapter -- see the
   discussion around equation (\ref{eq:numberofzm}) -- we thus find a total
   axial $R$-charge of
   \be
    6(g-1) - 2m(g-1).
   \ee
   From this, we see that the case of complex target space dimension 3 is very
   special: 
   here, the axial charge of the measure vanishes for any $g$, and hence the
   partition function is nonzero at every genus! If $m>3$ and $g>1$, the total
   axial charge of the measure is negative, and we have seen that we cannot
   cancel such a charge with local operators. Therefore, for these theories only
   the partition function at $g=1$ and a specific set of correlation functions
   at genus zero give nonzero results. Moreover, for $m=2$ and $m=1$, the
   results can be shown to be trivial by other arguments. Therefore, a
   Calabi-Yau threefold is by far the most interesting target space for a
   topological string theory. (This is thus the ``critical dimension'' we
   referred to earlier.) It is a very happy coincidence\footnote{Or, as one
   might very boldly conjecture, perhaps a clue as to why we live in four
   dimensions?} that this is exactly the dimension we are most interested in
   from the string theory perspective.

   \sk Finally, let us come back to the special cases of genus $0$ and $1$. At
   genus zero, the Riemann surface has a single point as its moduli space, so
   there are no extra integrals or $G$-insertions to worry about. Therefore, we can simply copy
   the topological field theory result saying that we have to introduce 
   local operators with total degree $(m,m)$ in the theory. The only remnant of
   the fact that we are integrating over metrics is that we should also somehow
   fix the remaining three symmetries of the sphere. The most straightforward
   way to do this is of course to consider three-point functions with
   insertions on three labelled points. As a gauge choice, we can then for
   example require these points to be at the points $0$, $1$ and $\infty$ in
   the compactified complex plane. For example, in the $A$-model on a
   Calabi-Yau threefold, the three-point function of three operators
   corresponding to $(1,1)$-forms would thus give a nonzero result.
   
   \sk In the case of the torus, we have seen that there is one ``unexpected''
   modular parameter over which we have to integrate. This means we have to
   insert one $G$- and one $\Gbar$-operator in the measure, which spoils the
   absence of the axial anomaly we had for $g=1$ in the topological field
   theory case. However, we also must fix the one remaining translational
   symmetry, which we can do by inserting a local operator at a labelled point.
   Thus, we can restore the axial $R$-charge to its zero value by choosing this
   to be an operator of degree $(1,1)$.
   
   \sk Summarizing, we have found that in topological string theory on a target
   Calabi-Yau three-fold, we have a nonvanishing three-point function of total
   degree (3,3) at genus zero; a nonvanishing one-point function of degree
   (1,1) at genus one, and a nonvanishing partition (``zero-point'') function
   at all genera $g>1$.

 \subsection{Nonlocal operators}
  In one respect, what we have achieved is great progress: we can now
  for any genus define a nonzero partition function (or for low genus a
  correlation function) of the topological string theory. On the other hand, we
  would also like to define correlation functions of an arbitrary
  number of operators at these genera. As we have seen, the insertion of extra
  local operators in the correlation functions is not possible, since any such
  insertion will spoil our carefully constructed absence of $R$-symmetry
  anomalies. Therefore, we have to introduce nonlocal operators\footnote{Here,
  we are ignoring the possibility of inserting operators that consist purely of
  $N=2$ superconformal partners of the worldsheet metric field. We will make
  some remarks on those at the end of this subsection.}.

  \sk There is one class of nonlocal operators which immediately comes to mind.
  In section \ref{sec:descenteqns} we saw, using the descent equations, that for
  every local operator we can define a corresponding one-form and a two-form
  operator. If we check
  the axial and vector charges of these operators, we find that if we start
  with an operator of degree $(1,1)$, the two-form operator we end up with
  actually has vanishing axial and vector charges! This has two important
  consequences. First of all, we can add the integral of this operator to our
  action,
  \be
   S[t] = S_0 + t^a \int \cO^{(2)}_a,
  \ee
  without spoiling the axial and vector symmetry of the theory. Secondly, we
  can insert the integrated operator into correlation functions,
  \be
   \langle \int \cO^{(2)}_1 \cdots \int \cO^{(2)}_n \rangle
  \ee
  and still get a nonzero result by the vanishing of the axial and vector
  charges. Of course, these two statements are related: one obtains such
  correlators by differentiating $S[t]$ with respect to the appropriate $t$'s,
  and then setting all $t^a=0$.

  \sk A few remarks are in place here. First of all, recall that the
  integration over the insertion points of the operators can be viewed as part
  of the integration over the moduli space of Riemann surfaces, where now we
  label a certain number of points on the Riemann surface. From this point of
  view, the 
  $g=0,1$ cases fit naturally into the same framework. We could unite
  the descendant fields into a worldsheet ``superfield'',
  \be
   \Phi_a = \cO^{(0)}_a + \cO^{(1)}_{a \ga} \gt^\ga + \cO^{(2)}_{a\ga \gb}
   \gt^\ga \gt^\gb
  \ee
  where we formally replaced each $dz$ and $d \zbar$ by corresponding fermionic
  coordinates $\gt^z$ and $\gt^{\zbar}$. (These objects $\gt^\ga$ should not be
  confused with the fermionic coordinates on $N=(2,2)$ superspace that we used
  before.) Now, one can write the above correlators as integrals over $n$
  copies of this ``superspace'':
  \be
   \int \prod_{s=1}^n d^2 z_s d^2 \gt_s \, \langle \Phi_{a_1}(z_1, \gt_1) \cdots
   \Phi_{a_n}(z_n, \gt_n)  \rangle
  \ee
  The integration prescription at genus $0$ and $1$ tells us to fix $3$
  and $1$ points respectively, so we need to remove this number of superspace
  integrals. Then, integrating over the other superspace coordinates, the genus
  $0$ correlators indeed become
  \be
   \langle \cO^{(0)}_{a_1} \cO^{(0)}_{a_2} \cO^{(0)}_{a_3} \int \cO^{(2)}_{a_4}
   \cdots \int \cO^{(2)}_{a_n} \rangle
  \ee
  From this prescription we note that these expressions are
  symmetric in the exchange of all $a_i$ and $a_j$. In particular, this means
  that the genus zero three-point functions at arbitrary $t$,
  \be
   c_{abc}[t] = \langle \cO^{(0)}_{a} \cO^{(0)}_{b} \cO^{(0)}_{c} \rangle[t]
  \ee
  have symmetric derivatives:
  \be
   \frac{\d c_{abc}}{\d t^d} = \frac{\d c_{abd}}{\d t^c},
  \ee
  and similarly with permuted indices. These equations can be viewed as
  integrability conditions, and using the Poincar\'e lemma we see that they
  imply that
  \be
   c_{ijk}[t] = \frac{\d Z_0[t]}{\d t^i \d t^j \d t^k}.
  \ee
  for some function $Z_0[t]$. Following the general philosophy that $n$-point
  functions are $n$th derivatives of the $t$-dependent partition function, we
  see that $Z_0[t]$ can be naturally thought of as the partition function at
  genus zero. Similarly, the partition function at genus 1 can be defined by
  integrating up the one-point functions once.
  
  \sk The quantities we have calculated above should of course be
  semi-topological invariants, meaning that they only depend on ``half'' of
  the moduli (either the K\"ahler ones or the complex structure ones) of the
  target space. For example, in the $A$-model we find the
  Gromov-Witten invariants that we already mentioned at the end of section
  \ref{sec:amodel}. In the $B$-model, it turns out that $F_0[t] = \ln Z_0[t]$
  is actually a quantity we already knew: it is the prepotential of the
  Calabi-Yau manifold! A discussion of why this is the case can be found in the
  paper \cite{Bershadsky:1993cx} by M.~Bershadsky, S.~Cecotti, H.~Ooguri and
  C.~Vafa. The higher genus partition functions can be thought of as ``quantum
  corrections'' to the prepotential.
 
  \sk Finally, there is a type of operator we have not discussed at all so far.
  Recall that in the topological string theory, the metric itself is now a
  dynamical field. Of course, we could not include the metric in our physical
  operators, since this would spoil the topological invariance. However, the
  metric is part of a $Q$-multiplet, and the highest field in this multiplet is
  a scalar field which is usually labelled $\varphi$. (It should not be
  confused with the fields $\phi^i$!) We can get more correlation functions by inserting
  operators $\varphi^k$ and the operators related to them by the descent
  equations into the correlation functions. These operators are called
  ``gravitational descendants''. Even the case where the power is $k=0$ is
  nontrivial; it does not insert any operator, but it does label a certain
  point, and hence changes the moduli space one integrates over. This operator
  is called the ``puncture operator''.

  \sk All of this seems to lead to an enormous amount of semi-topological
  target space invariants that can be calculated, but there are many recursion
  relations between the several correlators. This is similar to how we showed
  in section \ref{sec:cohomologicalft} that all 
  correlators for the cohomological field theories follow from the two-and
  three-point functions on the sphere. Here, it turns out that the set of all
  correlators has a structure which is related to the theory of integrable
  hierarchies. Unfortunately, a discussion of this is outside the scope of both
  these lectures and the author's current knowledge.

  \subsection{The holomorphic anomaly}
   \label{sec:holanom}
   We have now defined the partition function and correlation functions of
   topological string theory, but even though the expressions we obtained are
   much simpler than the path integrals for ordinary quantum field or string
   theories, it would still be very hard to explicitly calculate them.
   Fortunately, it turns out that the $t$-dependent partition and correlation
   functions are actually ``nearly holomorphic'' in $t$, and this is a great aid
   in exactly calculating these quantities.

   \sk Let us make this ``near holomorphy'' more precise. As we have seen,
   calculating correlation functions of primary operators in topological string
   theories amounts to taking $t$-derivatives of the corresponding perturbed
   partition function $Z[t]$ and consequently setting $t=0$. Recall that $Z[t]$
   is defined through adding terms to the action of the form
   \be
    t^a \int_\gS \cO^{(2)}_a,
    \label{eq:holterm}
   \ee
   Let us for definiteness consider the $B$-twisted model. We want to show that
   the above term is
   $\Qbar_B$-exact. For simplicity, we assume that $\cO^{(2)}_a$ is a bosonic
   operator, but what we are about to say can by inserting a few signs
   straightforwardly be generalized to the fermionic case. From the descent
   equations we studied in chapter \ref{sec:descenteqns}, we know that
   \be
    (\cO^{(2)}_a)_{+-} = - \{ G_+, [ G_-, \cO^{(0)}_a ] \},
    \label{eq:descentg}
   \ee
   where $G_+$ is the charge corresponding to the current $G_{zz}$, and $G_-$
   the one corresponding to $G_{\zbar \zbar}$. We can in fact express $G_\pm$
   in terms of the $N=(2,2)$ supercharges $\cQ$. 
   Recall from equation (\ref{eq:Qcomm}) that\footnote{The reader should not be
   confused by the fact that 
   here $L_0 - \Lbar_0$ seems to generate a translation whereas previously we
   identified it with the rotation generator. In chapter \ref{sec:twisting} we
   worked ``in cylindrical coordinates'' (even though for most of that chapter we
   did not compactify the $x^1$-direction), and a translation around the
   cylinder is a rotation if we conformally map the cylinder to the plane.
   Similarly, in chapter \ref{sec:twisting} the Hamiltonian $L_0 + \Lbar_0$
   generated an ordinary shift, whereas in the ``planar coordinates'' it
   generates a radial scaling. The cylindrical and planar coordinates are
   related by the conformal mapping $z_{plane} = \exp(i z_{cyl})$}
   \bea
    H & = & 2 \pi i (L_0 + \Lbar_0) \hspace{0.5em} = \hspace{0.5em} \half
    \{ \cQ_+, \cQbar_+ \} - \half \{ \cQ_-, \cQbar_- \} 
    \ret
    P & = & 2 \pi i (L_0 - \Lbar_0) \hspace{0.5em} = \hspace{0.5em} \half
    \{ \cQ_+, \cQbar_+ \} + \half \{ \cQ_-, \cQbar_- \}.
   \eea
   Thus, we find that the left- and right-moving energy-momentum charges
   satisfy
   \bea
    T_{+} & = &  2 \pi i L_0 \hspace{0.5em} = \hspace{0.5em} \phantom{-} \half
    \{ \cQ_+, \cQbar_+ \} \ret
    T_{-} & = &  2 \pi i \Lbar_0 \hspace{0.5em} = \hspace{0.5em} - \half
    \{ \cQ_-, \cQbar_- \}.
   \eea
   To find $G$ in the $B$-model, we should write these charges as
   commutators with respect to $Q_B = \cQbar_+ + \cQbar_-$, which using
   (\ref{eq:Qcomm}) is straightforward:
   \bea
    T_{+} & = & \phantom{-} \half \{ Q_B, \cQ_+ \} \ret
    T_{-} & = & - \half \{ Q_B, \cQ_- \},
   \eea
   so we arrive at the conclusion that for the $B$-model,
   \bea
    G_+ & = & \phantom{-}  \half \cQ_+ \ret
    G_- & = & - \half \cQ_-.
   \eea
   Now, we can rewrite (\ref{eq:descentg}) as
   \bea
    (\cO^{(2)}_a)_{+-} & = & - \{ G_+, [ G_-, \cO^{(0)}_a ] \} \ret
    & = & \frac{1}{4} \{ \cQ_+, [ \cQ_-, \cO^{(0)}_a ] \} \ret
    & = & \frac{1}{8} \{ \Qbar_B, [ (\cQ_- - \cQ_+), \cO^{(0)}_a ] \},
    \label{eq:Qbarexact}
   \eea
   which proves our claim that $\cO_a^{(2)}$ is $\Qbar_B$-exact. 
   
   \sk An $N=(2,2)$ sigma model with a real action does, apart from the term
   (\ref{eq:holterm}), also contain a term
   \be
    t^{\abar} \int_\gS \cObar^{(2)}_a,
    \label{eq:antiholterm}
   \ee
   where  $t^{\abar}$ is the complex conjugate of $t^a$. It is not immediately
   clear that $\cObar^{(2)}_a$ is a physical operator: we have seen that
   physical operators in the $B$-model correspond to forms that are
   $\delbar$-closed, but the complex conjugate of such a form is $\d$-closed.
   However, by taking the complex conjugate of (\ref{eq:Qbarexact}), we see
   that
   \be
    (\cObar^{(2)}_a)_{+-} = \frac{1}{8} \{ Q_B, [ (\cQbar_- - \cQbar_+),
    \cObar^{(0)}_a ] \}, 
   \ee
   so not only is the operator $Q_B$-closed, it is even $Q_B$-exact! This
   means that we can add terms of the form (\ref{eq:antiholterm}) to the
   action, and taking $t^{\abar}$-derivatives inserts $Q_B$-exact terms in the
   correlation functions. Naively, we would expect this to give a zero result,
   so all the physical quantities seem to be $\tbar$-independent, and thus
   holomorphic in $t$. We will see in a moment that this naive expectation
   turns out to be almost right, but not quite.
   
   \sk Before doing so, let us however comment briefly on the generalization of
   the above argument in the case of the $A$-model. It seems that a
   straightforward generalization of the argument fails, since $Q_A$ is its own
   complex conjugate, and the complex conjugate of the de Rham operator is also
   the same operator. However, note that the $N=(2,2)$-theory has a different
   kind of ``conjugation symmetry'': we can exchange the two supersymmetries,
   or in other words, exchange $\gt^+$ with $\gtbar^+$ and $\gt^-$ with
   $\gtbar^-$. This exchanges $Q_A$ with an operator which we might denote as
   $Q_{\Abar} \equiv \cQ_+ + \cQbar_-$. Using the above argument, we then find that
   the physical operators $\cO^{(2)}_a$ are $Q_{\Abar}$-exact, and that their
   conjugates in the new sense are $Q_A$-exact. We can now add these conjugates
   to the action with parameters $t^{\abar}$, and we again naively find
   independence of these parameters. In this case it is less natural to choose
   $t^a$ and $t^{\abar}$ to be complex conjugates\footnote{In fact, by mirror
   symmetry this choice is naturally related to the same choice in the
   $B$-model, as was pointed out by E.~Witten in \cite{Witten:1993ed}.}, but we
   are of course free to choose this particular ``background point'' and study
   how the theory behaves if we then vary $t^a$ and $t^{\abar}$ independently.
   
  \sk Now, let us see how the naive argument showing independence of the theory
  of $t^{\abar}$ fails. In fact, the argument above would certainly hold for
  topological {\em field} theories. However, in topological string theories, we
  have to worry about the insertions in the path integral of
  \be
   G \cdot \mu_i \equiv \int d^2 z \, G_{zz} \, (\mu_i)^z_{\zbar},
  \ee
  and their complex conjugates, when commuting the $Q_B$ towards
  the vacuum and making sure it gives a zero answer. Indeed, the
  $Q_B$-commutator of the above factor is not zero, but it gives
  \be
   \{ Q_B, G \cdot \mu_i \} = T \cdot \mu_i.
  \ee
  Now recall that
  \be
   T_{\ga \gb} = \frac{\d S}{\d h^{\ga \gb}}.
  \ee
  We did not give a very precise definition of $\mu_i$ above, but we know that
  it parameterizes the change in the metric under an infinitesimal change of
  the coordinates $m_i$ on the moduli space. One can make this intuition
  precise, and then finds the following ``chain rule'':
  \be
   T \cdot \mu_i = \frac{\d S}{\d m^i}.
  \ee
  Inserting this in the partition function, we find that
  \be
   \frac{\d F_g}{\d t^{\abar}} = \int_{M_g} \prod_{i=1}^{3g-3} dm^i d
   \mbar^i \, \sum_{j, k} \frac{\d^2}{\d m^j \d \mbar^k} \left\langle \left(
   \prod_{l \neq j} \int \mu_l \cdot G \right) \left( \prod_{l \neq k} \int
   \mubar_l \cdot \Gbar \right) \int \cObar_a^{(2)} \right\rangle,
  \ee
  where $F_g = \ln Z_g$ is the free energy at genus $g$, and the reason $F_g$
  appears in the above equation instead of $Z_g$ is, as usual in quantum field
  theory, that the expectation values in the right hand side are normalized
  such that $\langle 1 \rangle = 1$, and so the left hand side should be
  normalized accordingly and equal $Z_g^{-1} \d_{\abar} Z_g = \d_{\abar} F_g$.
  
\begin{figure}[ht]
   \begin{center}
    \includegraphics[height=4cm]{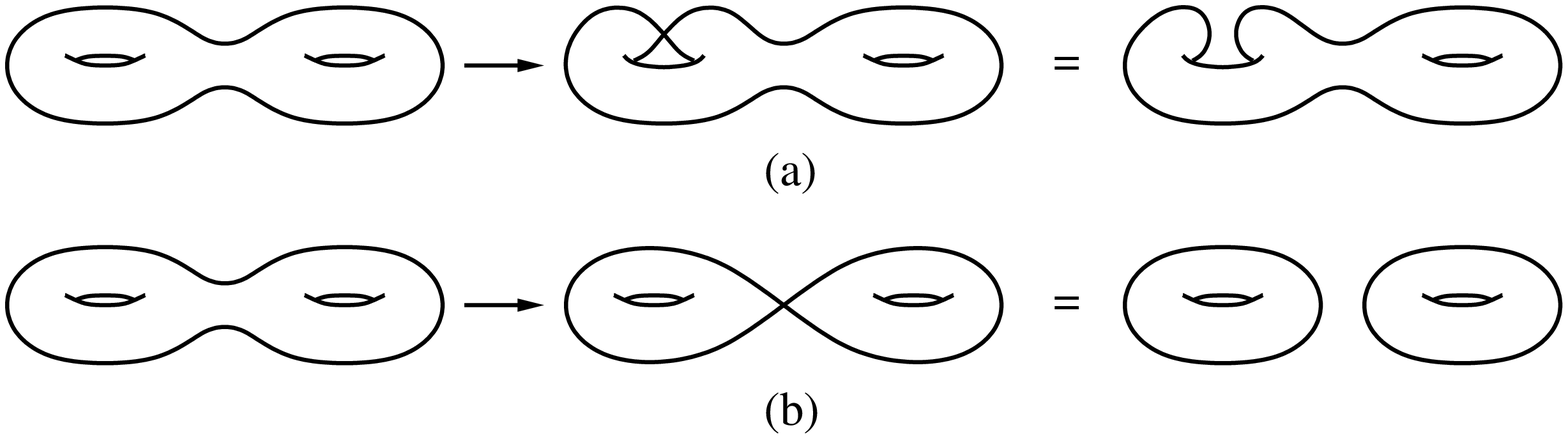}
   \end{center}
   \caption{At the boundary of the moduli space of genus $g$ surfaces, the
   surfaces degenerate because certain cycles are pinched. This either lowers
   the genus of the surface (a) or breaks the surface into two lower genus
   ones (b).}
   \label{fig:boundary}
  \end{figure}

  \sk Thus, as we have claimed before, we are integrating a total derivative over
  the moduli space of genus $g$ surfaces. If the moduli space did not have a
  boundary, this would indeed give zero, but in fact the moduli space {\em
  does} have a boundary. It consists of the moduli which make the genus $g$
  surface degenerate. This can happen in two ways: an internal cycle of the
  genus $g$ surface can be pinched, leaving a single surface of genus $g-1$, as
  in figure \ref{fig:boundary}a, or the surface can split up into two surfaces
  of genus $g_1$ and $g_2 = g - g_1$, as depicted in figure
  \ref{fig:boundary}b. By carefully considering the boundary contributions to
  the integral for these two types of boundaries, it was shown in
  \cite{Bershadsky:1993cx} by M.~Bershadsky, S.~Cecotti, H.~Ooguri and C.~Vafa
  that
  \be
   \frac{\d F_g}{\d t^{\abar}} = \half c_{\abar \bbar \cbar} e^{2K} G^{\bbar d}
   G^{\cbar e} \left( D_d D_e F_{g-1} + \sum_{r=1}^{g-1} D_d F_r D_e F_{g-r}
   \right),
  \ee
  where $G$ is a certain K\"ahler metric (the so-called ``Zamolodchikov
  metric'') on the space parameterized by the coupling constants $t^a,
  t^{\abar}$; $K$ is its K\"ahler potential, and the $D_a$ are covariant
  derivatives on this space\footnote{An important point
  which we have not stressed yet is the following. The coupling constants (or at
  least the holomorphic ones) parameterize the deformations of the topological
  theory. As we have seen in the previous chapter, the $A$-model only depends
  on the K\"ahler moduli of the target space, and the $B$-model only on its
  complex structure moduli. Therefore, deforming the $A$-model, for example,
  results in a change of the K\"ahler moduli. One can make this more precise,
  and see that with a good choice of coordinates, the space parameterized by the
  $t^a$ actually {\em is} the K\"ahler moduli space of the Calabi-Yau in the
  A-model case, and the complex structure moduli space in the $B$-model case.
  An extensive discussion of these facts can be found in the paper
  \cite{Bershadsky:1993cx} by Bershadsky et al.}. The coefficients $c_{\abar
  \bbar \cbar}$ are the three-point functions on the sphere of the operators
  $\cObar_a^{(0)}$. We will not derive the above formula in detail, but the
  reader should notice that the contributions from the two types of
  boundary are quite clear.

  \sk Using this formula, one can inductively determine the $t^{\abar}$
  dependence of the partition functions if the holomorphic $t^a$-dependence is
  known. Holomorphic functions on complex spaces (or more
  generally holomorphic sections of complex vector bundles) are quite rare:
  usually, there is only a finite-dimensional space of such functions. The same
  turns out to hold for our topological string partition functions: even though
  they are not quite holomorphic, their anti-holomorphic behavior is determined
  by the holomorphic dependence on the coordinates, and as a result there is 
  a finite number of coefficients which determines them.

  \sk Thus, just from the above structure and without doing any path integrals,
  one can already determine the topological string partition functions {\em up
  to a finite number of constants}! This leads to a feasible program for
  completely determining the topological string partition function for a given
  target space and at given genus. From the holomorphic anomaly equation, one
  first has to find the general form of the partition function. Then, all one
  has left to do is to fix the unknown constants. Here, the fact
  that in the $A$-model the partition function simply counts a number of
  points comes to our help: by requiring that the $A$-model partition functions
  are integral, one can often fix the unknown constants and completely
  determine the $t$-dependent partition function\footnote{In fact, 
  here mirror symmetry plays an important role as well: the holomorphy can most
  easily be used as a tool in the $B$-model, where the moduli space is simply
  $\cM$ itself. Then, one does a mirror map to apply the integrality argument in
  the corresponding $A$-model.}. In practice, the procedure is still quite
  elaborate, so we will not describe any examples here, but several have been
  worked out in detail in the literature. Once again, the pioneering work for
  this can be found in the paper \cite{Bershadsky:1993cx} by Bershadsky et al.

 \section{Applications}
  \label{sec:applications}
  Now that we have a basic understanding of how to construct (closed)
  topological string theories and their physical correlation functions, the
  next question is of course: what can we do with them? In the past ten years
  or so, it has turned out that topological strings have an enormous amount of
  applications. Their structure is complicated enough to relate them to
  physically interesting theories, yet simple enough to be able to
  obtain exact results.
  
  \sk As mentioned in the introduction, the present chapter is definitely not
  meant as a complete overview of the applications of topological strings that
  have been found over the past decade. The goal of these notes has been to
  reach a level where the reader can delve into the recent literature himself,
  with enough technical background to have a good understanding of what is
  going on. However, it would be too much of a cliffhanger if we would not at
  least give an impression of the type of results that one is able to obtain
  using topological strings, so we will try to give some glimpses of this in
  this chapter. Recently, a set of lecture notes by A.~Neitzke and C.~Vafa
  appeared \cite{Neitzke:2004ni}, giving a more detailed overview of the
  physical applications of topological strings.
  
  \subsection{$N=2$ F-terms}
   \label{sec:fterms}
   Ordinary (i.e.\ non-topological) string theories come in many flavors, with
   names such as ``type 0'', ``type I'', ``type IIA/B'', and ``heterotic''.
   Most of these theories are ten-dimensional, and can be viewed as special
   limits, compactifications or configurations of a universal
   eleven-dimensional theory called ``M-theory''. The best place to look for
   physical applications of topological string theories is in the two
   type II string theories on space-times of the form 
   \be
    M_4 \times CY_6,
   \ee
   where $M_4$ is four-dimensional Minkowski space and $CY_6$ is a Calabi-Yau
   threefold. The reason is that these theories have two ten-dimensional
   supersymmetry generators, which give us good control over their structure,
   and that their worldsheet description consists of two conformal field
   theories, which we may schematically write as
   \be
    CFT_{10} = CFT_4 \otimes CFT_6.
   \ee
   Here the first factor contains the four ``space-time'' coordinates of
   Minkowski space, and the second one is the sigma model mapping the string
   worldsheet into the Calabi-Yau. The worldsheet theory of the string is
   in general a locally $N=1$ supersymmetric theory, but as we have seen,
   because of the special properties of the Calabi-Yau, the $CFT_6$ in addition
   has an $N=(2,2)$ global worldsheet supersymmetry which we can twist.
   
   \sk Since the twisting is a relatively mild change of the theory, one might
   expect that some information of the $CFT_6$ is contained in the resulting
   topological string theory. In fact, the main restriction in going to the
   topological theory lies in the restriction of the set of physical states and
   operators,
   \be
    \{ \cO_{phys} \}_{top} \subset \{ \cO_{phys} \}_{CFT}.
   \ee
   Therefore, one would expect that quantities in the CFT which can be
   calculated using only the operators which are ``physical'' on the topological
   side have a good chance of being equal to their topological counterparts. As
   we will see, this heuristic argument turns out to be correct, and we can
   indeed use the topological string to calculate ordinary string quantities.

   \subsubsection{The low-energy theory}
    When a type II string theory is compactified on a Calabi-Yau manifold
    which is small compared to the distance scales (or inverse energy scales)
    that we are interested in, but large compared to the string length, it can
    be described very well by an ordinary four-dimensional field theory on
    $M_4$. It turns out that this theory is an $N=2$ supergravity theory, where
    the $N=2$ arises from the fact that the Calabi-Yau manifold has one
    covariantly constant spinor field, which can be used to construct a
    four-dimensional supersymmetry out of each ten-dimensional supersymmetry.
    Even though in the low energy limit the Calabi-Yau itself is no longer
    visible, many of its properties leave an imprint on the low-energy theory.

    \sk Most importantly, the choice of Calabi-Yau determines the field content
    of the low-energy theory. For example, the Calabi-Yau has certain moduli,
    which one can vary as a function of the position in $M_4$. As we
    have seen in section \ref{sec:modulispaces}, the number of moduli is as
    follows:
    \bea
     &&h^{2,1} \mbox{~complex structure moduli~} X^I(x^\mu) \ret
     &&h^{1,1} \mbox{~K\"ahler moduli~} Y^\ga(x^\mu),
    \eea
    where in the notation we indicated that these moduli now become {\em fields}
    on the four-dimensional space. Since the four-dimensional theory is a
    supersymmetric theory, these fields should be part of supermultiplets
    appearing as the components of superfields, similarly to the $N=(2,2)$ case
    in two dimensions. Several types of multiplets are known in the
    four-dimensional $N=2$ case, the most familiar ones being the so-called
    ``vector multiplet'' and the
    ``hypermultiplet''. By studying the compactification procedure for type IIB
    string theory, one finds that the moduli appear in the following types of
    multiplets:
    \bea
     X^I(x^\mu) & \rightarrow & \mbox{vector multiplets} \ret
     Y^\ga(x^\mu) & \rightarrow & \mbox{hypermultiplets}
    \eea
    For example, in type IIB string theory there is a ten-dimensional
    space-time four-form field with self-dual five-form field strength. After
    compactification, this 
    ten-dimensional four-form leads to a set of four-dimensional one-forms by
    writing its components with three ``internal'' indices as
    \be
     A_{aijk} = \sum_n A^n_a \go^n_{ijk}
    \ee
    Here, $\go^n$ form a basis of three-forms on the Calabi-Yau, and one can
    show (using the so-called Kaluza-Klein reduction) that the {\em harmonic}
    three-forms $\go^n$ give rise to {\em massless} one-forms (i.e.\ gauge
    fields) $A^m_a$ in four dimensions. We have seen that on a Calabi-Yau, there
    are $2 h^{2,1} + 2$ of those harmonic three-forms. 
    Therefore, we seem to obtain $2h^{2,1} + 2$ four-dimensional gauge fields,
    but in fact these are pairwise related because of the self-duality
    conditions on the field strength of $A_{aijk}$. So we end up with
    \be
     h^{2,1} + 1
    \ee
    gauge fields in four dimensions. One can show that $h^{2,1}$ of these are
    part of the vector multiplets together with the $X^I$ -- of course, the
    presence of a vector field in this multiplet is the reason for its
    name\footnote{Mathematically speaking, ``one-form field'' or ``covector
    field'' would perhaps be a better name for the gauge fields.}.
    The remaining ``universal'' vector field, which appears for any Calabi-Yau,
    is the so-called {\em graviphoton} field, which enters the supersymmetry
    multiplet of the metric. Continuing in this way, one can check that
    indeed all the low-energy fields fill out $N=2$ supergravity multiplets in
    the way we indicated.

    \sk The structure of supersymmetric field theory Lagrangeans is usually
    rather constrained. For example, as we saw in chapter \ref{sec:twisting},
    in two dimensions the $N=(2,2)$ Lagrangean could be built up from
    \begin{itemize}
     \item A K\"ahler potential $K(\Phi, \Phibar)$,
     \item A {\em holomorphic} superpotential $W(\Phi)$.
    \end{itemize}
    Similarly, in four-dimensional $N=2$ supergravity, if one restricts to a
    maximum of two derivatives, one has the following building blocks:
    \begin{itemize}
     \item A K\"ahler potential $K(X^I, \Xbar^I)$ for the vector multiplets,
     \item A K\"ahler\footnote{Actually, the manifold parameterized by the
     hypermultiplets turns out to have even more structure, turning it into
     what is called a ``hyperk\"ahler manifold''.} potential $\chi(Y^\ga,
     \Ybar^\ga)$ for the hypermultiplets,
     \item A {\em holomorphic} superpotential $F_0(X^I)$ for the
     vector multiplets,
     \item A {\em holomorphic} superpotential $W(Y^\ga)$ for the
     hypermultiplets.
    \end{itemize}
    As in the two-dimensional case, the K\"ahler potentials are called
    ``D-terms'', and the holomorphic quantities are called ``F-terms''. Of
    course, the corresponding anti-holomorphic quantities also appear. Let us
    focus on the superpotential $F_0$. Using the requirement of 
    supersymmetry, one can show that it has the following two properties:
    \begin{itemize}
     \item It must be homogeneous of degree $2$ in $X^I$ to ensure the
     conservation of $U(1)$ R-symmetry,
     \item The action written out in components has an $Sp(2n, \bR)$ symmetry
     (where $n = h^{2,1} + 1$) acting on $X^I$ and $F_I = \d_I F_0$. After
     quantization of the theory, this symmetry is broken to an $Sp(2n, \bZ)$
     symmetry. 
    \end{itemize}
    Both of these properties are very familiar: they are exactly the
    properties satisfied by the Calabi-Yau prepotential $F_0$ defined by
    \bea
     X^I & = & \int_{A^I} \gO \ret
     F_I & = & \int_{B_I} \gO
    \eea
    This is not a coincidence: one can show that, indeed, $F_0$ is nothing but
    the prepotential of the Calabi-Yau manifold! This relation between the
    geometry of the complex structure moduli space of the Calabi-Yau manifold
    and the supergravity Lagrangean was worked out in detail in the paper
    \cite{Strominger:1990pd} by A.~Strominger. 

    \sk Recall that the Calabi-Yau prepotential $F_0$ can also be calculated as
    the $B$-model genus 0 topological string free energy. Thus, we have found
    our first example of a quantity that appears in the ``physical'' theory
    which can be calculated using the topological theory!

   \subsubsection{Higher genus}
    Of course, in practice the above fact is not very useful: the
    prepotential can simply be calculated from geometrical data without the need
    to know about the topological string theory. However, encouraged by this
    result, one may ask whether the higher genus $B$-model partition
    functions also have some interpretation in terms of the $N=2$ low-energy
    supergravity theory. It was shown in \cite{Antoniadis:1993ze} by
    I.~Antoniadis, E.~Gava, K.~Narain and T.~Taylor that indeed they have.

    \sk In string theory, scattering amplitudes between space-time particles
    can be calculated from correlation functions of related operators -- the
    so-called ``vertex operators'' -- on the string worldsheet. In their
    paper, Antoniadis et al.\ calculated the following correlation 
    function in the compactified type IIB string theory:
    \be
     A_g = \langle T^{2g-2} R^2 \rangle_g.
    \ee 
    Here, $T$ is the vertex operator for the field strength of the graviphoton
    that we mentioned before, and $R$ is the vertex operator for the ``graviton
    field strength'', that is, the Riemann curvature. As we have seen, the
    vector fields in the supergravity theory are related to the complex
    structure moduli of the Calabi-Yau, and thus we may in particular expect
    the graviphoton field strength vertex operator to be a physical operator in
    the $B$-model. The same thing holds for the operator $R$, which is one of
    its superpartners. Hence, we may expect the above correlation function to
    be calculable using the topological string theory. Indeed, using a quite
    technical calculation, Antoniadis et al.\ show that it is given by
    \be
     A_g = (g!)^2 F_g.
    \ee
    Since this amplitude is nonzero it means that there is an interaction
    between $2g-2$ graviphotons and $2$ gravitons in the supergravity theory.
    One can show that the only possible terms in the $N=2$ supergravity action
    resulting in such interactions are
    \be
     S_{eff} \sim \int d^4 x \, g F_g(X) \left[ R^2 T^{2(g-1)} + 2 (g-1) (RT)^2
     T^{2(g-2)} \right],
    \ee
    where we denoted the four-dimensional fields by the same symbols as their
    two-dimensional vertex operators, and suppressed the contraction of their
    indices.
    
    \sk A much nicer expression is obtained when these terms are written in
    terms of the four-dimensional superfield $W$ containing both $T$ and $R$:
    they combine into 
    \be
     S_{eff} \sim \int d^4 x \, d^4 \gt \, W^{2g} F_g(X)
     \label{eq:fterms}
    \ee
    In other words, we can use the topological string theory to calculate these
    $F$-terms in the physical theory for all genera! Note in particular that the
    $g=0$ term is exactly the superpotential term we obtained before.

  \subsection{The NS 5-brane partition function}
   It turns out that the above story has the following interesting application,
   which was described by R.~Dijkgraaf, E.~Verlinde and the present author in
   \cite{Dijkgraaf:2002ac}. Besides strings, string theories contain several other extended
   objects of diverse dimensions. One of these is the so-called Neveu-Schwarz
   (or NS) five-brane; an object with a (5+1)-dimensional worldvolume.
   Let us consider type IIA string theory compactified on a certain Calabi-Yau
   manifold, and let us wrap an NS 5-brane on this Calabi-Yau. By this we mean
   that all of the six worldvolume dimensions wrap the Calabi-Yau, so we
   really have a Euclidean version of the NS 5-brane, and from the
   four-dimensional space-time point of view this object is an instanton.

   \sk Suppose now that we have $n$ of these NS five-brane instantons close to
   each other. Just like for strings, we can try to describe the dynamics of
   these objects by a field theory living on their worldvolume. The worldvolume
   theory on such a set of five-branes turns out to be a rather mysterious
   theory called ``little string theory''. (The ``string'' here is the boundary
   of a (2+1)-dimensional membrane which can be stretched between the different
   NS five-branes.) In a certain scaling limit this little string
   theory turns into a more ordinary gauge theory, where the only unusual fact
   is that the ``gauge field'' is a two-form field $B$ with a self-dual field
   strength.

   \sk One can now ask what the partition function of this two-form field
   theory is. One reason to be interested in this question is that this
   partition function encodes corrections, coming from these instantons, to the
   four-dimensional low-energy supergravity theory. It turns out that the
   semiclassical contribution to the partition function is quite
   straightforward to calculate. However, as usual, it is much harder to find
   quantum corrections to this result.

   \sk Fortunately, there is a trick which helps a great deal. It is a
   well-known fact from string theory that compactifying a type IIB string on a
   circle of radius $R$ leads to exactly the same theory as compactifying a
   type IIA string theory on a circle of radius $l_s^2 / R$, where $l_s$ is the
   string length. This relation is called ``T-duality'', and one can apply such
   a T-duality in one of the space-time directions perpendicular to the
   Calabi-Yau manifold. It turns out that the five-branes in the IIA theory
   have a completely different guise in the T-dual IIB theory. There, we don't
   see any five-branes, but instead there is a nontrivial four-dimensional
   metric background. That is, one is left with an ordinary compactified IIB
   string theory on this background! One may now ask how the contributions of
   the five-branes appear in the low-energy theory on this side of the duality,
   and it turns out that they are exactly encoded in the terms
   (\ref{eq:fterms}). In other words, {\em the quantum corrections 
   to the NS five-brane partition function can be calculated using the B-model
   topological string}!

   \sk An interesting remark to make is the following. As we saw in section
   \ref{sec:holanom}, the topological string partition function satisfies a
   holomorphic anomaly equation. Surprisingly, one can also show that the {\em
   semiclassical} part of the partition
   function satisfies such an equation! To be precise, this part of the
   partition function satisfies the conjugate equation. The precise meaning of
   this observation is an intriguing open issue.

  \subsection{Geometric transitions}
   One of the most interesting results in string theory of the past few years
   is the discovery of the so-called ``AdS/CFT-correspondence''. In this
   correspondence, ten-dimensional closed string theories are shown to be
   equivalent to theories of open strings which have their
   end points on lower-dimensional objects called D-branes. The first
   example of this behavior was discovered by J.~Maldacena in
   \cite{Maldacena:1997re}. It relates type IIB string theory on $AdS_5 \times
   S^5$ to a theory of open strings ending on a set of (3+1)-dimensional
   D-branes. Here, $AdS_5$ stands for ``five-dimensional anti-de Sitter
   space-time'', which is a maximally symmetric (4+1)-dimensional space-time
   with a negative cosmological constant. The worldvolume theory, describing
   the low-energy behavior of the open strings ending on the D3-branes, is a
   conformal field theory, which explains the name of the correspondence.
   
   \sk A geometric transition is the topological equivalent of an
   AdS/CFT-correspon\-dence. In this case, a closed topological string
   theory is related to a theory of open topological strings, which can again
   be described by a related worldvolume theory. The simplest example, 
   which is the only one we will discuss here, is the geometric transition
   between Chern-Simons theory on $S^3$ and the closed $A$-model topological
   string on the resolved conifold. It was discovered by R.~Gopakumar and
   C.~Vafa in \cite{Gopakumar:1998ki}.

   \subsubsection{Conifolds}
\begin{figure}[ht]
     \begin{center}
      \includegraphics[height=4cm]{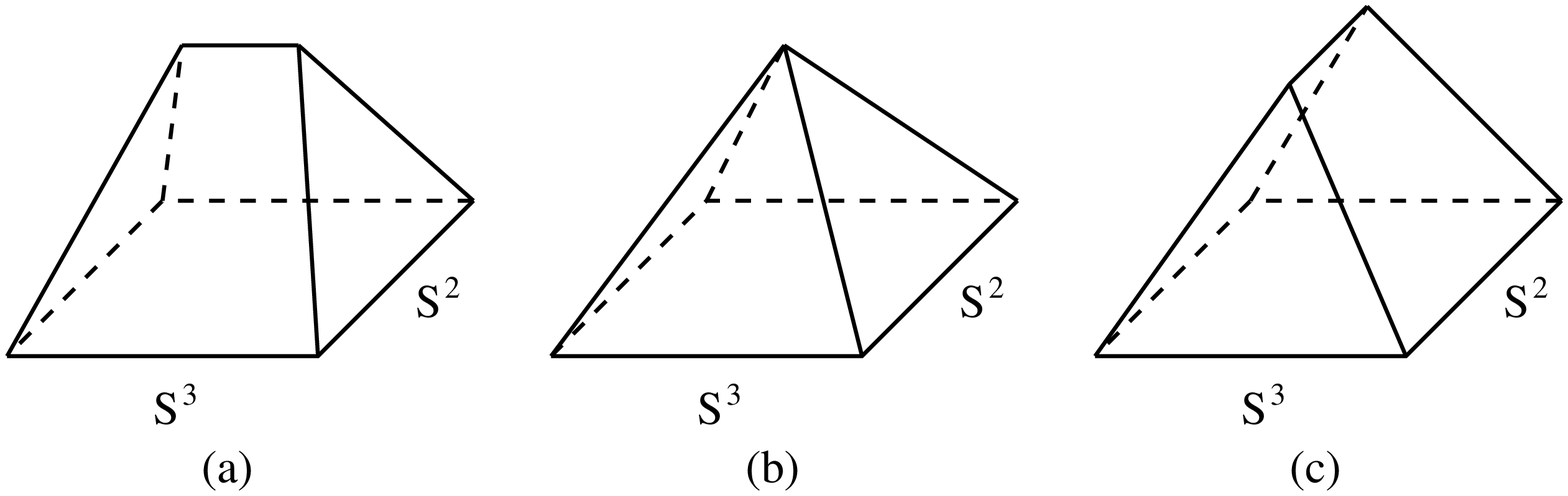}
     \end{center}
     \caption{(a) The deformed conifold, (b) the ordinary conifold, (c) the
     resolved conifold. In each case, the square at the base represents the
     slice $S^3 \times S^2$. In (b), the tip of the cone is a singularity; in
     (a) it is replaced by an $S^3$, and in (c) by an $S^2$.}
     \label{fig:conifold}
    \end{figure}

    The conifold is the simplest example of a non-compact Calabi-Yau
    threefold: it is the set of solutions to the equation
    \be
     x_1 x_2 - x_3 x_4 = 0
    \ee
    in $\bC^4$. The resulting manifold is a cone, meaning in this case that any
    real multiple of a solution to this equation is again a solution.
    The point $(0,0,0,0)$ is the ``tip'' of this cone, and it is a singular
    point of the solution space. Note that by writing
    \bea
     x_1 & = & \phantom{-} z_1 + i z_2 \ret
     x_2 & = & \phantom{-} z_1 - i z_2 \ret
     x_3 & = & \phantom{-} z_3 + i z_4 \ret
     x_4 & = & - z_3 + i z_4,
    \eea
    where the $z_i$ are still {\em complex} numbers, one can also write the
    equation as
    \bea
     z_1^2 + z_2^2 + z_3^2 + z_4^2 = 0.
    \eea
    Writing each $z_i$ as $a_i + i b_i$, with $a_i$ and $b_i$ real, we
    obtain the two equations
    \bea
     |a|^2 - |b|^2 & = & 0 \ret
     a \cdot b & = & 0.
     \label{eq:conifold}
    \eea
    Here $a \cdot b = \sum_i a_i b_i$ and $|a|^2 = a \cdot a$. Since the
    geometry is a cone, let us focus on a ``slice'' of this cone given by
    \be
     |a|^2 + |b|^2 = 2 r^2,
    \ee
    for some $r \in \bR$. On this slice, the first equation in
    (\ref{eq:conifold}) becomes
    \be
     |a|^2 = r^2,
     \label{eq:sphere}
    \ee
    which is the equation defining a three-sphere of radius $r$. The same holds
    for $b$, so both $a$ and $b$ lie on three-spheres. However, we also have to
    take the second equation in (\ref{eq:conifold}) into account. Let us
    suppose that we fix an $a$ satisfying (\ref{eq:sphere}). Then $b$ has to
    lie on a three-sphere, but also on the plane
    through the origin defined by $a \cdot b = 0$. That is, $b$ lies on a
    two-sphere. This holds for every $a$, so the slice we are considering is a
    fibration of two-spheres over the three-sphere. With a little more work,
    one can show that this fibration is trivial, so the conifold is a cone over
    $S^2 \times S^3$. This is depicted in figure \ref{fig:conifold}b.

    \sk Since the conifold is a singular geometry, we would like to find
    geometries which approximate it, but which are non-singular. There are two
    interesting ways in which this can be done. The simplest way is to replace
    the defining equation by
    \be
     x_1 x_2 - x_3 x_4 = \mu^2.
     \label{eq:defconifold}
    \ee
    From the two equations constraining $a$ and $b$, we now see that $|a|^2 \geq
    \mu^2$. In other words, the parameter $r$ should be at least $\mu$. At
    $r=\mu$, the $a$-sphere still has finite radius $\mu$, but the $b$-sphere
    shrinks to zero size. In other words, we have deformed the geometry to the one
    depicted in figure \ref{fig:conifold}a. This geometry is called the
    ``deformed conifold''. Even though this is not clear from the picture, from
    the equation (\ref{eq:defconifold}) one can straightforwardly show that it
    is nonsingular. One can also show that it is topologically equivalent to
    the cotangent bundle on the three-sphere,
    \be
     T^* S^3.
    \ee
    Here, the $S^3$ on which the cotangent bundle is defined is exactly the
    $S^3$ at the ``tip'' of the deformed conifold.

    \sk The second way to change the conifold geometry arises from studying
    the two equations
    \bea
     x_1 A + x_3 B & = & 0 \ret
     x_4 A + x_2 B & = & 0.
     \label{eq:rescon}
    \eea
    Here, we require $A$ and $B$ to be homogeneous complex coordinates on a
    $\bC P^1$, i.\ e.\ 
    \bea
     (A, B) & \neq & (0,0) \ret
     (A, B) & \sim & (\gl A, \gl B)
    \eea
    where $\gl$ is any nonzero complex number. If one of the $x_i$ is nonzero,
    say $x_1$, one can solve for $A$ or $B$, e.\ g.\
    \be
     A = - \frac{x_3 B}{x_1}
    \ee
    and insert this in the other equation to obtain
    \be
     x_1 x_2 - x_3 x_4 = 0
    \ee
    which is the conifold equation. However, if all $x_i$ are zero, any $A$ and
    $B$ solve the system of equations (\ref{eq:rescon}). In other
    words, we have constructed a geometry which away from the former
    singularity is completely the same as the conifold, but the singularity
    itself is replaced by a $\bC P^1$, which topologically is the same as an
    $S^2$. From 
    the defining equations one can again easily show that the resulting
    geometry is nonsingular, so we have now replaced our conifold geometry by
    the so-called ``resolved conifold'' which is depicted in figure
    \ref{fig:conifold}c.

   \subsubsection{Topological D-branes}
    Since topological string theories are in many ways similar to an ordinary
    string theories, one natural question which arises is: are there also open
    topological strings which can end on D-branes? So far in these notes, we
    have restricted our attention to closed topological strings, whose
    worldsheets are closed Riemann surfaces. To answer the above question
    rigorously, we would have to study boundary conditions on worldsheets with
    boundaries which preserve the $Q$-symmetry. This subject would take us at
    least an extra lecture to explore, so we will limit ourselves here to
    mentioning the results.

    \sk In the $A$-model, one can only construct three-dimensional $D$-branes
    wrapping so-called ``Lagrangean'' submanifolds of $\cM$. Here,
    ``Lagrangean'' simply means that the K\"ahler form $\go$ vanishes on this
    submanifold. In the $B$-model, one can construct $D$-branes of any even
    dimension, as long as these branes wrap {\em holomorphic} submanifolds of
    $\cM$.
    
    \sk Just like in ordinary string theory, when we consider open topological
    strings ending on a D-brane, there should be a field theory on the brane
    worldvolume describing the low-energy physics of the open strings.
    Moreover, since we are studying topological theories, one may expect such
    a theory to inherit the property that it only depends on a restricted
    amount of data of the manifolds involved. A key example is the case
    of the $A$-model on the deformed conifold,
    \be
     \cM = T^* S^3,
    \ee
    where we wrap $N$ D-branes on the $S^3$ in the base. (One can show that this
    is indeed a Lagrangean submanifold.) In ordinary string theory, the
    worldvolume theory on $N$ D-branes has a $U(N)$ gauge symmetry, so putting
    the ingredients together we can make the guess that the worldvolume theory
    is a three-dimensional, topological field theory with $U(N)$ gauge symmetry.
    There is really only one candidate for such a theory: the Chern-Simons
    theory which we discussed in section \ref{sec:chernsimons}. Recall that it
    consists of a single $U(N)$ gauge field, and has the action
    \be
     S = \frac{k}{4 \pi} \int_{S^3} \Tr \left( A \wedge d A + \frac{2}{3} A
     \wedge A \wedge A \right).
    \ee
    In \cite{Witten:1992fb} (before the invention of D-branes!), E.~Witten
    showed that this is indeed the theory one obtains. In fact, he showed even
    more: this theory actually describes the {\em full} topological string field
    theory on the D-branes, even without going to a low-energy limit.
    
    \sk Let us briefly outline the argument that gives this result. In his
    paper, Witten derived the open string field theory action for the open
    A-model topological string; it reads
    \be
     S = \int \Tr \left( \cA * Q_A \cA + \frac{2}{3} \cA * \cA * \cA
     \right).
    \ee
    (We will not worry about the precise prefactor or its quantization here.)
    The form of this action is of course very similar to Chern-Simons theory,
    but its interpretation is completely different: $\cA$ is a ``string
    field'' (a wave function on the space of all maps from an open string to
    the space-time  manifold), $Q_A$ is the topological symmetry generator,
    which has a natural action on the string field, and $*$ is a certain
    noncommutative product. In a way analogous to what we discussed in chapter
    \ref{sec:twisting}, Witten shows that the topological properties of the
    theory imply that only the constant maps contribute, so $\cA$ becomes  
    a field on $\cM$ -- and since open strings can only end on D-branes, it
    actually becomes a field on $S^3$. Moreover, recall that $Q_A$ can be
    interpreted as a de Rham differential. Using these observations and the
    precise definition of the star product one can indeed show that the string
    field theory action reduces to Chern-Simons theory on $S^3$.

   \subsubsection{General idea of open/closed dualities}
    \label{sec:AdSCFT}
\begin{figure}[ht]
     \begin{center}
      \includegraphics[height=5cm]{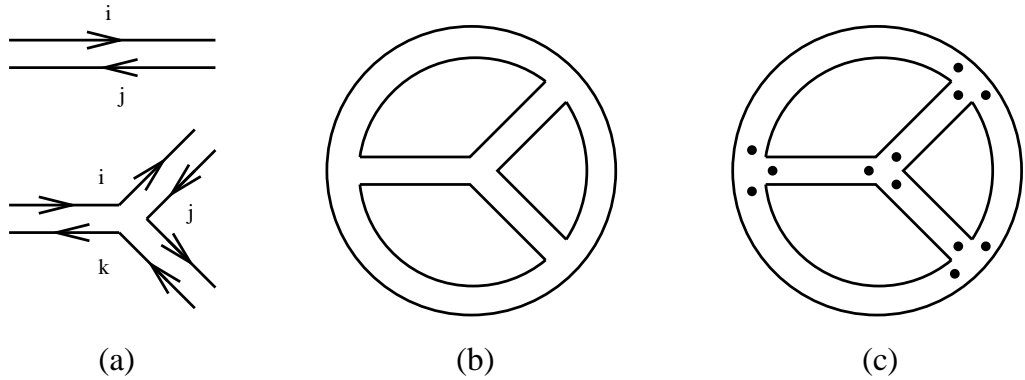}
     \end{center}
     \caption{(a) The propagator for a component of a matrix-valued field is
     usually drawn as a double line, where each line is labelled by one of the
     two matrix indices and the two indices are distinguished by giving the lines an
     orientation. Using this notation, the three-point coupling following from
     matrix multiplication can be drawn as in the bottom of this figure. (b) A
     Feynman diagram appearing in the perturbative expansion of the partition
     sum. The indices and orientations are left out here, but one should sum
     over all four labels associated with the closed lines. (c) By drawing a
     dot at the end of each edge, one sees that in this case $2E = 3V_3$.}
     \label{fig:doubleline}
    \end{figure}

    It was pointed out by G.~'t~Hooft in \cite{'tHooft:1973jz} that $U(N)$
    gauge theories, when we expand their results in $1/N$, are very similar
    to closed string theories. Let us very briefly recall his argument. If we
    have a gauge
    theory with only adjoint fields (i.e.\ where all fields are matrices with
    two indices on which the gauge group acts), one can draw Feynman diagrams
    for its partition function using the double-line notation, where each index
    is drawn as an oriented line -- see figure \ref{fig:doubleline}a. In figure
    \ref{fig:doubleline}b, we draw an example of such a diagram.
    
\begin{figure}[ht]
     \begin{center}
      \includegraphics[height=4cm]{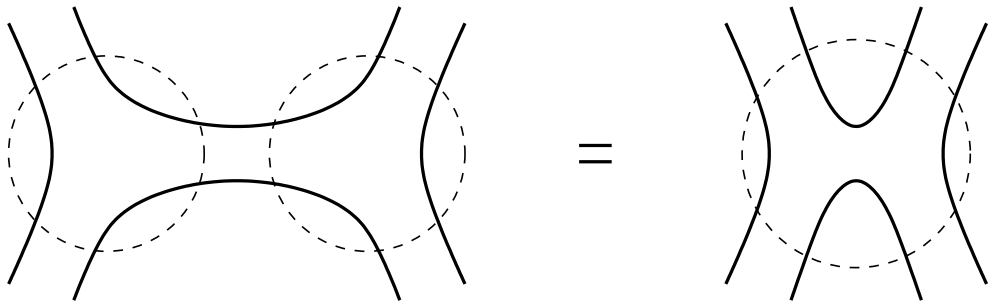}
     \end{center}
     \caption{In open string theory, two three-point vertices are conformally
     the same thing as a single four-point vertex.}
     \label{fig:fourpoint}
    \end{figure}

    \sk In the example we have only drawn three-point vertices, but of course the
    theory may contain $p$-point vertices for $p > 3$ as well. (We will not
    worry about renormalizability here.) Each of these vertices will come with
    its own coupling constant $\gk_p$. Now, we want to think of diagrams such
    as figure \ref{fig:doubleline}b as open string worldsheets. In particular,
    this means that we want to think of two nearby three-point vertices and a
    single four-point vertex as the same, since they are the same thing up to a
    conformal transformation -- see figure \ref{fig:fourpoint}. Similarly,
    three connected three-point vertices should be the same as a five-point
    vertex, and so on. In terms of the coupling constants, this means that we
    have to require that
    \be
     \gk_p = g^{p-2}
    \ee
    for some ``universal'' coupling constant $g$. With this notation, a diagram
    has an overall factor
    \be
     N^F g^{\sum (p-2) V_p},
    \ee
    where $V_p$ is the number of $p$-point vertices in the diagram, and $F$ the
    number of closed lines. The factor of $N^F$ arises because each closed line
    in the diagram represent an index which must be summed over, and these
    indices can take $N$ values. Note that we can draw the diagram of figure 
    \ref{fig:doubleline}b on a sphere, and then color the interior of all
    of the closed lines, including the one which is on the outside of the
    diagram. This can be viewed as a triangulation of the sphere, with $V =
    \sum V_p$ vertices, $E$ edges and $F$ faces. The same thing can be done for
    any diagram that can be drawn in a plane. For non-planar diagrams -- one
    where one edge crosses over another one, for example -- we can do the same
    thing, but for this we need to draw the diagram on a Riemann surface of
    higher genus.
    
    \sk There is a very simple relation between the numbers $V_p$ and $E$ that
    always holds:
    \be
     \sum p V_p = 2 E.
    \ee
    This relation can be easily verified by drawing a dot at the ends of
    each edge in a diagram, as in figure \ref{fig:doubleline}c. The two sides
    in the above equation then simply correspond to two different ways of
    counting the total number of dots -- either by counting them edge by edge,
    or by counting them vertex by vertex. Using this formula, the overall
    prefactor of the diagram becomes
    \bea
     N^F g^{2E - 2V} & = & N^{F-E+V} (g^2 N)^{E-V} \ret
     & = & N^{2-2g} \gl^{2g-2+F}.
    \eea
    In the last line, we used the famous formula due to Euler saying that the genus
    of a Riemann surface is related to the number of vertices, edges and faces
    of a triangulation of this surface, by
    \be
     F-E+V = 2 - 2g,
    \ee
    Moreover, we defined the so-called ``'t Hooft coupling'' $\gl = g^2 N$.
    
    \sk We thus find that the free energy (unfortunately also conventionally
    denoted by $F$), which is the sum of connected Feynman diagrams, has the
    structure
    \bea
     F & = & \sum_{g,F} C_{g,F} N^{2-2g} \gl^{2g-2+F} \ret
     & \equiv & \sum_g N^{2-2g} F_g(\gl),
    \eea
    where the last line defines the quantity $F_g(\gl)$. For fixed $\gl$, this
    expression looks exactly like the partition function of a {\em
    closed} string theory with string coupling constant $1/N$ -- provided the
    resummation we have done leads to finite results, of course. In the large
    $N$ limit with $\gl$ fixed, one may therefore conjecture that a gauge
    theory coming from an {\em open} string theory can equally well be
    described by a weakly coupled {\em closed} string theory. The $AdS/CFT$ 
    correspondence is an example where this indeed turns out to be the case.

   \subsubsection{The transition}
    So, the natural question is: can a similar thing be done in the case of the
    Chern-Simons theory on $S^3$? We have seen that it is indeed related to an
    open topological string theory, so what would the corresponding closed
    string theory be? We can again get a hint from the $AdS/CFT$
    correspondence. This correspondence can be derived by studying the
    backreaction of a set of D3-branes on the space-time geometry: one can 
    then describe the same system {\em either} as a field theory on the
    D-branes, {\em or} as a closed string theory in the created background 
    without the D-branes. In the A-model, the target space geometry is encoded
    by the K\"ahler form $\go$. Since $T^* S^3$ has no nontrivial two-cycles, one
    has
    \be
     \int_{S^2} \go = 0
     \label{eq:kahlervolume}
    \ee
    for any two-dimensional submanifold. However, as soon as we introduce
    D-branes on the $S^3$, one can consider a two-sphere ``surrounding'' this
    three-sphere, and this can no longer be shrunk to a point because of the
    presence of the D-branes. Therefore, this cycle could have nonzero
    K\"ahler volumes (\ref{eq:kahlervolume}), and in fact it turns out that it
    does. This is very similar to the case of ordinary D-branes, which create a
    flux through the $n$-spheres surrounding them. Here, one can argue that
    \be
     \int_{S^2} \go = N g_s
     \label{eq:nzkahlervol}
    \ee
    for a two-sphere surrounding $N$ branes. Therefore, if we would want to
    describe the same physics in terms of a geometry without branes, we would
    have to find a geometry with a nontrivial two-cycle which
    satisfies (\ref{eq:nzkahlervol}). Far away from the location of the (former)
    branes, the geometry would however still have to be the same as the
    conifold geometry. The reader must have guessed the result by now: one
    expects that this closed string geometry is the {\em resolved} conifold
    geometry of figure \ref{fig:conifold}c. In \cite{Gopakumar:1998ki},
    R.~Gopakumar and C.~Vafa argued that this is indeed the case by comparing
    several quantities on both sides of this duality. They also gave a sketch
    of a proof of this duality, which was worked out in detail by H.~Ooguri and
    C.~Vafa in \cite{Ooguri:2002gx}.

  \subsection{Topological strings and matrix models}
   \label{ssec:matrixmodels}
   Several relations between topological string theories and matrix models have
   been found. Here, we will mention the so-called Dijkgraaf-Vafa
   correspondence, and the relation between $c=1$ strings, matrix models and
   topological strings -- both of which have been of much interest lately. Very
   recently, an excellent review by M.~Mari\~no \cite{Marino:2004eq} on the first
   subject appeared, so we refer the reader to his lecture notes for more
   information.

   \subsubsection{Zero-dimensional matrix models}
   \label{sec:dijkgraafvafa}
   The simplest example of a matrix model is a ``zero-dimensional quantum field
   theory'' which has a partition function of the form
   \be
    Z = \int D M e^{-\frac{1}{g^2} \mbox{\scriptsize Tr~} W(M)}
   \ee
   where $M$ is an $N \times N$ matrix, $W$ is a function of $M$ (usually a
   polynomial), and the integral is over a certain ``ensemble'' of matrices,
   such as the set of all Hermitean matrices. In the case of a polynomial $W$,
   this theory is a zero-dimensional example of a $U(N)$ gauge theory, since we
   have the symmetry
   \be
    M \to U M U^\dagger
    \label{eq:gaugetrans}
   \ee
   where $U$ is a unitary $N \times N$ matrix. Thus, the arguments
   we used in section \ref{sec:AdSCFT} hold here as well, and one may expect
   this theory in the large $N$ limit to be described by a string theory. Since
   the above ``gauge theory'' is very simple, we expect that the resulting
   string theory will also be very simple. It turns out that for some matrix
   models this simple string theory actually is a topological string theory.

   \sk If $M$ is Hermitean\footnote{Actually, in many of the examples that we
   will mention, $M$ is a more general complex matrix. This leads to some extra
   subtleties in the calculations, but the general form of the results does not
   change.}, we can diagonalize it using a gauge transformation
   (\ref{eq:gaugetrans}). Therefore, it would seem that after gauge fixing,
   one could write the matrix model as
   \be
    Z \buildrel{?}\over{=} \int \prod_{i=1}^N d \gl_i \, e^{-\frac{N}{\gl}
    \sum_{i=1}^{N} W(\gl_i)}
   \ee
   where $\gl_i$ for $i=1 \cdots N$ are the matrix eigenvalues, and we wrote
   the expression in terms of the ``large $N$'' variables $N$ and the 't Hooft
   coupling $\gl = g^2 N$. Again, we used conventional notations from the
   literature, but be aware that $\gl$ and $\gl_i$ are completely unrelated
   objects! The above formula turns out to be not quite true, since one has to
   include the Jacobian of the change of variables in the path integral
   measure. Doing this carefully, one finds
   \be
    Z = \int \prod_{i=1}^N d \gl_i \, \gD^2 ( \vec{\gl} ) \, e^{-\frac{N}{\gl}
    \sum_{i=1}^N W(\gl_i)}
   \ee
   where $\gD$ is the so-called Vandermonde determinant,
   \be
    \gD(\vec{\gl}) = \prod_{i<j} (\gl_i - \gl_j).
   \ee
   Note that in the large $N$ limit, the prefactor of the action grows, and
   therefore the semiclassical approximation should be quite good. By writing
   the Vandermonde determinant as $\exp (\ln \gD)$ and considering it as an
   additional term in the action we see that the equation of motion for $\gl_i$
   is
   \be
    W'(\gl_i) - \frac{2 \gl}{N} \sum_{j \neq i} \frac{1}{\gl_i - \gl_j} = 0.
   \ee
   If we only had the first term, all eigenvalues would sit at extrema of the
   potential. Due to the small second term, the eigenvalues will slightly repel
   (though with infinite strength if their distance becomes zero!), and
   will in the large $N$ limit fill out ``cuts'' around these extrema.

   \sk What is the closed topological string theory corresponding to this
   model? It can be constructed by looking at the noncompact Calabi-Yau
   manifold
   \be
    y^2 + u^2 + v^2 + (W'(x))^2 = 0
   \ee
   inside $\bC^4$. Note that near the extrema of $W(x)$, this manifold looks
   like the conifold we have considered before. (We are assuming that $W'$ does
   not have double zeroes.) In particular, if $W$ is a
   polynomial of degree $n+1$, it has $n$ extrema, and there are $n$
   conifold singularities at these extrema. Again, we can choose to
   get rid of these singularities by either resolving or deforming them. When
   we resolve them, they get replaced by $\bC P^1$s, and we can study the
   $B$-model topological string theory on this resolved geometry. The reason for
   choosing the resolved geometry is that now we can wrap $N$ $B$-type D2-branes
   on the $\bC P^1$s, $N^i$ on each one such that
   \be
    \sum_{i=1}^n N^i = N.
   \ee
   Now, one can again write down the topological string field theory on this
   target space, this time for the $B$-model, and see what it results in. One
   might expect the theory to reduce to a two-dimensional theory now, but in
   fact it was shown by R.~Dijkgraaf and C.~Vafa in \cite{Dijkgraaf:2002fc}
   that the theory reduces even further, and that the string field can be
   reduced to an ordinary matrix. That is, the ``string field theory'' for this
   model is nothing but a matrix model! Dijkgraaf and Vafa show that it is
   precisely the matrix model we have constructed before, expanded around the
   extremum where $N_i$ eigenvalues sit at the $i$th minimum. On the other
   hand, we can perform a geometric transition at each point, which replaces
   the $\bC P^1$s with branes by $S^3$s, and the open topological string theory
   by a closed one. Thus, we find that this closed topological string theory
   should be holographically dual to the matrix model we discussed. One can
   in fact view the deformed singularities as a direct analogue of the ``cut''
   on which the matrix model eigenvalues sit.

   \subsubsection{Matrix quantum mechanics}
   \label{sec:mqm}
   Besides the matrix integrals we mentioned above, the next simplest example
   of a matrix model is a matrix quantum mechanics, with a path integral of the
   form
   \be
    Z = \int D M(t) e^{- \frac{1}{g^2} \mbox{\scriptsize Tr} (\half \dot{M}^2 - W(M))}
   \ee
   where now $M(t)$ is a {\em time-dependent} matrix. Surprisingly enough, by
   diagonalizing the matrices as in the previous example, such matrix models
   can sometimes still be solved exactly, and again they turn out to be
   holographically dual to simple string theories. To be precise, one finds
   that the dual string theories are strings with a two-dimensional target
   space! These two-dimensional strings, called\footnote{The name arises
   because one can also construct these strings by starting from ``ordinary''
   strings in {\em one} dimension -- which have central charge $c=1$ -- and
   remove the conformal anomaly by making the overall scale of the metric
   dynamical.} $c=1$ strings, are not quite the ordinary strings that 
   we are used to -- for example, the string coupling in these theories is
   position-dependent. This difference turns out to be crucial, since it
   ensures that, even though the theory is not ten-dimensional, there is no
   conformal anomaly. A good review on the relation between matrix quantum
   mechanics and two-dimensional strings is the 1991 paper by I.~Klebanov
   \cite{Klebanov:1991qa}. The recent interest in the subject started with the
   work of J.~McGreevy and H.~Verlinde \cite{McGreevy:2003kb}, T.~Takayanagi
   and N.~Toumbas \cite{Takayanagi:2003sm} and M.~Douglas, I.~Klebanov,
   D.~Kutasov, J.~Maldacena, E.~Martinec and N.~Seiberg \cite{Douglas:2003up},
   where it was shown how to generalize the old results to the case of
   supersymmetric strings, and how to give them a holographic interpretation.

   \sk Here we are not really interested in two-dimensional strings, since
   however simple they may be, they are not topological. There is however an
   exception to this, as was shown in \cite{Ghoshal:1995wm} by D.~Ghoshal and C.~Vafa. The
   two-dimensional target space has one space- and one time-direction, which is
   usually taken to be Euclidean, and one may compactify this Euclidean
   time-direction on a circle. One then finds that the theory has T-duality-like
   properties, and in particular there is a self-dual radius. Ghoshal and Vafa
   showed that at this radius, the $c=1$ string is in fact equivalent to
   a topological string! However, this topological string is not just some
   twisted version of the $c=1$ string. This should not come as a surprise:
   first of all the $c=1$ string has no supersymmetry, and thus no
   $R$-symmetry, so there is nothing to twist with, and secondly we saw that
   only topological strings with a six-dimensional target space give really
   interesting physics. In fact, what Ghoshal and Vafa showed was that the
   $c=1$ string at self-dual radius was equivalent to the topological $B$-model
   string on our by now familiar deformed conifold! Unfortunately, their
   precise arguments are somewhat beyond the scope of these lectures.

   \sk As for the matrix quantum mechanics, we can of course apply the same
   procedure there, and compactify its time direction on a circle with the
   self-dual radius. Thus, we find that there are actually two different matrix
   models describing topological strings on the conifold: the ordinary matrix
   integral that we discussed in the previous subsection, and the matrix quantum
   mechanics that we discussed here. Both of these descriptions have led to
   interesting further applications, as we will see in the next two sections.

  \subsection{$N=1$ F-terms}
   The relations between topological strings and matrix models are
   very interesting from a mathematical point of view, but as physicists we
   would also like to know whether we can extract any physical information from
   these correspondences. For the models of section \ref{sec:dijkgraafvafa},
   the answer to this question was given in a beautiful series of papers by
   R.~Dijkgraaf and C.~Vafa \cite{Dijkgraaf:2002fc, Dijkgraaf:2002vw,
   Dijkgraaf:2002dh}: one can actually use the correspondence to compute 
   $F$-terms in $N=1$ supersymmetric gauge theories!
   
   \sk That the theories constructed above are related to $N=1$ theories can be
   seen as follows. We can ``geometrically engineer'' such theories by taking
   the $N=2$ theory we obtained after compactifying type II string theories on
   a Calabi-Yau, and then breaking half of the supersymmetry. A well-known way
   to break half of the supersymmetries in string theory is by including
   properly aligned D-branes. In particular, if we use D5-branes in type IIB
   theory which wrap Minkowski space and the non-trivial two-cycles in the
   geometry, the full theory will be $N=1$ supersymmetric, and the internal
   theory will exactly be the untwisted version of the open topological string
   theories appearing in the Dijkgraaf-Vafa correspondence.
   
   \sk So what does the topological string theory compute for us? To find this
   out, we go through the geometric transition and replace the branes in the
   topological theory by fluxes through three-cycles. How do we see these
   fluxes in the four-dimensional theory? For a similar case, we know the
   answer: the flux of the holomorphic three-form $\gO$ through the
   A-cycles is given by the values of the four-dimensional scalar fields $X^I$.
   It turns out that the other fluxes similarly turn on background values for
   the superpartners of $X^I$. In particular, in the case we are interested in,
   the $N^I$ units of flux coming from the $N^I$ D-branes wrapped around the
   $I$th $\bC P^1$ in the dual theory are encoded in four dimensions as a
   background value for a $\gt^2$ component in the superfield corresponding to
   $X^I$. Recall from section \ref{sec:fterms} that in the $N=2$ theory we had the
   term
   \be
    \int d^4 x d^4 \gt F_0(X^I).
   \ee
   Now that we have broken supersymmetry down to $N=1$, it is no longer natural
   to write the fields in terms of $N=2$ superfields. Instead, we should write
   them in terms of $N=1$ superfields which have only two $\gt$-components. In
   terms of the action, this is very easy to do: one simply integrates over two
   of the four $\gt$-components. One can choose these components to be exactly
   the ones that appear with the superpartner mentioned above, and thus we find
   that we can rewrite the above term in ``$N=1$ language'' as
   \be
    \int d^4 x d^2 \gt N^I \frac{\d F_0}{\d X^I}.
   \ee
   If we would only have fluxes through the $A$-cycles, this would be the full
   answer, but there may of course be fluxes through the $B$-cycles as well. In
   the noncompact geometries we study, these cycles are noncompact as well, and
   as a result the fluxes $\tau_I$ through them are not quantized. Since the
   $\tau_I$ are fluxes through the ``conjugate'' cycles, one may expect that
   they appear in the low-energy theory multiplying the ``conjugate'' variables
   as well -- that is,
   \be
    W_{flux}(X^I) = N^I \frac{\d F_0}{\d X^I} + \tau_I X^I.
    \label{eq:superpot1}
   \ee 
   One can indeed show that this equation is correct. Now, we would like to
   translate this back to the theory before the geometric transition. As we
   explained above, here the effective four-dimensional theory is an $N=1$
   gauge theory. It is known for such theories that, for physical calculations,
   the ``microscopic'' definition of the theory is not the most useful one.
   Instead, one should use the theory of renormalization to obtain an
   ``effective'' field theory in terms of different fields. This field theory
   can then be used to describe the physics at the energy scales we are really
   interested in. Going through this whole procedure would put us on a very
   long side track, so let us just mention the most important results one
   finds. The effective theory contains so-called ``glueball fields'' $S^I$,
   where $I$ labels the different gauge groups of the theory. In the effective
   action, there is a potential for this glueball field of the form
   \be
    W_{eff}(S^I) = N^I \frac{\d F}{\d S^I} + \tau_I S^I,
    \label{eq:superpot2}
   \ee
   where now $N^I$ is the rank of the corresponding gauge group, $\tau_I$ is
   the complexified coupling constant of this gauge group, and $F$ is the
   prepotential of the microscopic theory. It seems very natural
   to identify the different quantities in (\ref{eq:superpot1}) with the ones in
   (\ref{eq:superpot2}). For the label $I$, for example, we can immediately see
   that this identification is correct. In (\ref{eq:superpot2}), $I$ labels the
   different gauge groups. These gauge groups arise from the stacks of branes
   we wrapped around the $\bC P^1$s in the microscopic theory. But each of these
   $\bC P^1$s becomes an $S^3$ $A$-cycle after the geometric transition. In the
   closed topological string, the number of $A$-cycles is precisely the number
   of variables $X^I$, so we see that the label $I$ in (\ref{eq:superpot1})
   indeed is the same label. Moreover, the rank $N^I$ of the gauge groups in
   (\ref{eq:superpot2}) equals the number of $D$-branes wrapping the $I$th $\bC
   P^1$ in the microscopic theory, which after the geometric transition is the
   amount of flux through the $A^I$-cycle -- which is precisely $N^I$ in
   (\ref{eq:superpot1})! Showing that the $\tau_I$  in the two equations are the
   same is more difficult, but can also be done. The conjecture by Dijkgraaf
   and Vafa (which has been proved since then) is now obtained by putting all
   of the pieces together: they claim that {\em the planar matrix model free
   energy equals the genus zero topological string free energy and can be
   used to calculate the effective glueball superpotential in the corresponding
   $N=1$ gauge theory according to (\ref{eq:superpot2})}. They also predict
   similar relations between the higher genus topological string free energies
   and terms in the effective $N=1$ action. A general proof of these relations
   has not been given yet, but a lot of evidence for their correctness has been
   found.
   
   \sk One important check on the above proposal is that it should reproduce
   the so-called Veneziano-Yankielowicz term in the glueball superpotential.
   This is a universal term of the form
   \be
    W \sim NS \log S
   \ee
   that appears in any glueball superpotential. From the geometric point of
   view, one can indeed show that if
   \be
    \int_A \gO = S
   \ee
   is the flux through some compact $A$-cycle, then
   \be
    \int_B \gO \sim \frac{1}{2 \pi i} S \log S
   \ee
   is one term in the flux through the dual noncompact $B$-cycle. One reason for
   this is that if we change $S$ by a phase factor $e^{2 \pi i}$, the $B$-cycle
   should at most change by an integer number of $A$-cycles, so that its
   intersection properties are not spoiled. As we know, the above term should
   be identified with $\d F / \d S$, and inserting this in
   (\ref{eq:superpot2}), we indeed see the Veneziano-Yankielowicz term
   appearing in the superpotential\footnote{We have not been careful with the 
   numerical prefactors here. The reason for this is that in our story, the
   overall normalization of $\gO$ has not been fixed.}. Exactly the same term
   can be found directly from the matrix model as well.

  \subsection{Topological strings and black holes}
   There have been applications of topological string theories to black holes in
   both 4 and 5 dimensions. The 5-dimensional relation comes from
   compactifying M-theory on a Calabi-Yau manifold; we will not discuss it here,
   but focus on the more recent 4-dimensional case which arises when we
   compactify type II string theories on a Calabi-Yau.
   
   \sk Up till now, the four-dimensional space in our compactifications has 
   been a flat Minkowski space. However, one can just as well choose a
   different four-dimensional space, and one very interesting possibility
   is to study a four-dimensional black hole background. That is, we take
   our ten-dimensional space-time to be of the form
   \be
    BH_4 \times CY_6.
   \ee
   Because of the supersymmetry of the problem, the four-dimensional black
   hole turns out to be extremal, meaning that its mass is given by a specific
   function of its charges, and that any other (non-supersymmetric)
   four-dimensional black hole with the same charges would have a larger mass.
   A physically very interesting quantity to calculate for such an extremal black
   hole is its entropy as a function of its charges. 
   
   \sk As we have seen, there are $2h^{2,1} + 2$ four-dimensional gauge fields,
   where half of these fields are the magnetic duals of the other half, and the
   black hole can have a charge for each of these fields. We denote the
   electric and magnetic charges by
   \be
    (P^I, Q_I).
   \ee
   Just like in the Minkowski case, the four-dimensional effective theory is an
   $N=2$ supergravity theory. An important fact about $N=2$ black holes is
   that they exhibit a so-called ``attractor mechanism'' which determines the
   values of the scalar fields $X^I$ at the horizon in terms of the charges.
   That is, even though we can of course choose the $X^I(x^\mu)$ (which are the
   position-dependent moduli of the Calabi-Yau) freely at infinity, the $N=2$
   supergravity equations of motion will ``attract'' them to certain specific
   values at 
   the black hole horizon. To be precise, one finds the following relation:
   \bea
    P^I & = & \int_{A^I} \Re \gO = \Re X^I \ret
    Q_I & = & \int_{B_I} \Re \gO = \Re F_I.
   \eea
   A well-known result by J.~Bekenstein and S.~Hawking states that the entropy
   of a black hole is proportional to the area of its horizon. This area is a
   function of the black hole mass, or in the extremal case, of its charges.
   With the above equations, we can in turn express these charges in terms of
   the moduli at the horizon, and finally, these moduli are given by the
   periods of the holomorphic three-form $\gO$. Putting all of these steps
   together, one finds a simple formula for the entropy in terms of the
   three-form,
   \be
    S_{BH} = \frac{i \pi}{4} \int_\cM \gO \wedge \gObar.
    \label{eq:bhentropy}
   \ee
   The holomorphic three-form is one of the key ingredients of the $B$-model
   topological string. The immediate question raised by this formula is thus:
   can we calculate the entropy using topological string theory? The answer is
   positive, and turns out to be relatively simple: using the Riemann bilinear
   identity on finds that the above quantity is the Legendre transform of the
   Calabi-Yau prepotential $F_0$! In other words, one can express the entropy
   of the black hole in terms of the genus zero free energy of the $B$-model
   topological string.
      
   \sk What about higher genera? It turns out that one step in the derivation
   of (\ref{eq:bhentropy}) is not quite precise: the Bekenstein-Hawking
   relation between the black hole entropy and its mass has quantum
   corrections. These corrections were incorporated in the paper
   \cite{Ooguri:2004zv} by H.~Ooguri, A.~Strominger and C.~Vafa. There, it was
   shown that they can be exactly expressed in terms of the higher loop topological
   string amplitudes! The simplest way to write the result is in terms of the
   black hole partition function (from which one can easily calculate its
   entropy): for this partition function Ooguri, Strominger and Vafa reached
   the conclusion that
   \be
    Z_{BH} = |Z_{top}|^2.
   \ee
   Thus, the $N=2$ black hole partition function is simply the absolute value
   squared of the $B$-model partition function on the Calabi-Yau at the horizon!
   
  \subsubsection{Black holes and matrix models}
   Since topological strings describe black holes, and topological strings are
   often equivalent to matrix models, it is natural to ask if black holes can
   also be described directly by matrix models. In \cite{Danielsson:2004ti},
   U.~Danielsson, M.~Olsson and the present author have shown that this is
   indeed the case. In fact, here one finds a nice physical application of the
   matrix quantum mechanics models that we introduced in section \ref{sec:mqm}.

   \sk In particular, in \cite{Danielsson:2004ti} the matrix quantum mechanics corresponding to
   the type 0A string theory in two dimensions is studied. This theory is a close
   cousin of the $c=1$ string theory, but it has some additional fields, and
   the corresponding matrix model has the non-polynomial potential $W(x) = -x^2
   + 1/x^2$. In general, it is known that a matrix model with a Hamiltonian
   $H(p,x)$ is related to a topological string on a noncompact manifold of
   the form
   \be
    uv + H(p,x) = 0,
   \ee
   as was shown by M.~Aganagic, R.~Dijkgraaf, A.~Klemm, M.~Marino and C.~Vafa
   in \cite{Aganagic:2003qj}. Actually, it turns out that because of the
   nonpolynomial potential, the above expression is not quite correct, but it
   is not hard to correct it. In \cite{Danielsson:2004ti} this was worked out in
   detail, and it was found that the partition function of this corrected
   topological string theory indeed equals that of the matrix model, and that
   both theories reproduce the entropy of the black hole that was conjectured
   by Ooguri, Strominger and Vafa:
   \be
    Z_{BH} = |Z_{top}|^2 = Z_{MM}.
   \ee
   An interesting consequence is that, as we mentioned in section \ref{sec:mqm}, the
   matrix model which describes the black hole has a certain radius in the
   (Euclidean) time-like direction; this can be interpreted as the fact that it
   has a certain temperature. On the other hand, the black hole is extremal,
   which means that it does not radiate and hence does 
   not have a temperature. For this reason, thermodynamic quantities of one
   theory cannot be directly compared to thermodynamic quantities of the other
   theory, even though the partition functions are the same. For example, by
   careful consideration of the thermodynamics in both theories, it turns out
   that the crucial relation between the matrix model free energy and the black
   hole entropy is
   \be
    S_{BH} = - \frac{F_{MM}}{T_{MM}}.
   \ee

  \subsection{Topological M-theory?}
   As we have seen in this chapter, just like in ordinary string theory, there
   are many relations between apparently different topological string theories.
   There is mirror symmetry, relating the $A$-model on one manifold to the
   $B$-model on another; there are geometric transitions relating the same
   types of theories on different manifolds, and there are holographic
   relations between topological string theories and field theories. There is
   even a conjectured $S$-duality, relating the topological $A$- and $B$-models
   on the {\em same} manifold, that we have not discussed above.
   
   \sk In ordinary string theory, the main reason for the presence of all these
   dualities is the existence of M-theory: the eleven-dimensional theory of
   which all lower-dimensional theories can be derived. A good microscopic
   definition of M-theory has not been found yet, but there is an overwhelming
   amount of evidence for its existence.
   
   \sk Thus, it has been conjectured by many people that there is also such a
   thing as ``topological M-theory'' -- a possibly seven-dimensional
   topological theory that encodes all of the lower-dimensional topological
   theories and their relations. Very recently, using mathematical ideas of
   N.~Hitchin, a microscopic definition of such a theory was proposed by
   R.~Dijkgraaf, S.~Gukov, A.~Neitzke and C.~Vafa in \cite{Dijkgraaf:2004te}.
   It is still too early to tell whether this theory will indeed give a unified
   description of the different topological string and field theories. Perhaps
   this will be a good subject for a future set of lecture notes.

\end{document}